\documentclass[%
 reprint,
nofootinbib,
 amsmath,amssymb,
 aps,
]{revtex4-2}

\usepackage{graphicx}
\usepackage{dcolumn}
\usepackage{bm}

\usepackage{graphics}
\usepackage{amsmath,amssymb,braket}
\usepackage{mathtools}
\usepackage{fancyvrb,comment,array}
\usepackage{booktabs,multirow}

\allowdisplaybreaks[1]
\usepackage{color}
\newcommand{\Tr}{\mathrm{Tr}}
\newcommand{\diag}{\mathrm{diag}}

\newcommand{\LC}{\mathrm{LC}}
\newcommand{\HM}{\mathrm{HM}}
\newcommand{\La}{\mathcal{L}}
\newcommand{\OC}{\mathcal{O}}

\newcommand{\Slash}[1]{\ooalign{\hfil/\hfil\crcr$#1$}}

\def\simge{\mathrel{
\rlap{\raise~0.511ex \hbox{$>$}}{\lower~0.511ex \hbox{$\sim$}}}}
\def\simle{\mathrel{
\rlap{\raise~0.511ex \hbox{$<$}}{\lower~0.511ex \hbox{$\sim$}}}}
\def\bigs{\mathrel{
\rlap{\raise~0.531ex \hbox{$>$}}{\lower~0.531ex \hbox{$<$}}}}

\usepackage{booktabs}
\usepackage[normalem]{ulem}
\usepackage{multirow}

\usepackage{hyperref}

\begin{document}

\preprint{APS/123-QED}

\title{Analysis of bound and resonant states of 
doubly heavy tetraquarks with spin $J\leq 2$}

\author{Manato~Sakai}
\email[]{msakai@hken.phys.nagoya-u.ac.jp}
\affiliation{Department of Physics, Nagoya University, Nagoya 464-8602, Japan}
\author{Yasuhiro~Yamaguchi}
\email[]{yamaguchi@hken.phys.nagoya-u.ac.jp}
\affiliation{Department of Physics, Nagoya University, Nagoya 464-8602, Japan}
\affiliation{Kobayashi-Maskawa Institute for the Origin of Particles and the Universe, Nagoya University, Nagoya, 464-8602, Japan}

\date{\today}

\begin{abstract}

Recently, a number of exotic hadrons have been reported in the experiments, and most of these states lie slightly below the threshold. 
Therefore, these states are considered to be hadronic molecules composed of mesons or baryons. 
{In 2022, the doubly charmed tetraquark $T_{cc}$ was reported by the LHCb experiment, which is considered to be composed of two heavy quarks and two light antiquarks, and located slightly below the  $DD^\ast$ threshold. 
The observation motivates us to} study the bound and resonant states of the doubly heavy tetraquarks as two meson systems. 
We employ the one boson exchange potential as an interaction between two mesons, where one free parameter has been determined to reproduce $T_{cc}$ reported by the LHCb experiment. 
{In our previous work, bound states of {$D^{(\ast)}D^{(\ast)}$} and {$B^{(\ast)}B^{(\ast)}$} molecules were studied, where we respected the heavy quark spin symmetry (HQS)}. Using the same framework, we study not only bound but also resonant states with $J\leq 2$ in this paper. 
Furthermore, we discuss the HQS partner structures of $T_{cc}$ and $T_{bb}$ obtained in our study by introducing the light cloud basis.

\end{abstract}

\maketitle


\section{\label{sec:intro}Introduction}

Some experiments, such as Belle, BaBar, BESIII and LHCb, have reported many exotic hadrons, which cannot be explained by the ordinary hadron pictures, {three quark states as baryons and quark-antiquark states as mesons}.
For instance, in the heavy quark sector, $XYZ$, $P_c$ and $T_{cc}$ {have been} reported, since the discovery of $X(3872)$ by the Belle experiment in 2003~\cite{Belle:2003nnu,BaBar:2005hhc,Belle:2007dxy,Belle:2007hrb,BaBar:2008bxw,LHCb:2015yax,LHCb:2019kea,LHCb:2020bwg,BESIII:2020qkh,LHCb:2020jpq,LHCb:2021chn,LHCb:2021vvq,LHCb:2021auc}. 
{Structures of exotic hadrons have attracted a lot of interest. In particular, there have been many works on multiquark structures. As a hadronic state including four or five quark systems, compact tetraquarks and pentaquarks have been studied using the few-body calculations, effective models, etc. Near thresholds, the appearance of a hadron composite state, called hadronic molecules, has been discussed. Hadronic molecules are analog states to the deuteron as a proton-neutron bound state with the small binding energy, bound by the colorless nuclear force. Hadrons with gluons as explicit constituents have also been studied as hybrid states and glueballs. In contrast to such hadronic states, peak structures on experimental data, indicating the existence of exotic hadrons, would also be explained by the kinematical effects such as thresholds cusp and triangle singularity. Studying the exotic hadron structures and the origin of the peak structures in the experimental data would be related to understanding phenomena of the low-energy Quantum Chromodynamics (QCD), such as the color confinement and the hadron interactions.}

{The doubly charmed tetraquark $T_{cc}$ has properties that differ from other exotic hadrons: it is an open charm state composed of two charm quarks and two light antiquarks. 
{It is obvious that this tetraquark state cannot be explained by the ordinary hadron pictures.}
Thus, $T_{cc}$ is a genuine exotic hadron, which is why it has been extensively studied for a long time using several models such as the quark model~\cite{Ader:1981db,Ballot:1983iv,Zouzou:1986qh,Meng:2020knc,Meng:2021yjr,Yan:2018gik,Mutuk:2023oyz,Mutuk:2024vzv}, hadronic molecule~\cite{Tornqvist:1993ng,Ohkoda:2012hv,Li:2012ss,Chen:2022asf,Wang:2021yld,Wang:2021ajy,Ren:2021dsi,Asanuma:2023atv,Sakai:2023syt,Ren:2023pip,Cheng:2022qcm,Lin:2022wmj}, Lattice QCD~\cite{Ikeda:2013vwa,Padmanath:2022cvl,Lyu:2023xro}, heavy quark spin symmetry~\cite{Eichten:2017ffp,Cheng:2020wxa,Tanaka:2024siw}, string model~\cite{Andreev:2021eyj}, and QCD sum rules~\cite{Navarra:2007yw,Du:2012wp,Agaev:2021vur,Xin:2021wcr,Xin:2022bzt}. 
}

For experimental studies, 
the doubly charmed tetraquark ${T_{cc}(cc\bar{u}\bar{d})}^+$ was reported by the LHCb experiment in 2022~\cite{LHCb:2021auc,LHCb:2021vvq}. 
This state is consistent with an isoscalar state having $J^P = 1^+$. 
The mass differences {(binding energies)} measured from $D^{\ast +}D^0$ threshold and decay widths {have been analyzed by two different methods,   
the Breit-Wigner fitting with~\cite{LHCb:2021vvq}
\begin{align}
    \delta m_{\mathrm{BW}} &= 
    -273 \pm 61 \pm 5^{+11}_{-14}\ \mathrm{keV},  \\
    \Gamma_\mathrm{BW} &= 
    410 \pm 165 \pm 43 ^{+18}_{-38}\ \mathrm{keV}, 
\end{align}
and the pole analysis with~\cite{LHCb:2021auc} 
\begin{align}
    \delta m_\mathrm{pole} &= 
    -360 \pm 40 ^{+4}_{-0}\ \mathrm{keV}, \\
    \Gamma_\mathrm{pole} &= 
    48 \pm 2^{+0}_{-14} \ \mathrm{keV}. 
\end{align}
Both values of the binding energies measured from $D^{\ast+}D^0$ threshold are small, which {leads} us to analyze this state using the hadronic molecular model.

Many exotic hadrons including $T_{cc}$ are composed of heavy quarks such as $c$ and $b$, which motivates us to respect the heavy quark spin symmetry (HQS)~\cite{Neubert:1993mb,Grinstein:1995uv,Casalbuoni:1996pg}. 
This symmetry {emerges because} the spin interaction terms in the heavy quark Lagrangian are suppressed by $\mathcal{O}(1/m_Q)$. 
For a heavy meson with a heavy quark and a light antiquark, the mass of a pseudoscalar meson and that of a vector meson are degenerate {in the heavy quark limit due to the HQS}.  
In fact, in the heavy quark sector, the mass difference between $D$ and $D^\ast$ is about $140$ MeV, and that between $B$ and $B^\ast$ is about $45$ MeV, which is smaller than that in the light quark sector. 

{The HQS is considered to help to generate hadronic molecules. Approximate mass degeneracy of heavy mesons induces the coupled channel effects. For $D^{(\ast)}D^{(\ast)}$ systems considered to be constituents of the doubly charmed tetraquarks, thresholds of $DD$, $DD^{\ast}$ and $D^{\ast}D^{\ast}$ are very close to each other. The level repulsion caused by the coupled channel effect produces an attraction to the lower channels. In addition, mixture of $D$ and $D^*$ mesons provides the one pion exchange interaction. The pion exchange is forbidden in the single $DD$ channel because the $DD\pi$ vertex violates the parity conservation. However, the coupled channel $D^{(\ast)}D^{(\ast)}$ systems possess the pion exchange via the $DD^*\pi$ coupling. The one pion exchange potential (OPEP) has been known as a driving force to bind atomic nuclei. In particular, the tensor force of the OPEP produces a strong attraction. In the $D^{(\ast)}D^{(\ast)}$ systems, the OPEP tensor force is considered to be enhanced by the coupled channel effect. These mechanisms of generating attractions are more induced in the bottom sector, because the HQS is realized well.}

{

As mentioned above, the HQS induces mass degeneracies between the heavy hadrons with different angular momenta in the heavy quark limit because the heavy quark spin is a conserved quantity. 
Thus, the division of the total spin into the heavy quark spin $S_Q$ and the light cloud spin $j_\ell$, which is everything but the heavy quark spin, clarifies where the hadrons belong. 
For heavy mesons, the masses of pseudoscalar and vector mesons with $(S_Q,j_\ell) = (1/2,1/2)$ are degenerate, which means that these states belong to the same HQS doublet. 
In fact, the charmed mesons $D$ and $D^\ast$ approximately belong to the HQS doublet. 
Similarly, for baryons, the baryons with spin $1/2$ and spin $3/2$ whose spin structure is $(S_Q,j_\ell)=(1/2,1)$ belongs to the HQS doublet in the heavy quark limit. 
In our world, $\Sigma_c$ and $\Sigma^\ast_c$ baryons approximately belong to the HQS doublet. 
Indeed the mass difference is $65\,\mathrm{MeV}$, which is smaller than that of light sector baryons. 
On the other hand, $\Lambda_c$ with spin $1/2$ has the spin structure $(S_Q,j_\ell)=(1/2,0)$ in the heavy quark limit, then approximately belongs to the HQS singlet.

In our study, we apply the above discussion to the hadronic molecules $D^{(\ast)}D^{(\ast)}$ and $B^{(\ast)}B^{(\ast)}$ using the light cloud basis. 
Here, the light cloud basis for a hadronic molecule is characterized by the spin of two heavy quarks and light cloud spin. 
The HQS multiplet structures of $\bar{D}^{(\ast)}N$, $P_c$, $T_{cc}$ and $T_{bb}$ have been studied in Refs.~\cite{Yasui:2013vca,Yamaguchi:2014era,Shimizu:2018ran,Sakai:2023syt}.

}

In this study, we investigate the bound and resonant states of the doubly heavy tetraquarks, $T_{cc}$ and $T_{bb}$ with spin $J\leq 2$, while we have studied the bound states of $T_{cc}$ and $T_{bb}$ with spin $J=0,1$ in our previous study~\cite{Sakai:2023syt}. 
We consider the doubly heavy tetraquarks as $D^{(\ast)}D^{(\ast)}$ or $B^{(\ast)}B^{(\ast)}$ molecules, where we employ the one boson ($\pi, \rho, \omega, \sigma$) exchange potential as an interaction between two heavy mesons $P^{(\ast)}=D^{(\ast)}, B^{(\ast)}$. 
This potential respects the heavy quark spin and chiral symmetries and includes the cutoff parameter $\Lambda$ which is determined to reproduce the experimental value of $T_{cc}$ with $0(1^+)$ in our previous study~\cite{Sakai:2023syt}. 
Moreover, we discuss the HQS multiplet structures of $T_{cc}$ and $T_{bb}$ using the light cloud basis. 

This paper is organized as follows. 
In Sec.~\ref{Sec;Formalism}, we construct the one boson exchange potential. 
In Sec.~\ref{Sec;Numerical_Result}, we show the numerical results of the doubly heavy tetraquarks with spin $J\leq 2$. 
We investigate the bound and resonant states of $T_{cc}$ and $T_{bb}$ as hadronic molecules and their HQS multiplet structures. 
Finally, we summarize our results in Sec.~\ref{Sec;Summary}.

\section{Formalism\label{Sec;Formalism}}

In this study, we {analyze bound and} resonant states of {two heavy mesons as} $T_{cc}$ and $T_{bb}$. Interactions between heavy mesons are given by the one boson exchange potentials with respect to the HQS. In this section, first we derive the one boson exchange model from the effective Lagrangians. Using the obtained model, bound and resonant states are studied by solving the Schr\"odinger equation. For analyzing resonances,  we apply the complex scaling method (CSM). 
{which is also reviewed briefly in this section}. 

\subsection{One boson exchange potential}

We construct the effective Lagrangian for the heavy-light mesons by respecting the HQS~\cite{Neubert:1993mb,Grinstein:1995uv,Casalbuoni:1996pg}. 
First, we define the heavy meson field $H_a$ which is written as 
\begin{align}
    H_a &= 
    \frac{1+\Slash{v}}{2}[
    \gamma_\mu P^{\ast\mu} _a - \gamma_5 P_a
    ],  \\
    \bar{H}_a &= 
    \gamma^0 H_a^\dagger \gamma^0 \nonumber \\ &=
    [
    \gamma_\mu P^{\ast\mu\dagger} _a + \gamma_5 P^\dagger _a 
    ] \frac{1+\Slash{v}}{2},
\end{align}
where 
$P$ and $P^\ast$ are the pseudoscalar and vector meson fields, 
$a$ is the isospin index and $v^\mu$ is the four velocity of the heavy meson satisfying $v^2=1$, $v^0>0$. 
Using the heavy meson field, 
we derive the effective Lagrangian for a {light} pseudoscalar meson which is invariant under the HQS and the flavor symmetry: 
\begin{align}
    \La_\pi =&
    ig_\pi \Tr[
    H_b \gamma_\mu \gamma_5 A^\mu_{ba} \bar{H}_a
    ] \nonumber \\ =&
    -\frac{g_\pi}{f_\pi}(P^{\dagger\mu}_a P_b + 
    P^\dagger_{a\mu} P^{\ast\mu}_b
    )\partial_\mu 
    {(\vec{\tau}\cdot\vec{\pi})_{ba}}
    \nonumber \\ &- 
    i\frac{g_\pi}{f_\pi}\epsilon^{\mu\nu\alpha\beta}v_\mu P^{\ast\dagger}_{a\nu}P^\ast_{b\alpha}\partial_\beta(\vec{\pi}\cdot\vec{\tau})_{ba}\label{eq;pi_Lag}.
\end{align}
Here, $A^\mu$ is the axial current written by  
\begin{equation}
    A^\mu = \frac{1}{2}\left[
    \xi^\dagger(\partial^\mu \xi) - 
    \xi(\partial^\mu \xi ^\dagger)
    \right],
\end{equation}
where 
\begin{align}
    \xi &= 
    \exp\left(
    i\frac{\hat{\pi}}{\sqrt{2}f_\pi}
    \right) =
    \exp\left(
    i\frac{\vec{\pi}\cdot\vec{\tau}}{2f_\pi}
    \right), \\ 
    \hat{\pi} &= \frac{1}{\sqrt{2}}\begin{pmatrix}
        \pi^0 & \sqrt{2}\pi^+ \\
        \sqrt{2}\pi^- & - \pi^0
    \end{pmatrix} = 
    \frac{1}{\sqrt{2}}\vec{\pi}\cdot\vec{\tau},
\end{align}
with $f_\pi=93$ MeV and $\vec{\tau}$ being the pion decay constant and the Pauli matrices,  respectively. 
The interaction effective Lagrangian is separated into terms for $P P^\ast\pi$ and $P^\ast P^\ast\pi$ as shown {in} Eq.~\eqref{eq;pi_Lag}.

We also obtain the interaction Lagrangians for the {light} vector mesons ($\rho$ and $\omega$) and $\sigma$ meson~\cite{Casalbuoni:1996pg}: 
\begin{align}
    \La_v =& 
    -i\beta \Tr[
    H_b v^\mu \rho_\mu \bar{H}_a
    ]  + 
    i\lambda \Tr[
    H_b \sigma^{\mu\nu}F_{\mu\nu}(\rho)_{ba}\bar{H}_a
    ] \nonumber \\ =&
    \sqrt{2}\beta g_V (
    P^\ast_b \cdot P^{\ast\dagger}_a - P_b P_a^\dagger
    )v\cdot(\hat{\rho})_{ba} \nonumber \\ & 
    -2\sqrt{2}g_V\lambda \epsilon^{\mu\nu\alpha\beta}v_\mu (
    P^{\ast\dagger}_{\nu a}P_b + P^\dagger_a P^\ast _{\nu b}
    )\partial_\alpha (\hat{\rho}_\beta)_{ba} \nonumber \\ &
    +i2\sqrt{2}\lambda g_V P^{\ast\mu\dagger}_a P ^{\ast\nu}_b (
    \partial_\mu(\hat{\rho}_\nu)_{ba} - \partial_\nu (\hat{\rho}_\mu)_{ba}
    ), \\
    \La_\sigma &= 
    g_\sigma \Tr[
    H\sigma\bar{H}
    ]\nonumber \\ &= 
    2g_\sigma (
    P^{\ast\dagger}_\mu P^{\ast\mu} - P^\dagger P
    )\sigma, 
\end{align}
respectively, where 
\begin{align}
    F_{\mu\nu}(\rho) &= 
    \partial_\mu \rho_\nu - \partial_\nu \rho_\mu +
    [\rho_\mu,\rho_\nu], \\
    g_V &= 
    \frac{m_\rho}{\sqrt{2}f_\pi}, \\
    \rho_\mu &= 
    \frac{ig_V}{\sqrt{2}}\hat{\rho}_\mu,
\end{align}
and $\hat{\rho}_\mu$ is the vector meson fields defined by 
\begin{align}
    \hat{\rho}_\mu &= 
    \frac{1}{\sqrt{2}}\begin{pmatrix}
        \rho^0 + \omega & \sqrt{2}\rho^+ \\
        \sqrt{2}\rho^- & -\rho^0 + \omega
    \end{pmatrix}_\mu \nonumber \\ &= 
    \frac{1}{\sqrt{2}}(
    \vec{\tau}\cdot\vec{\rho}_\mu + \omega_\mu \mathbf{1}).
\end{align}

We derive the one boson exchange potential by calculating the scattering amplitude and using the relation of this amplitude $i\mathcal{M}$ and potentials in momentum space $V(\vec{q})$ as 
\begin{equation}
    V(\vec{q}) =i \frac{i\mathcal{M}}{\sqrt{\Pi_i 2m_i \Pi_f 2m_f}}, 
\end{equation}
where $m_i$ is the initial state mass and $m_f$ is the final state one.  
Therefore we obtain the one boson exchange potential in the coordinate space by the Fourier transformation:
\begin{equation}
    V({r};m) = \int \frac{d^3q}{(2\pi)^3}V(\vec{q}) {F(\vec{q};m)}^2e^{i\vec{q}\cdot\vec{r}}, 
\end{equation}
where we attach the form factor $F(\vec{q};m)$ to each vertex in order to consider the hadronic size: 
\begin{equation}
    F(\vec{q};m ) =
    \frac{\Lambda^2 - m^2}{\Lambda^2 + \vec{q}^{\,\,2}}. 
\end{equation}
$\Lambda$ is the cut off parameter which is the only free parameter in our study. 
In addition, we adopt the static approximation, which indicates the energy transfers are neglected, because of the small mass differences between a heavy-light pseudoscalar and vector mesons. 
Here, we show the one boson exchange potentials explicitly:
\begin{align}
    V_\pi(r;m_\pi) &= 
    k_\pi[
    \vec{\OC}_1\cdot\vec{\OC}_2 C(r;m_\pi) + 
    S_{\OC_1 \OC_2} T(r;m_\pi) 
    ]\vec{\tau}_1\cdot\vec{\tau}_2, \\
    V_v (r;m_v) &= 
    k_\beta C(r;m_v)  \nonumber \\ &+ 
    k_\lambda[
    \vec{\OC}_1\cdot\vec{\OC}_2 C(r;m_v) + 
    S_{\OC_1 \OC_2} T(r;m_v) 
    ]\vec{\tau}_1\cdot\vec{\tau}_2, \\ 
    V_\sigma (r;m_\sigma) &= 
    k_\sigma C(r;m_\sigma),
\end{align}
where $k_\pi$, $k_\beta$, $k_\lambda$, $k_\sigma$ are coefficients:  
{
\begin{align}
    k_\pi &= \pm \frac{1}{3} \left(\frac{g_\pi}{2f_\pi}\right)^2, \\
    k_\beta &= \left( \frac{\beta g_V}{2m_v}\right)^2,\\
    k_\lambda &= \pm\frac{1}{3} (\lambda g_V)^2,\\
    k_\sigma &= - \left(\frac{g_\sigma}{m_\sigma}\right)^2,
\end{align}
which signs depend on the channel. 
}
$C(r;m)$ and $T(r;m)$ are the central and tensor potentials, respectively, {given in Ref.~\cite{Sakai:2023syt}}. 
}
For the $\omega$ meson exchange, $\vec{\tau}_1\cdot\vec{\tau}_2$ is removed because an $\omega$ meson is isoscalar. 
$\vec{\OC}$ is {the spin operator, and depending on interaction processes, either} the polarization vector or spin-one operator is chosen. $S_{\OC_1 \OC_2}(r)$ is the tensor operator defined by 
\begin{equation}
    S_{\OC_1 \OC_2}(\hat{r}) = 
    3(\vec{\OC}_1\cdot\hat{r})(\vec{\OC}_2\cdot \hat{r}) - 
    \vec{\OC}_1 \cdot \vec{\OC}_2, 
\end{equation}
with $\hat{r}$ being a unit vector of the position vector.

Here, we show the coupling constants in Tabel~\ref{tab:coupling_constant}. 
\renewcommand{\arraystretch}{1.25}
    \begin{table}[htb]
        \caption{Coupling constants of the effective Lagrangians~\cite{CLEO:2001foe,Liu:2019stu,Isola:2003fh}.}
        \centering
        \begin{tabular}{cc} \toprule[0.3mm]
        Coupling constants     &  Values \\ \midrule[0.1mm]
        $g_\pi$     &  0.59 \\
        $g_V$ & $\frac{m_\rho}{\sqrt{2}f_\pi}$ \\
        $\beta$  & 0.9 \\
        $\lambda$ & $0.56\,\mathrm{GeV^{-1}}$ \\
        $g_\sigma$ & 3.4 \\ \bottomrule[0.3mm]
        \end{tabular}
        \label{tab:coupling_constant}
    \end{table}
    \renewcommand{\arraystretch}{1.00}
The coupling constant of the pion exchange $g_\pi$ is determined by the $D^\ast \to D\pi$ decay~\cite{CLEO:2001foe}. 
The coupling constants $\beta$, $\lambda$ are done by the Lattice QCD and $B$ meson decay, respectively~\cite{Liu:2019stu,Isola:2003fh}. 
On the other hand, the value of the $\sigma$ coupling $g_\sigma$ is uncertain, then we employ $g_\sigma = g_{NN\sigma}/3 = 3.4$, which is one-third of a coupling constant of a nucleon and a $\sigma$ meson denoted by $g_{\sigma NN}${~\cite{Sakai:2023syt}}. 

\subsection{Complex scaling method}

In our study, we solve the coupled channel Schr\"{o}dinger equation using the Gaussian expansion method (GEM) and complex scaling method (CSM)~\cite{Hiyama:2003cu,Hiyama:2018ivm,Suzuki:2005wv,Myo:2014ypa,Myo:2020rni}. 
{In this framework, the Schr\"odinger equation is given by}: 
\begin{equation}
    H^\theta \chi^\theta = E_\theta \chi^\theta,
    \label{eq;Shrodinger_eq}
\end{equation}
where $H^\theta$, $\chi^\theta$ are the complex scaled Hamiltonian matrix and radial wavefunction with {an angle} $\theta$, {respectively.} 
In Eq.~\eqref{eq;Shrodinger_eq}, the position vector $\vec{r}$ is rotated on the complex plane to apply CSM: 
\begin{equation}
    \vec{r} \to \vec{r}e^{i\theta}.
\end{equation}
{$E_\theta$ is a complex energy eigenvalue obtained} by solving Eq.~\eqref{eq;Shrodinger_eq}. {The eigenstates are classified into} bound, resonant and continuum states, where the {corresponding} energy eigenvalue is represented as $-B$ with $B$ the binding energy for bound states and $E_r - i\frac{\Gamma}{2}$ with $E_r$ and $\Gamma$ the resonant energy and decay width, respectively, for resonances~\cite{Aguilar:1971ve,Balslev:1971vb}.


\section{Numerical result\label{Sec;Numerical_Result}}

In this section, we show the numerical results of the bound and resonant states of $T_{cc}$ and $T_{bb}$ with spin $J\leq 2$. 
As shown in the previous section, we have one free parameter $\Lambda$. 
In our previous study, we tuned $\Lambda=1069.8$ MeV, which reproduces the empirical binding energy of $T_{cc}$ with $0(1^+)$~\cite{Sakai:2023syt}. 
Therefore, we also employ this cut off $\Lambda$ in this study. 

We consider the doubly heavy tetraquarks $T_{cc}$ and $T_{bb}$ as the $D^{(\ast)}D^{(\ast)}$ and $B^{(\ast)}B^{(\ast)}$ molecules respectively, where we employ the coupled channel analysis because of the HQS. 
Let us show the possible channels of $P^{(\ast)}P^{(\ast)}$ in Table~\ref{table;Coupled_channel}.
We solve the coupled channel Schr\"{o}dinger equation using the complex scaling method~\cite{Suzuki:2005wv,Myo:2014ypa,Myo:2020rni} with the Gaussian expansion method~\cite{Hiyama:2003cu,Hiyama:2018ivm} in order to obtain the energy eigenvalues and decay widths.

\renewcommand{\arraystretch}{1.25}
\begin{table}[tbp]
  \caption{Possible channels of $P^{(\ast)}P^{(\ast)}$ with spin $J\leq 2$~\cite{Ohkoda:2012hv}. We use the notation $[PP^\ast]_\pm = \frac{1}{\sqrt{2}}(PP^\ast \pm P^\ast P)$
  and $^{2S+1}L_J$ with $S$, $L$, $J$ being the spin, angular momentum and total angular momentum, respectively. }
  \centering\begin{tabular}{cc}
    \toprule[0.3mm]
    $I(J^P)$ & Channels \\ \midrule[0.1mm]
    $0(0^-)$ & $[PP^\ast]_+(^3P_0)$ \vspace{1mm}\\ 
    $0(1^+)$ & $[PP^\ast]_- (^3S_1, ^3D_1)$, $P^\ast P^\ast (^3S_1, ^3D_1)$ \vspace{1mm}\\
    $0(1^-)$ & $PP(^1P_1)$, $[PP^\ast]_+ (^3 P_1)$, $P^\ast P^\ast (^1 P_1, ^5P_1, ^5F_1)$ \vspace{1mm}\\
    $0(2^+)$ & $[PP^\ast]_- (^3D_2)$, $P^\ast P^\ast(^3D_2)$ \vspace{1mm}\\
    $0(2^-)$ & $[PP^\ast]_+ (^3P_2, ^3F_2)$, $P^\ast P^\ast (^5P_2, ^5F_2)$ \vspace{1mm}\\
    $1(0^+)$ & $PP(^1S_0)$, $P^\ast P^\ast (^1S_0,^5D_0)$ \vspace{1mm}\\
    $1(0^-)$ & $[PP^\ast]_- (^3P_0)$, $P^\ast P^\ast (^3P_0)$ \vspace{1mm}\\
    $1(1^+)$ & $[PP^\ast]_+ (^3S_1, ^3D_1)$, $P^\ast P^\ast(^5D_1)$ \vspace{1mm}\\
    $1(1^-)$ & $[PP^\ast]_- (^3P_1)$, $P^\ast P^\ast (^3P_1)$ \vspace{1mm}\\
    $1(2^+)$ & $PP(^1D_2)$ , $[PP^\ast]_+(^3D_2)$  $P^\ast P^\ast (^1D_2, ^5S_2, ^5D_2, ^5G_2)$ \vspace{1mm}\\
    $1(2^-)$ & $[PP^\ast]_- (^3P_2, ^3F_2)$, $P^\ast P^\ast (^3P_2, ^3F_2)$\\
    \bottomrule[0.3mm]
 \end{tabular}\label{table;Coupled_channel}
\end{table}
\renewcommand{\arraystretch}{1.00}

\subsection{$T_{cc}$ and $T_{bb}$ with spin $J\leq 2$}
In our previous research~\cite{Sakai:2023syt}, 
we {searched only} bound states of $T_{cc}$ and $T_{bb}$ as the bottom counterparts of $T_{cc}$ with $J=0,1$. 
In this work, we extend our model analysis to $J=2$ bound states and resonances for $J\leq 2$.

\begin{figure*}[htb]
    \begin{tabular}{cc}
    \centering
    \includegraphics[scale = 0.5]{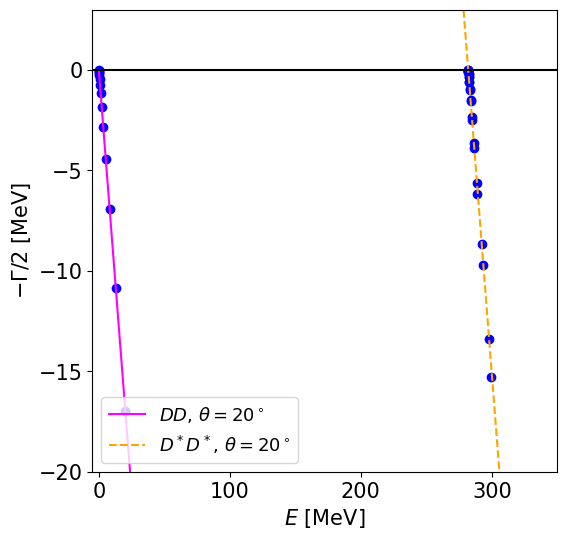} 
    &
    \centering
    \includegraphics[scale=0.5]{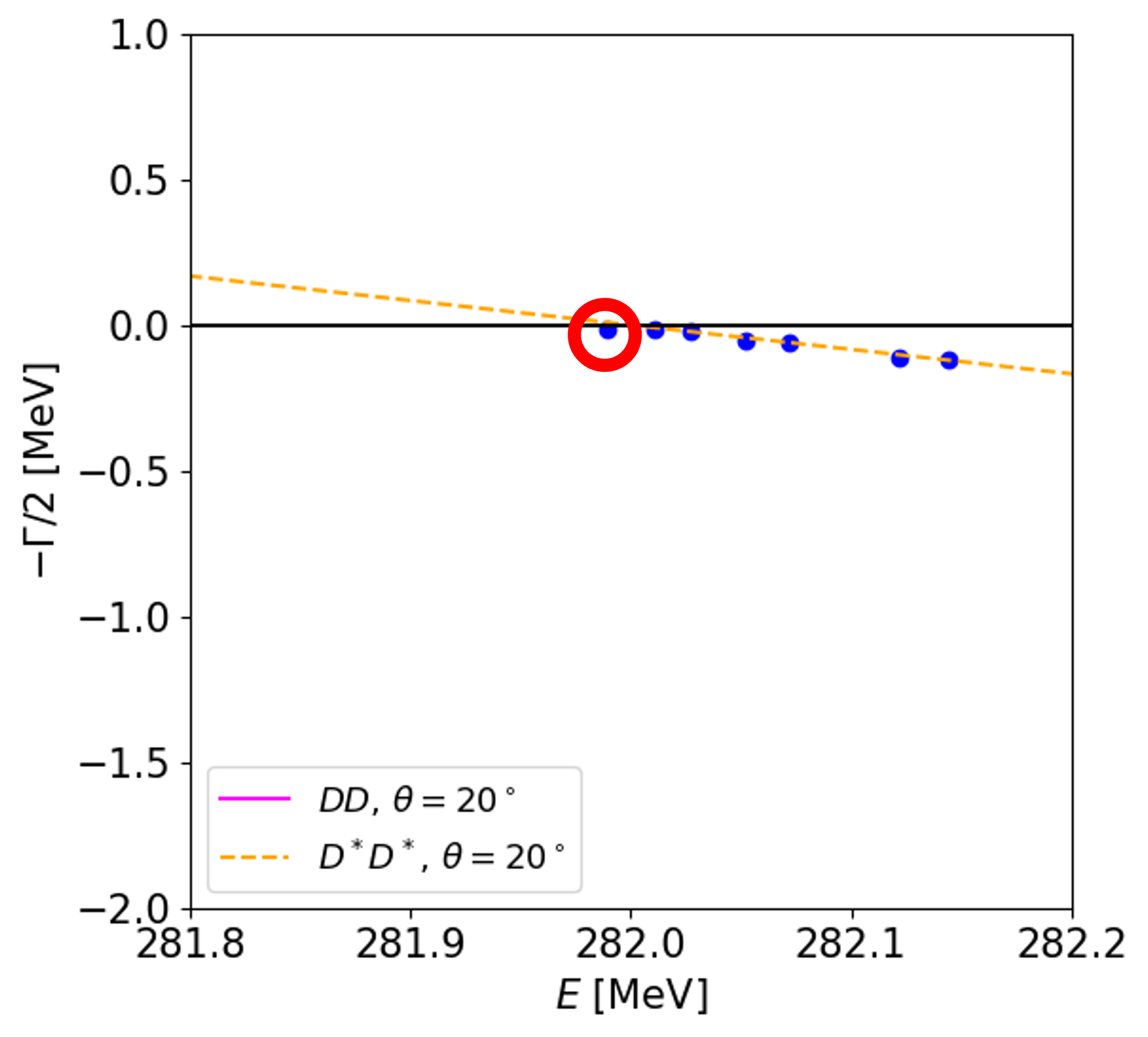}
    \end{tabular}
    \caption{Riemann sheets for the resonance of $T_{cc}(1(0^+))$. The left panel shows the overall Riemann sheet and right one shows the Riemann sheet in the neighborhood of the $D^\ast D^\ast$ threshold. 
    The red circled dot shows the resonance of $T_{cc}(1(0^+))$. 
    }
    \label{fig;Tcc_10pl_re}
\end{figure*}

\renewcommand{\arraystretch}{1.25}
\begin{table}[htb]
  \caption{The energy eigenvalues of bound and resonant states of $T_{bb}$. The energies $E_B$ of bound states are real and measured from the lowest  thresholds. On the other hand, those of resonant states are complex. Their real parts $E_r$ are resonance energies measured from the lowest threshold and their imaginary parts are half of the decay widths $\Gamma$. {The last column shows to which HQS multiplet the state belongs, as discussed in Ref.~\cite{Sakai:2023syt} and in the present work~(\ref{subsubsec;obtained_Tbb}).}The values are given in the units of MeV. }
  \centering\begin{tabular}{cccc}
    \toprule[0.3mm]
    \multirow{2}{*}{$I(J^P)$} & \multirow{2}{*}{lowest threshold} & $E_B$ & \multirow{2}{*}{HQS multiplet structure}  \\
     && $E_r - i \frac{\Gamma}{2}$ &\\ \midrule[0.1mm]
     $0(0^-)$ & $BB^\ast$ & $-24.4$ & triplet-1\vspace{1mm}\\ 
     \multirow{2}{*}{$0(1^+)$} & \multirow{2}{*}{$BB^\ast$} & $-46.0$ & singlet-1 \\
     && $35.9 - i \frac{8.02}{2}$  & singlet-2 \vspace{1mm}\\ 
     $0(1^-)$ & $BB$ & $1.52 - i \frac{1.13}{2}$  & triplet-1\vspace{1mm}\\ 
     $0(2^+)$ & $BB^\ast$ & - \vspace{1mm} & - \\ 
     \multirow{2}{*}{$0(2^-)$} & \multirow{2}{*}{$BB^\ast$} & $-6.11$ & triplet-1\\ 
     && $33.1 - i\frac{8.74}{2}$ \vspace{1mm} & triplet-2\\ 
     \multirow{2}{*}{$1(0^+)$} & \multirow{2}{*}{$BB$} & $-7.23$ & singlet-3 \\
     && $76.4 - i\frac{4.99}{2}$ \vspace{1mm} & singlet-4\\
     $1(0^-)$ & $BB^\ast$ & - \vspace{1mm} & -\\
     $1(1^+)$ & $BB^\ast$ & $-2.46$ \vspace{1mm} & triplet-3\\
     $1(1^-)$ & $BB^\ast$ & - \vspace{1mm} & -\\
     $1(2^+)$ & $BB$ & $89.2 - i\frac{3.09}{2}$ & triplet-3\vspace{1mm} \\
     $1(2^-)$ & $BB^\ast$ & - \vspace{1mm} & -\\
    \bottomrule[0.3mm]
 \end{tabular}\label{table;energy_bound_resonant}
\end{table}
\renewcommand{\arraystretch}{1.00}

\begin{figure*}
    \centering
    \includegraphics[scale=0.4]{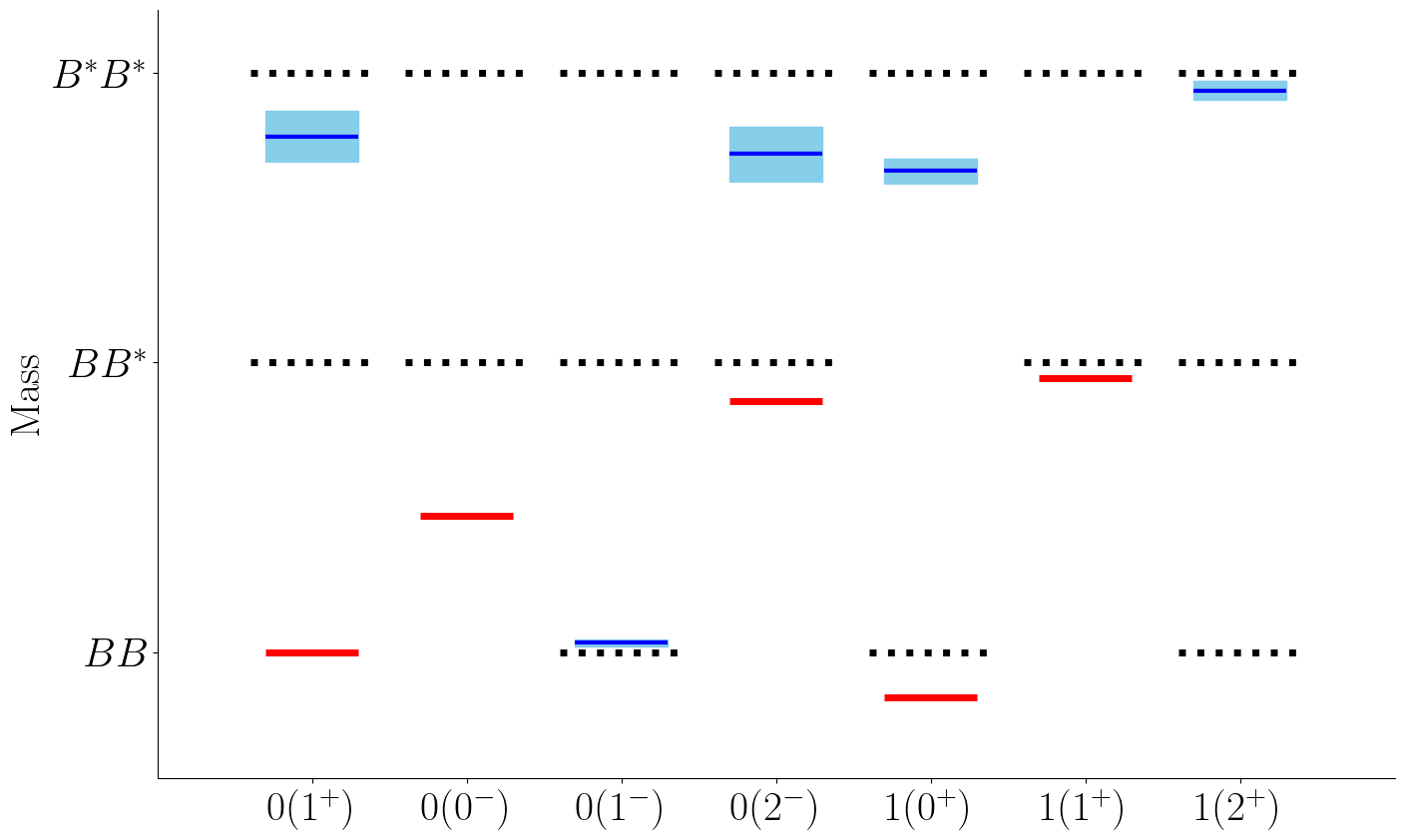}
    \caption{$T_{bb}$ obtained in our analysis. The red lines show the {mass of the bound state} of $T_{bb}(I(J^P))$ measured from the lowest threshold. The blue lines in the box shows the {mass of resonance} of $T_{bb}(I(J^P))$ and the widths of the blue box are the decay width. 
    The black dotted lines show the thresholds of each $I(J^P)$ state. 
    }
    \label{fig;Tbb_all_plot}
\end{figure*}

\begin{figure*}
    \begin{tabular}{ccc}
       (a): $T_{bb}(0(1^+))$  & \multicolumn{2}{c}{(b): $T_{bb}(0(1^-))$}
        \\
       \includegraphics[width=0.33\linewidth]{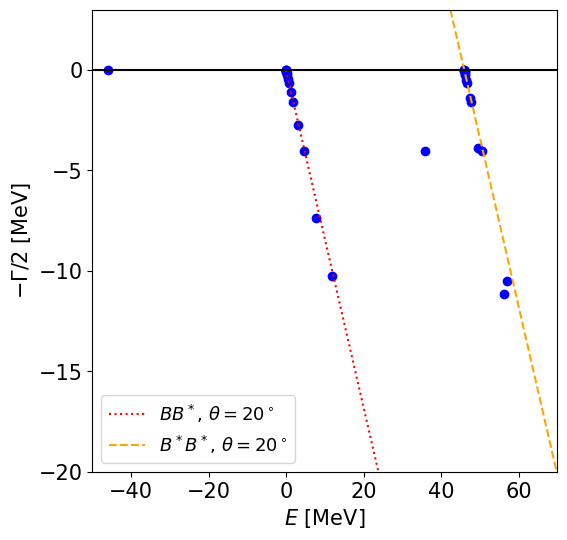}
         & 
        \includegraphics[width=0.33\linewidth]{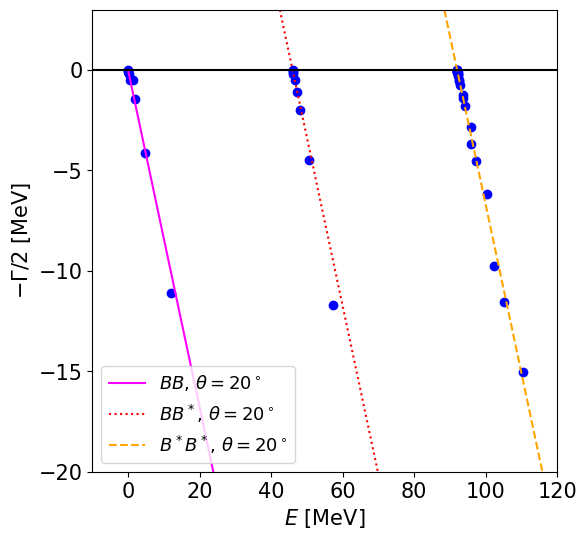}
        &
        \includegraphics[width=0.33\linewidth]{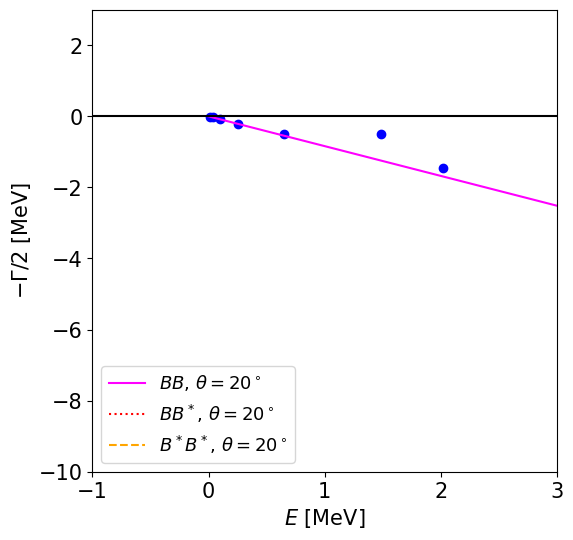}\\
        (c): $T_{bb}(0(2^-))$ & (d): $T_{bb}(1(0^+))$ & (e): $T_{bb}(1(2^+))$  \\
        \includegraphics[width=0.33\linewidth]{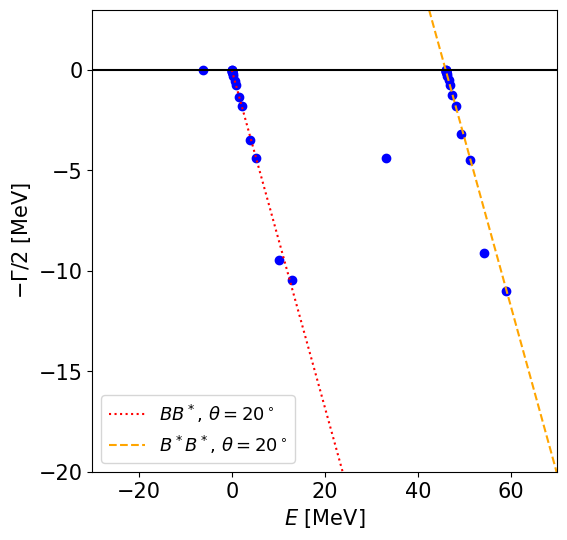}
        &
        \includegraphics[width=0.33\linewidth]{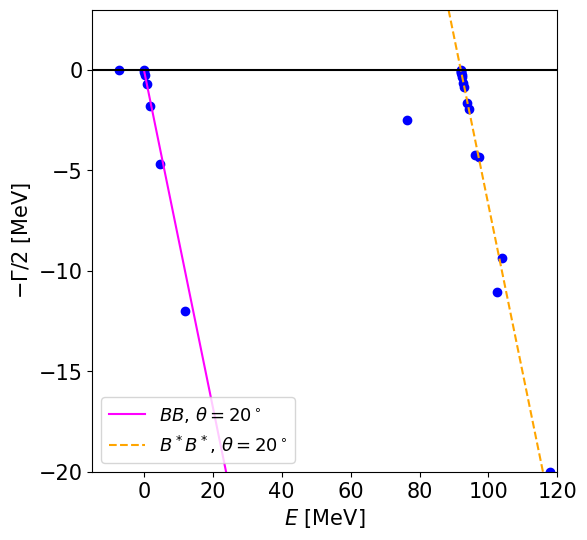}
        &
        \includegraphics[width=0.33\linewidth]{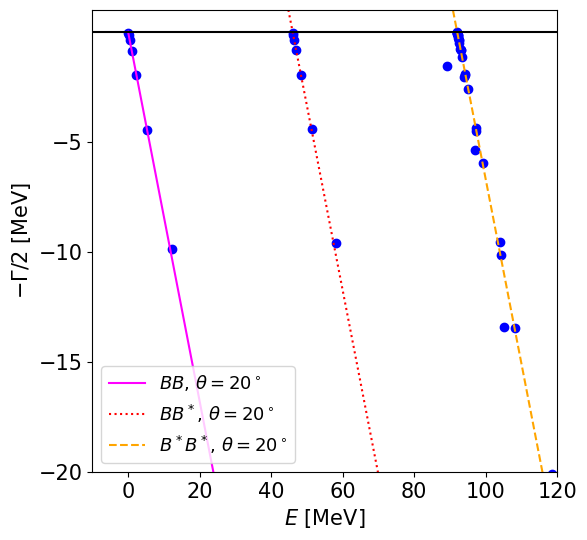}
    \end{tabular}
       \caption{Riemann sheets in the energy plane. (a), (b), (c), (d) and (e) show the Riemann sheets of $0(1^+)$, $0(1^-)$, $0(2^-)$, $1(0^+)$ and $1(2^+)$ states. 
       {As for (b), the overall Riemann sheet (left panel) and Riemann sheet in the neighborhood of $E=0$ (right panel) are shown. }
    The horizontal and vertical axes are real and imaginary parts of energies, respectively. The imaginary part of energies corresponds to a half-decay width $\Gamma/2$. {The magenta solid, red dotted and orange dashed lines represent the $BB$, $BB^\ast$, $B^\ast B^\ast$ thresholds rotated by $2\theta$ {($\theta = 20^\circ$)}, respectively, {where scattering states lie on the lines. }}}
    \label{fig;Riemann_sheet}
\end{figure*}

For $T_{cc}$ states, 
{
we found the bound state of $T_{cc}(0(1^+))$, which corresponds to the state reported by LHCb in our previous study~[28]. 
In additon, we find new resonance of $T_{cc}(1(0^+))$ with an energy eigenvalue being $E = 282 - i \frac{0.0280}{2}\,\mathrm{MeV}$. The Reimann sheet in the energy plane is shown in Fig.~\ref{fig;Tcc_10pl_re}.
The Riemann sheet shows the bound, resonant and continuous states in the framework of CSM. 
In the Riemann sheet, dots on the horizontal axis below the lowest threshold show bound states, while isolated dots with an imaginary part of energy do resonances. Dots stand in the lines rotated by $2\theta$ are discretized continuous states. In Fig.~\ref{fig;Tcc_10pl_re}, the red circled dot show the resonance of $T_{cc}(1(0^+))$. 
Moreover, this resonance is the Feshbach resonance because a bound state of $D^\ast D^\ast$ appears when the lowest channel $DD$ is cut. 
Finally, let us discuss why only $T_{cc}$ with $0(1^+)$ and $1(0^+)$ are found. 
The $0(1^+)$ state has the $S$-wave {component} and is isoscalar. Thus, the centrifugal barrier is absent in the dominant channel, and the magnitude of the $\pi$, $\rho$ exchange forces of an isoscalar state is three times larger than that of an isovector state due to the isospin factor. 
In particular, the tensor force of the $\pi$ exchange in the $S-D$ mixing produces an attraction as discussed in Ref.~\cite{Sakai:2023syt}.
{Therefore, this state earns the most attractive force.}
As for the $1(0^+)$ state, $D^\ast D^\ast $ with $1(0^+)$ has also the $S-D$ mixing induced by the tensor forces, so that this quantum number earns a strong attractive force. 
Moreover, the kinetic energy of $D^\ast D^\ast$ is smaller than that of $DD^{(\ast)}$. 
That is why the bound state of $T_{cc}(0(1^+))$ corresponding to the state reported by LHCb and the resonance of $T_{cc}(1(0^+))$, which is a quasi-bound state of $D^\ast D^\ast$, are found. 
}

{In the bottom sector,} 
however, we obtain the bound and resonant states of $T_{bb}$ with some quantum numbers. 
The energy eigenvalues and decay widths are summarized in Table~\ref{table;energy_bound_resonant} {and plotted as shown in Fig.~\ref{fig;Tbb_all_plot}}. 
We also show the Riemann sheet in the energy plane in Fig.~\ref{fig;Riemann_sheet}.
In this work, we find new bound states with $I(J^P)=0(2^-)$ that were not studied in our previous work~[28]. In addition, by solving the complex-scaled Schr\"odinger equations, resonant states are obtained for $I(J^P)=0(1^+), 0(1^-), 0(2^-), 1(0^+), 1(2^+)$. 
{Since the kinetic energy of $T_{bb}$ is smaller than that of $T_{cc}$, it is considered that $T_{bb}$ is more likely to be bound than $T_{cc}$ discussed in our previous study~\cite{Sakai:2023syt}. 
For resonances of $T_{bb}$ except for $0(1^-)$ state, it turns out that these states are realized as the Feshbach resonances because they become bound states of upper channel{, $B^\ast B^\ast$,} when the lower ones are switched off.
{However, the $0(1^-)$ resonance disappears when the $BB$ channel is omitted. Thus, this state is generated by the centrifugal barrier.}}



We consider that although the $DD\pi$ and $BB\pi$ thresholds are close to $D^{(*)}D^{(*)}$ and $B^{(*)}B^{(*)}$ thresholds, respectively, contributions from the $DD\pi$ and $BB\pi$ effects are not large. 
In the charm sector, the $DD\pi$ channel can be included effectively by not using the static approximation, i.e., by considering the energy transition between $D$ and $D^\ast$ at the $DD^*\pi$ vertex in the OPEP, because $D^\ast$ can decay into $D\pi$ and thus the exchanged pion can be on shell, as done in ~Refs.~\cite{Asanuma:2023atv,Cheng:2022qcm,Lin:2022wmj}. 
The $DD\pi$ threshold is located slightly below the $DD^\ast$ threshold, and thus the $DD\pi$ effect is expected to play an important role. 
However, the result in Ref.~\cite{Asanuma:2023atv} {including the $DD\pi$ channel effectively} is the almost same as our result, {and hence the impact of the $DD\pi$ channel on the $DD^\ast$ molecule is not so large.}
In the bottom sector, the same approach done in the $D^{(*)}D^{(*)}$ cannot be used because the $B^* \to B\pi$ decay is forbidden kinamatically. However, {we consider that the} $BB\pi$ {channel} is also not important because the mass difference between $BB\pi$ and $B^\ast B^\ast$ is larger than that between $DD^\ast$ and $DD\pi$. 
If the $BB\pi$ channel can be included, the energy eigenvalue of the resonances, which are quasi-bound states of $B^\ast B^\ast$, are just push down because of the level repulsion.
Thus, the $BB\pi$ channel provides an attractive interaction, and its inclusion does not eliminate the bound and resonant states predicted in this study.



\subsection{HQS partners}

{
In this section, we investigate the heavy quark spin structures and HQS partners of bound and resonant states of $T_{cc}$ and $T_{bb}$ found in this study. 
In order to achieve this purpose, we introduce the light-cloud basis (LCB) as discussed in Refs.~\cite{Yasui:2013vca,Yamaguchi:2014era,Shimizu:2018ran,Sakai:2023syt}. 
First, we give formulation of the LCB for the two meson system, and show that the HQS multiplet structures appear. After that, we apply the LCB to the $T_{cc}$ and $T_{bb}$ states obtained in this study. 
}

\subsubsection{Heavy quark spin multiplet structure}

In this subsection, the formulation of the LCB for two meson molecules, introduced in Ref.~\cite{Sakai:2023syt}, is reviewed.
In this basis, we can describe a hadronic molecule with spin-parity $J^P$ in terms of a heavy diquark with spin $S_Q$ and a light-cloud component with spin $j_\ell$ in the spin space as 
\begin{equation}
    \Ket{
    \Big[
    [
    Q Q
    ]_{S_Q}\ 
    \big[
    L \ [
    \bar{q}\bar{q}
    ]_{S_q}  
    \big]_{j_\ell}
    \Big]_J
    }
\end{equation}
{In the spin wave function, $Q$ ($\bar{q}$) denotes the heavy quark spin (the light antiquark spin).} 
{$S_q$ is the total spin of two light antiquarks of heavy pseudo or vector mesons.} 
This basis enables us to classify $T_{cc}$ and $T_{bb}$ obtained {in our studies~\cite{Sakai:2023syt}} by the heavy quark spin and light-cloud spin. 
{For the bound states of $T_{cc}$ and $T_{bb}$ with spin $J = 0,1$, we classified these states using the heavy quark spin structures~\cite{Sakai:2023syt}.} 

Here, taking the states with $1(J^+)$ ($J=0,1,2$) as examples, we explain the LCB in detail. 
As shown in Table.~\ref{table;Coupled_channel}, the possible channels of $1(J^+)$ states $\psi_{1(J^P)}^\HM$ are 
\begin{align}
    \psi^\HM_{1(0^+)} &= \begin{pmatrix}
        \ket{PP (^1S_0)} \\ 
        \ket{P^\ast P^\ast (^1S_0)} \\
        \ket{P^\ast P^\ast (^5D_0)} 
    \end{pmatrix}, \\
    \psi^\HM_{1(1^+)} &= \begin{pmatrix}
        \ket{[PP^\ast]_+(^3S_1)} \\
        \ket{[PP^\ast]_+(^3D_1)} \\
        \ket{P^\ast P^\ast (^5D_1)}
    \end{pmatrix},\\ \vspace{1mm}
    \psi^\HM_{1(2^+)} &= \begin{pmatrix}
        \ket{PP (^1D_2)} \\
        \ket{[PP^\ast]_+ (^3 D_2)} \\
        \ket{P^\ast P^\ast (^1D_2)} \\
        \ket{P^\ast P^\ast (^5 S_2)} \\
        \ket{P^\ast P^\ast (^5 D_2)} \\
        \ket{P^\ast P^\ast (^5G_2)} 
    \end{pmatrix},
\end{align}
where $\HM$ means the hadronic molecule basis (HMB). 
The one pion exchange potential in HMB are
\begin{widetext}
    \begin{align}
        V^{\mathrm{HM}}_{\pi,1(0^{+})} &= \begin{pmatrix}
            0&\!\!-\sqrt{3}C_{\pi}&\!\!\sqrt{6}T_{\pi}\\\vspace{1mm}
            -\sqrt{3}C_{\pi}&\!\!-2C_{\pi}&\!\!-\sqrt{2}T_{\pi}\\\vspace{1mm}
            \sqrt{6}T_{\pi}&\!\!-\sqrt{2}T_{\pi}&\!\!C_{\pi}-2T_{\pi}
          \end{pmatrix}, \\
        V^{\mathrm{HM}}_{\pi,1(1^{+})} &= \begin{pmatrix}
              C_{\pi}&\!\!-\sqrt{2}T_{\pi}&\!\!-\sqrt{6}T_{\pi}\\\vspace{1mm}
              -\sqrt{2}T_{\pi}&\!\!C_{\pi}+T_{\pi}&\!\!-\sqrt{3}T_{\pi}\\\vspace{1mm}
              -\sqrt{6}T_{\pi}&\!\!-\sqrt{3}T_{\pi}&\!\!C_{\pi}-T_{\pi}
            \end{pmatrix},\\
        V^\HM_{\pi, 1(2^+)} &= \begin{pmatrix}
                0 &\!\! 0 &\!\! -\sqrt{3}C_\pi &\!\! \sqrt{\frac{6}{5}}T_\pi &\!\! -2 \sqrt{\frac{3}{7}}T_\pi &\!\! 6\sqrt{\frac{3}{35}}T_\pi \\\vspace{1mm} 
                0 &\!\! C_\pi - T_\pi &\!\! 0 &\!\! 3\sqrt{\frac{2}{5}}T_\pi &\!\! -\frac{3}{\sqrt{7}}T_\pi &\!\! - \frac{12}{\sqrt{35}}T_\pi \\ \vspace{1mm} 
                -\sqrt{3}C_\pi &\!\! 0 &\!\! 2C_\pi &\!\! -\sqrt{\frac{2}{5}}T_\pi &\!\! \frac{2}{\sqrt{7}}T_\pi &\!\! -\frac{6}{\sqrt{35}}T_\pi \\ \vspace{1mm}
                \sqrt{\frac{6}{5}}T_\pi &\!\! 3\sqrt{\frac{2}{5}}T_\pi &\!\! -\sqrt{\frac{2}{5}}T_\pi &\!\! C_\pi &\!\! \sqrt{\frac{14}{5}}T_\pi &\!\! 0 \\ \vspace{1mm} 
                -2\sqrt{\frac{3}{7}}T_\pi &\!\! -\frac{3}{\sqrt{7}}T_\pi &\!\! \frac{2}{\sqrt{7}}T_\pi &\!\! \sqrt{\frac{14}{5}}T_\pi &\!\! C_\pi + \frac{3}{7}T_\pi &\!\! \frac{12}{7\sqrt{5}}T_\pi \\ \vspace{1mm} 
                6\sqrt{\frac{3}{35}}T_\pi &\!\! -\frac{12}{\sqrt{35}}T_\pi &\!\! -\frac{6}{\sqrt{35}}T_\pi &\!\! 0 &\!\! \frac{12}{7\sqrt{5}}T_\pi &\!\! C_\pi - \frac{10}{7}T_\pi
            \end{pmatrix},
    \end{align}
\end{widetext}
where $C_\pi$ and $T_\pi$ are written by 
\begin{align}
    C_\pi &= \frac{1}{3}\left(
    \frac{g_\pi}{2f_\pi}
    \right)^2 C(r;m_\pi), \\
    T_\pi &= \frac{1}{3}\left(
    \frac{g_\pi}{2f_\pi}
    \right)^2 T(r;m_\pi).
\end{align}
Next, we implement a basis transformation from the HMB to the LCB. 
This is an unitary transformation and the potential matrices are transformed into the block-diagonal matrices: 
\begin{widetext}
\begin{align}
    \psi^{\mathrm{LC}}_{1(0^{+})} &= U^{-1}_{1(0^{+})}\psi^{\mathrm{HM}}_{1(0^{+})}\notag\\
      &=\begin{pmatrix}
        \Ket{{\Big[{\big[QQ\big]}_{0}\ {\big[S\ {[\bar{q}\bar{q}]}_{0}\big]}_{0}\Big]}_{0}}\vspace{1mm}\\
        \Ket{{\Big[{\big[QQ\big]}_{1}\ {\big[S\ {[\bar{q}\bar{q}]}_{1}\big]}_{1}\Big]}_{0}}\vspace{1mm}\\
        \Ket{{\Big[{\big[QQ\big]}_{1}\ {\big[D\ {[\bar{q}\bar{q}]}_{1}\big]}_{1}\Big]}_{0}}
      \end{pmatrix}, \label{eq:LCB10+} \\
      U_{1(0^{+})} &= 
        \begin{pmatrix}
          \frac{1}{2}&\frac{\sqrt{3}}{2}&0\vspace{1mm}\\
          \frac{\sqrt{3}}{2}&-\frac{1}{2}&0\vspace{1mm}\\
          0&0&1
        \end{pmatrix},\\
    V^{\mathrm{LC}}_{\pi,1(0^{+})} 
      &= \left(\begin{array}{c|cc}
        -3C_{\pi}&0&0\\\hline
        0&C_{\pi}&2\sqrt{2}T_{\pi}\\
        0&2\sqrt{2}T_{\pi}&C_{\pi}-2T_{\pi}
      \end{array}\right),\\
    \psi^{\mathrm{LC}}_{1(1^{+})} &= U^{-1}_{1(1^{+})}\psi^{\mathrm{HM}}_{1(0^{+})}\notag\\
      &=\begin{pmatrix}
        \Ket{{\Big[{\big[QQ\big]}_{1}\ {\big[S\ {[\bar{q}\bar{q}]}_{1}\big]}_{1}\Big]}_{1}}\vspace{1mm}\\
        \Ket{{\Big[{\big[QQ\big]}_{1}\ {\big[D\ {[\bar{q}\bar{q}]}_{1}\big]}_{1}\Big]}_{1}}\vspace{1mm}\\
        \Ket{{\Big[{\big[QQ\big]}_{1}\ {\big[D\ {[\bar{q}\bar{q}]}_{1}\big]}_{2}\Big]}_{1}}
      \end{pmatrix}, \label{eq:LCB11+} \\
      U_{1(1^{+})} &= 
        \begin{pmatrix}
          1&0&0\vspace{1mm}\\
          0&-\frac{1}{2}&-\frac{\sqrt{3}}{2}\vspace{1mm}\\
          0&-\frac{\sqrt{3}}{2}&\frac{1}{2}
        \end{pmatrix},\\
    V^{\mathrm{LC}}_{\pi,1(1^{+})} 
      &= \left(\begin{array}{cc|c}
        C_{\pi}&2\sqrt{2}T_{\pi}&0\\
        2\sqrt{2}T_{\pi}&C_{\pi}-2T_{\pi}&0\\\hline
        0&0&C_{\pi}+2T_{\pi}
      \end{array}\right),\\
    \psi^{\mathrm{LC}}_{1(2^+)} 
      &= U^{-1}_{1(2^+)}\psi^{\mathrm{HM}}_{1(2^+)}\\
      &= \begin{pmatrix}
        \Ket{{\Big[{\big[QQ\big]}_{0}\ {\big[D\ {[\bar{q}\bar{q}]}_{0}\big]}_{2}\Big]}_{2}}\vspace{1mm}\\
        \Ket{{\Big[{\big[QQ\big]}_{1}\ {\big[S\ {[\bar{q}\bar{q}]}_{1}\big]}_{1}\Big]}_{2}}\vspace{1mm}\\
        \Ket{{\Big[{\big[QQ\big]}_{1}\ {\big[D\ {[\bar{q}\bar{q}]}_{1}\big]}_{1}\Big]}_{2}}\vspace{1mm}\\
        \Ket{{\Big[{\big[QQ\big]}_{1}\ {\big[D\ {[\bar{q}\bar{q}]}_{1}\big]}_{2}\Big]}_{2}}\vspace{1mm}\\
        \Ket{{\Big[{\big[QQ\big]}_{1}\ {\big[D\ {[\bar{q}\bar{q}]}_{1}\big]}_{3}\Big]}_{2}} \vspace{1mm}\\
        \Ket{{\Big[{\big[QQ\big]}_{1}\ {\big[G\ {[\bar{q}\bar{q}]}_{1}\big]}_{3}\Big]}_{2}}
      \end{pmatrix}, \label{eq:LCB12+}\\
      U_{1(2^+)} &= \begin{pmatrix}
        \frac{1}{2} &\!\! 0 &\!\! \frac{\sqrt{15}}{10} &\!\! \frac{1}{2} &\!\! \frac{\sqrt{35}}{10} &\!\! 0 \\ \vspace{1mm}
        0 &\!\! 0 &\!\! \frac{3\sqrt{5}}{10} &\!\! \frac{\sqrt{3}}{6} &\!\! - \sqrt{\frac{7}{15}} &\!\! 0\\ \vspace{1mm}
        \frac{\sqrt{3}}{2} &\!\! 0 &\!\! - \frac{\sqrt{5}}{10} &\!\! -\frac{\sqrt{3}}{6} &\!\! -\frac{\sqrt{105}}{30} & 0 \\ \vspace{1mm} 
        0 &\!\! 1 &\!\! 0 &\!\!0 &\!\!0 &\!\!0 \\ \vspace{1mm}
        0 &\!\!0 &\!\!\frac{\sqrt{35}}{10} &\!\! - \frac{\sqrt{21}}{6} &\!\! \frac{1}{\sqrt{15}} &\!\! 0 \\ \vspace{1mm} 
        0 &\!\! 0 &\!\!0 &\!\!0 &\!\!0 &\!\!1
      \end{pmatrix},\\
      V^{\mathrm{LC}}_{\pi,1(2^+)} 
      &= \left(\begin{array}{c|cc|c|cc}
        -3C_{\pi} & 0 & 0 & 0 & 0 & 0\\
        \hline
        0 & C_{\pi}& 2\sqrt{2}T_\pi & 0 & 0 & 0\\
        0 & 2\sqrt{2}T_\pi & C_{\pi}-2T_{\pi} & 0 & 0 & 0\\
        \hline
        0 & 0 & 0 & C_\pi+2T_\pi &0 &0 \\
        \hline
        0 &\!\! 0 &\!\!0 &\!\! 0 &\!\! C_\pi - \frac{4}{7}T_\pi &\!\! \frac{12\sqrt{3}}{7}T_\pi \\ 
        0 &\!\! 0 &\!\!0 &\!\! 0 &\!\! \frac{12\sqrt{3}}{7}T_\pi &\!\! C_\pi - \frac{10}{7}T_\pi
      \end{array}\right)
\end{align}
\end{widetext}
We note the reason why the potential matrices in the LCB is block diagonal. 
{
In the heavy quark limit (HQL), the mixing  of states with the different spin strucuture is surpressed by $\mathcal{O}(1/m_Q)$. 
}
Thus, the off-diagonal components mixing the states with the different $S_Q$ or $j_\ell$ must vanish.

As you can see {in Eqs.~\eqref{eq:LCB10+}, \eqref{eq:LCB11+} and \eqref{eq:LCB12+}}, $\psi^\LC_{1(0^+),1(1^+),1(2^+)}$ have the same component 
\begin{equation}
    \begin{pmatrix}
        \Ket{{\Big[{\big[QQ\big]}_{1}\ {\big[S\ {[\bar{q}\bar{q}]}_{1}\big]}_{1}\Big]}_J}\vspace{1mm}\\
        \Ket{{\Big[{\big[QQ\big]}_{1}\ {\big[D\ {[\bar{q}\bar{q}]}_{1}\big]}_{1}\Big]}_J}
    \end{pmatrix},
\end{equation}
{with $S_Q=1$ and $j_\ell=1$,} and the potential matrices of $1(0^+),1(1^+),1(2^+)$ also have the same component
\begin{equation}
    \begin{pmatrix}
        C_\pi & 2\sqrt{2}T_\pi \\
        2\sqrt{2}T_\pi & C_\pi - 2 T_\pi 
    \end{pmatrix}.\label{eq;example_potential_LC}
\end{equation}
Thus, in the HQL, the energy eigenstates with Eq.~\eqref{eq;example_potential_LC} are degenerate and the corresponding {components of} $1(0^+)$, $1(1^+)$, {and} $1(2^+)$ belong to the same HQS triplet. 
{The spin combination} $S_Q \otimes j_\ell = 0 \oplus 1 \oplus 2$ {also indicates the appearance of the HQS triplet.}

In addition, {another component} of the ${1(1^+)}$ state {in the LCB}
\begin{equation}
    \Ket{{\Big[{\big[QQ\big]}_{1}\ {\big[D\ {[\bar{q}\bar{q}]}_{1}\big]}_{2}\Big]}_{2}}\label{eq;LC_11pl_12pl}
\end{equation}
coincides with that of the ${1(2^+)}$ state, and the corresponding {potentials in the LCB are also the same,}  
\begin{equation}
    \begin{pmatrix}
        C_\pi + 2\sqrt{2}T_\pi 
    \end{pmatrix}.
\end{equation}
Therefore, the energy eigenvalues of $1(1^+)$ and $1(2^+)$ states are degenerate. 
{Since $S_Q\otimes j_\ell= 1 \oplus 2 \oplus 3$ with $S_Q=1$ and $j_\ell=2$, the $1(3^+)$ state is also expected to exist, and the $1(1^+)$, $1(2^+)$ and $1(3^+)$ states belong the HQS triplet. In this work, however, we do not analyze the spin $J=3$ states.}

\subsubsection{HQS partners of {$T_{cc}$ and} $T_{bb}$ obtained in our study\label{subsubsec;obtained_Tbb}}

\begin{figure*}
    \begin{tabular}{ccc}
       (a): $0(1^+)$  &  (b): $0(1^-)$ & (c): $ 0(2^-)$\\
       \includegraphics[width=0.33\linewidth]{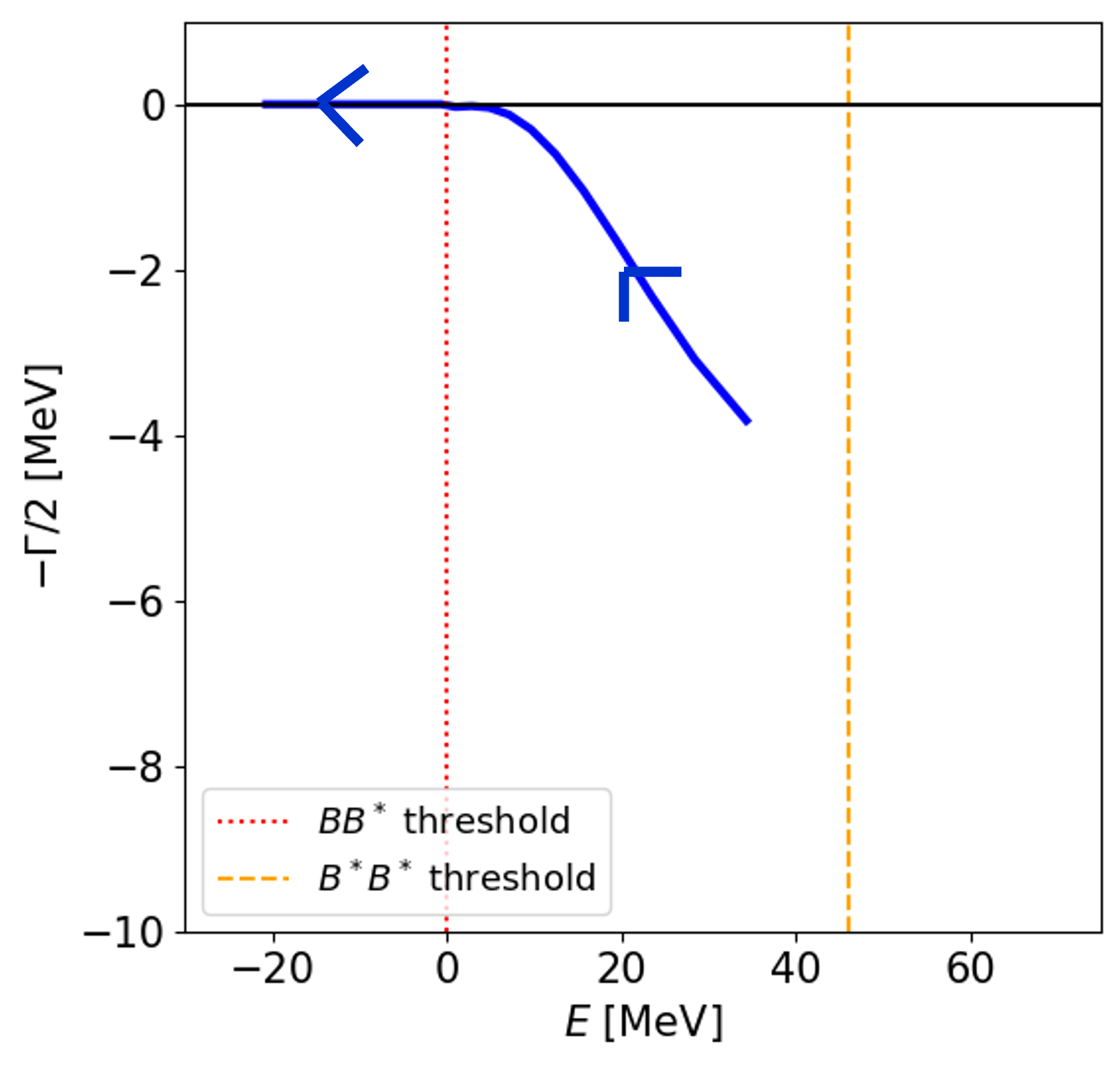}
         & 
        \includegraphics[width=0.33\linewidth]{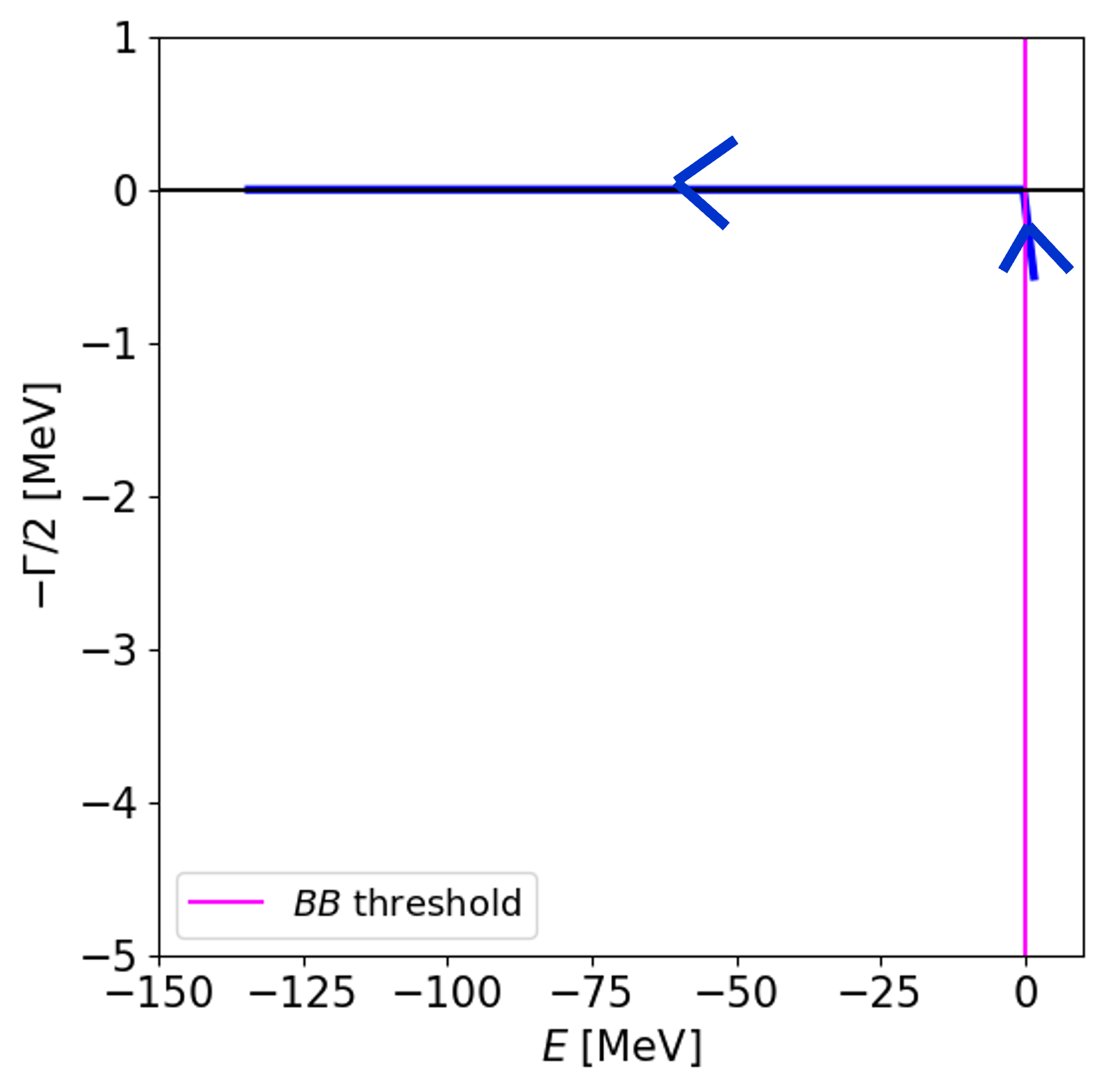}
        &
        \includegraphics[width=0.33\linewidth]{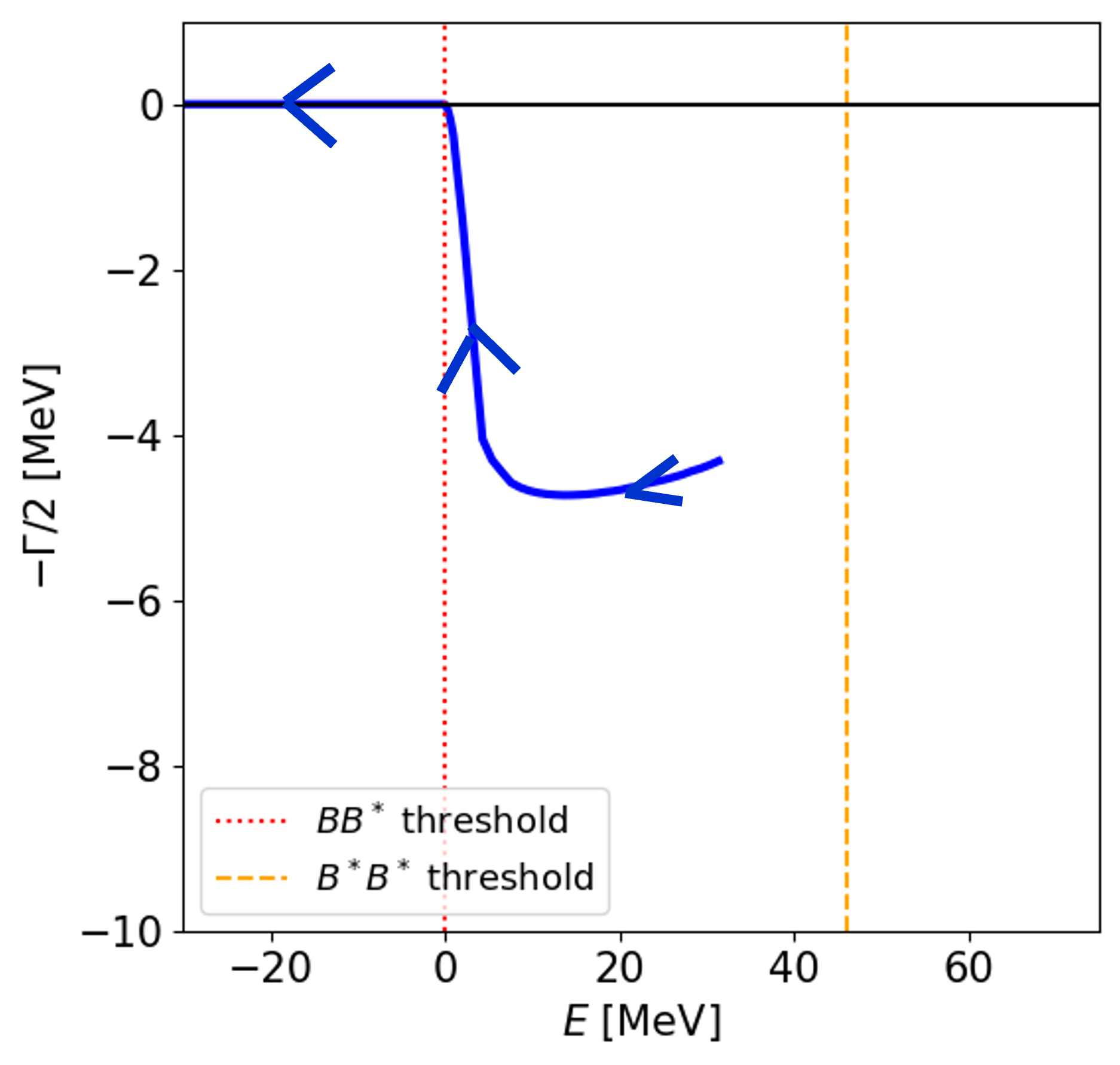}\\
        (d): $1(0^+)$ & (e): $1(2^+)$ & \\
        \includegraphics[width=0.33\linewidth]{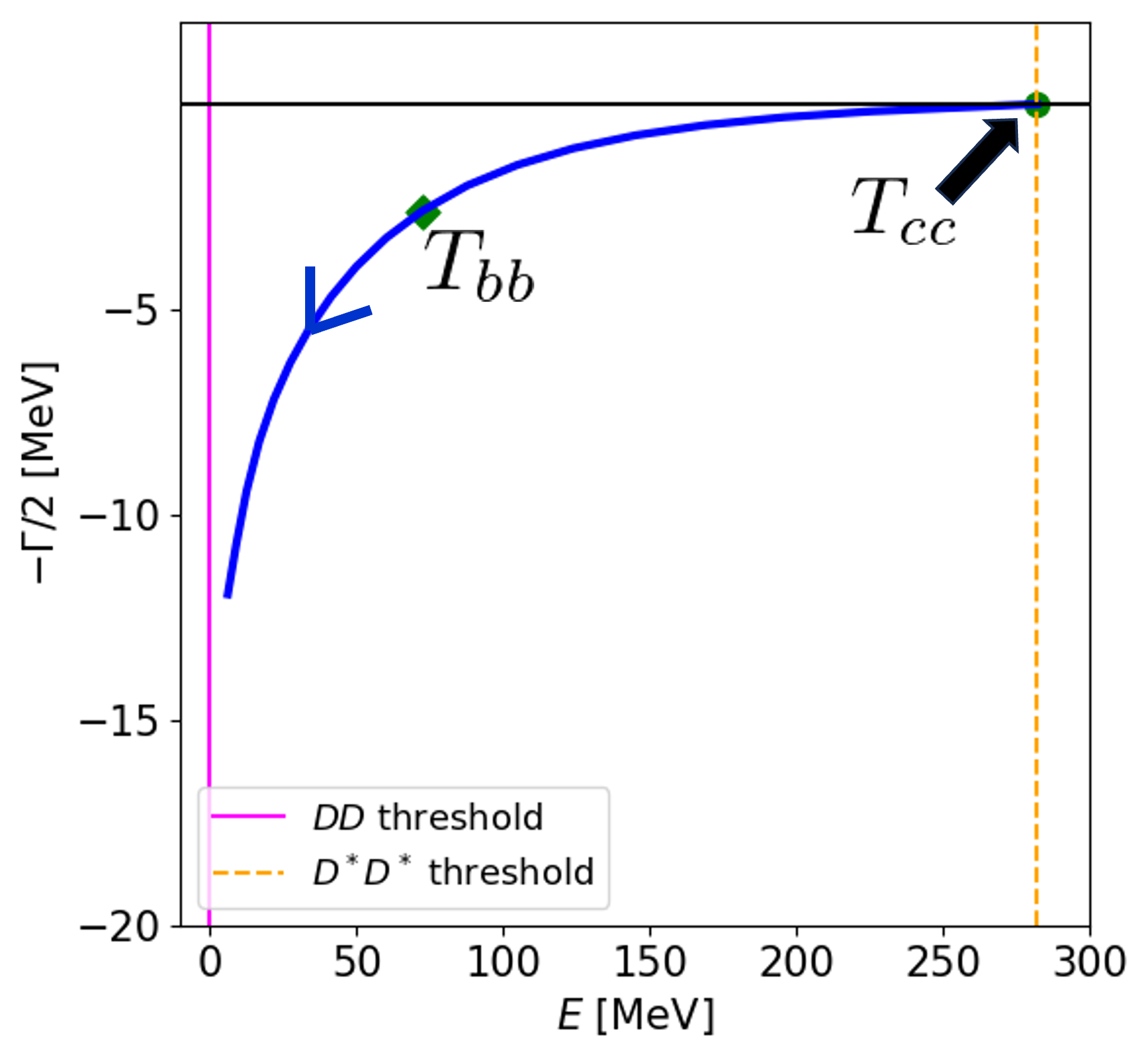}
        &
        \includegraphics[width=0.33\linewidth]{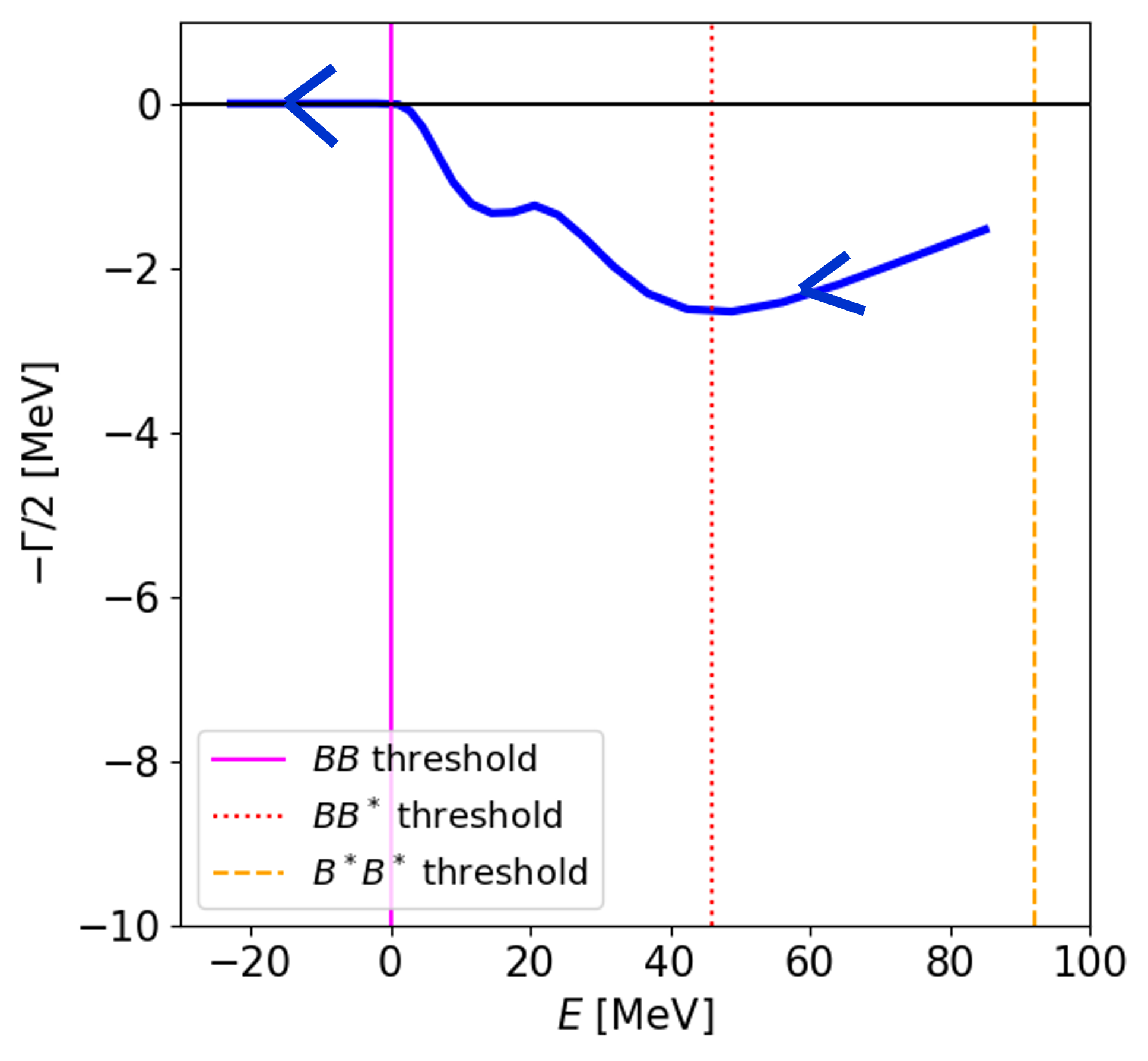}
    \end{tabular}
       \caption{{The mass dependence of the the energy eigenvalues of the resonances of {$T_{cc}$ and} $T_{bb}$. (a)-(e) show the change of the energy eigenvalues as increasing the mass of the heavy meson composing a hadronic molecule. The arrows show the direction of change of the energy eigenvalues. We note that, as the masses of the heavy mesons of a hadronic molecule increase, the differences between $PP$, $PP^\ast$ and $P^\ast P^\ast$ become smaller, but we show {the $DD$, $D D^\ast$ and $D^\ast D^\ast$ thresholds for (d) and $BB$, $BB^\ast$ and $B^\ast B^\ast$ thresholds for (a)-(c) and (e) in these figures}.}
    }
    \label{fig;massdiff_Riemann}
\end{figure*}


In this section, the spin structure of the obtained bound and resonant states of {$T_{cc}$ and} $T_{bb}$ is studied. In order to do it, we increase masses of constituent {heavy} mesons of the hadronic molecular states, and obtain properties of the states in the {HQL}, which are connected with the $T_{bb}$ states in the {charm and bottom sectors}. Therefore, a state in the {HQL}, connected with {$T_{cc}$ and} $T_{bb}$, is regarded as an origin of the resonance of {$T_{cc}$ and} $T_{bb}$.

First, we study the spin structure of the resonant state of $T_{bb}$ with $0(1^+)$. 

The mass dependence of the energy eigenvalue with $0(1^+)$ is shown in Fig.~\ref{fig;massdiff_Riemann}~(a). 
As can been seen from Fig.~\ref{fig;massdiff_Riemann}~(a), the larger the mass of a vector meson $P^\ast$ is, the smaller the energy and decay width are, and the resonant state continuously transitions to the bound state below the threshold. This is due to the centrifugal force of the $D$-wave states. Otherwise, the resonant state should transition through the virtual state{~\cite{Taylor:1972pty}}. 
In the heavy quark limit, however, the mixing ratio of the bound state is obtained as 
\begin{align}
    &f([PP^\ast]_-(^3S_1)) = f([P^\ast P^\ast](^3S_1)) = 50\, \%,\\
    &f([PP^\ast]_-(^3D_1)) = f([P^\ast P^\ast](^3D_1)) = 0\, \%,
\end{align}
where $f(\text{channel})$ is the mixing ratio of a channel.
This result shows that the obtained bound state has only $S$-wave components, while $D$-wave ones are suppressed. Furthermore, this mixing ratio indicates that the bound state  in the heavy quark limit corresponds to
\begin{equation}
    \begin{pmatrix}
        {\Ket{{\Big[{\big[QQ\big]}_{1}\ {\big[S\ {[\bar{q}\bar{q}]}_{0}\big]}_{0}\Big]}_{1}}}
    \end{pmatrix} 
\end{equation}
{with $S_Q=1$ and $j_\ell=0$ in the light cloud basis, 
as seen in the unitary matrix $U_{0(1^+)}^{-1}$ in Eq.~\eqref{eq:U_01+}. 
Thus, the origin of the resonance of $T_{bb}$ with $0(1^+)$ belongs to the HQS singlet {(singlet-2 in Table~\ref{table;energy_bound_resonant})}.

Second, let us discuss the HQS multiplet structures of $I(J^P) = 0(J^-)$ states. 
As mentioned below, the bound states of $T_{bb}$ with $0(0^-)$ and $0(2^-)$ and resonant state of $T_{bb}$ with $0(1^-)$ in Table~\ref{table;energy_bound_resonant} belong to the same HQS triplet {(triplet-1 in Table~\ref{table;energy_bound_resonant})}, which means these states are HQS partners with each other. 
{On the other hand, the resonance of $T_{bb}(0(2^-))$ belongs to the other HQS triplet (triplet-2 in Table~\ref{table;energy_bound_resonant}).}

In our previous study, we examine the heavy quark spin structure of $T_{bb}$ with $0(0^-)$~\cite{Sakai:2023syt}. 
This state corresponds to the {one} composed of 
\begin{equation}
    \begin{pmatrix}
        {\Ket{{\Big[{\big[QQ\big]}_{1}\ {\big[P\ {[\bar{q}\bar{q}]}_{1}\big]}_{1}\Big]}_0}}
    \end{pmatrix}
\end{equation}
with $S_Q = 1$ and $j_\ell = {1}$ in the LCB in the HQL. 
Thus, this state is a member of the HQS triplet because $S_Q\otimes j_\ell = 0 \oplus 1 \oplus 2$. 
Here, we also consider the spin structures of the resonant state of $T_{bb}$ with $0(1^-)$ and bound state of $T_{bb}$ with $0(2^-)$. 
By increasing the masses, it is found that these {$T_{bb}$} states are smoothly connected with the bound states of $T_{QQ}(0(J^-))$ ($J=1,2$) in the HQL whose mixing ratios are 
\begin{align}
    &f(PP(^1P_1)) = f([PP^\ast]_+ (^3P_1)) = 25 \,\% \\
    &f(P^\ast P^\ast (^1P_1)) = 8.33 \,\% \\
    &f(P^\ast P^\ast (^5P_1)) = 41.7\, \% \\
    &f(P^\ast P^\ast (^5F_1)) = 0\,\%
\end{align}
for the $0(1^-)$ state and 
\begin{align}
    &f([PP^\ast]_+ (^3P_2)) :f(P^\ast P^\ast (^5P_2)) = 1 : 3 \\
    &f([PP^\ast]_+ (^3F_2)) = f(P^\ast P^\ast (^5F_2)) = 0\, \%
\end{align}
for the $0(2^-)$ state, respectively. 
These mixing ratios show that the components of the resonance with $0(1^-)$ and bound state with $0(2^-)$ in the LCB in HQL are
\begin{equation}
    \begin{pmatrix}
        {\Ket{{\Big[{\big[QQ\big]}_{1}\ {\big[P\ {[\bar{q}\bar{q}]}_{1}\big]}_{1}\Big]}_{J=1,2}}}
    \end{pmatrix}
\end{equation}
with $S_Q = 1, j_\ell = 1$ as seen in the unitary matrix $U^{-1}_{0(1^-)}$ in Eq.~\eqref{eq:U_01-} and $U^{-1}_{0(2^-)}$ in Eq.~\eqref{eq:U_02-}. 
Thus, the bound states of the $0(0^-)$ and $0(2^-)$ states and resonance of the $0(1^-)$ state belong to the same HQS triplet. 

{As for the $0(1^-)$ state, 
we note that the {$T_{bb}$} resonance connects with {the $T_{QQ}$} bound state directly shown in Fig.~\ref{fig;massdiff_Riemann}~(b) because of the centrifugal force of the $P$-wave in the lowest channel. }

In addition to the bound state, $T_{bb}(0(2^-))$ also has a resonance. The mass dependence of the complex energy is shown in Fig.~\ref{fig;massdiff_Riemann}~(c). 
{In this case,} 
the real part of the energy eigenvalue becomes smaller as increasing the mass of the vector meson. 
On the other hand, the decay width initially becomes larger, then decreases. 
The decrease in the decay width is {understood} due to the transition from the resonance to the bound state, 
similar to the case of the resonance of $T_{bb}(0(1^+))$ in Fig.~\ref{fig;massdiff_Riemann}~(a). 
{
On the other hand, the {initial} increase in the decay width occurs as $m_{P^\ast}$ increases because the real part of the energy becomes {lower} and the wave function narrows. 
This increases the overlap with the wave function of the final state in the lower channel, and hence a larger decay width is obtained. 
}

Moreover, we discuss the HQS structure of the $T_{bb}(0(2^-))$ resonance {(triplet-2 in Table~\ref{table;energy_bound_resonant})}. 
As $m_{P^\ast}$ increases to the HQL, this state goes to a bound state whose mixing ratio is 
\begin{align}
    &f([PP^\ast]_-(^3P_2)): f([PP^\ast]_-(^3F_2)) = 3:1,\\
    &f(P^\ast P^\ast(^3P_2)): f(P^\ast P^\ast(^3F_2)) = 1:2. 
\end{align}
{
It indicates that the bound state in the HQL corresponds to the one} composed of 
\begin{equation}
    \begin{pmatrix}
        {\Ket{{\Big[{\big[QQ\big]}_{1}\ {\big[P\ {[\bar{q}\bar{q}]}_{1}\big]}_{2}\Big]}_{2}}}\vspace{1mm}\\
        {\Ket{{\Big[{\big[QQ\big]}_{1}\ {\big[F\ {[\bar{q}\bar{q}]}_{1}\big]}_2\Big]}_{2}}}
    \end{pmatrix}.
\end{equation}
Thus, this spin structure is $S_Q=1$ and $j_\ell=2$, which generates the HQS triplet with $0(1^-)$, $0(2^-)$ and $0(3^-)$ states. Hence, we expect that $T_{bb}(0(1^-))$ and $T_{bb}(0(3^-))$ states are also found as a HQS partner of the $T_{bb}(0(2^-))$ resonance in the bottom sector. However, a $T_{bb}(0(1^-))$ state belonging to this HQS triplet is not found due to the lack of an attraction in the present model.

Finally, we discuss the HQS multiplet strucrures of the isovector state with $P=+$. 
First, we study {the $1(0^+)$ state.} 
The mass dependence of {the {$T_{cc}(1(0^+))$ and} $T_{bb}(1(0^+))$ spectra} is shown in Fig.~\ref{fig;massdiff_Riemann}~(d). 
The solid line in this figure is the energy {eigenvalues} where {the mass of the vector meson} is {varied} from {$m_{D^{\ast}}$} to $2m_{B^\ast}$. 
In this mass region, we find that as the vector meson mass becomes larger, the real part of the energy eigenvalue becomes smaller, while the decay width becomes larger. 
{This trajectory structure is different from the ones of the others in Fig.~\ref{fig;massdiff_Riemann}. 
We consider that the $1(0^+)$ resonance becomes a virtual state though a state whose pole is located in the third-quadrant of the complex energy plane for $m_{P^*} \gtrsim 2m_{B^\ast}$ ~\cite{Masui:2000mug}. 
{However, the available region of the angle $\theta$ of the complex scaling method is between $0^\circ$ and $45^\circ$, and hence a virtual state is not accessible in this study~\cite{Suzuki:2005wv,Myo:2014ypa,Myo:2020rni}. }
{For $m_{P^*}\geq {3m_{B^\ast}}$, the $1(0^+)$ state transitions into a bound state which 
is corresponding to the second excited state in HQL~\cite{Sakai:2023syt} whose {component in LCB} is 
\begin{equation}
    \begin{pmatrix}
        {\Ket{{\Big[{\big[QQ\big]}_{0}\ {\big[P\ {[\bar{q}\bar{q}]}_{0}\big]}_{0}\Big]}_{0}}}
    \end{pmatrix}
    .
\end{equation}
This state with $S_Q=0$ and $j_\ell=0$ belongs to the HQS singlet {(singlet-4 in Table~\ref{table;energy_bound_resonant})}. 
We note that for $m_{P^\ast} > {1.9} m_{B^{\ast}}$, another bound state appears below the lowest threshold, which becomes} the first excited state of $T_{QQ}$ in the HQL~\cite{Sakai:2023syt}.}

Second, we discuss the HQS multiplet structures of $1(1^+)$ and $1(2^+)$ states. 
Based on the following discussion, the bound state of $T_{bb}$ with $1(1^+)$ and bound state of $T_{bb}$ with $1(2^+)$ belong to the HQS triplet {(triplet-3 in Table~\ref{table;energy_bound_resonant})}. 
Here, we review the HQS multiplet structures of $T_{bb}$ with $1(1^+)$ discussed in our previous study~\cite{Sakai:2023syt}. 
As the mass of the heavy meson increases, the bound state of $T_{bb}$ with $1(1^+)$ connects to the state whose component in the LCB is 
\begin{equation}
    \begin{pmatrix}
        \Ket{{\Big[{\big[QQ\big]}_{1}\ {\big[S\ {[\bar{q}\bar{q}]}_{1}\big]}_{1}\Big]}_{1}}\vspace{1mm}\\
        \Ket{{\Big[{\big[QQ\big]}_{1}\ {\big[D\ {[\bar{q}\bar{q}]}_{1}\big]}_{1}\Big]}_{1}}
    \end{pmatrix},
        \label{eq:LCB11+Bound}
\end{equation}
{with $S_Q=1$ and $j_\ell=1$,} smoothly. 
Thus, it turns out that $T_{bb}(1(1^+))$ is the member of the HQS triplet structure, 
{i.e. $S_Q\otimes j_\ell=0\oplus 1\oplus 2$. Hence, we expected an existence of a $T_{bb}(1(2^+))$ state as an HQS partner of the $T_{bb}(1(1^+))$ bound state\footnote{We note that the $1(0^+)$ state is also a member of the same HQS triplet and in fact it was found as the first excited state in the HQL in Ref.~\cite{Sakai:2023syt}. However, in the bottom sector, we do not obtain a $T_{bb}$ state corresponding to the member, while the obtained $T_{bb}(1(0^+))$ bound and resonant states in Table~\ref{table;energy_bound_resonant} belong to other HQS multiplets as discussed above. }. 

Next, we discuss the heavy quark spin structure of the $T_{bb}(1(2^+))$ resonance obtained in Table~\ref{table;energy_bound_resonant}. 
{Increasing the mass of the heavy meson,} the energy eigenvalue transitions from a resonant state to a bound state as shown in Fig.~\ref{fig;massdiff_Riemann}~(e). 
{
Basically, the decay width initially increases, while at some point, begins to decrease. This behavior is similar to that of the $0(2^-)$ state discussed above. There is also a bump structure of the trajectory in Fig.~\ref{fig;massdiff_Riemann}~(e). As the mass increases, the resonance energy decreases and simultaneously, an energy difference between first and second thresholds also decreases. Initially, the resonance energy is higher than the second threshold. However, at the point, the resonance comes below the threshold. 
}

The bound state corresponding to the origin of the $T_{bb}(1(2^+))$ resonance in the HQL has the following mixing ratio: 
\begin{align}
    &f(P^\ast P^\ast(^5S_2)) = 77.9\,\%,\\ 
    &f(PP(^1D_2)) = 3.31\, \%,\\
    &f([PP^\ast]_+(^3D_2)) = 9.93\, \%,\\
    &f(P^\ast P^\ast (^1D_2)) = 1.10\, \%,\\
    &f(P^\ast P^\ast (^5D_2)) = 7.72\, \%, \\
    &f(P^\ast P^\ast(^5G_2)) = 0\,\%,
\end{align}
and
$f(PP (^1D_2)) : f([PP^\ast]_+(^3D_2):f(P^\ast P^\ast (^1D_2)):f(P^\ast P^\ast (^5D_2))=3:9:1:7$. 
This mixing ratio indicates that the bound state is 
composed of  
\begin{equation}
    \begin{pmatrix}
        \Ket{{\Big[{\big[QQ\big]}_{1}\ {\big[S\ {[\bar{q}\bar{q}]}_{1}\big]}_{1}\Big]}_{2}}\vspace{1mm}\\
        \Ket{{\Big[{\big[QQ\big]}_{1}\ {\big[D\ {[\bar{q}\bar{q}]}_{1}\big]}_{1}\Big]}_{2}}
    \end{pmatrix}
\end{equation}
with $S_Q = 1$ and $j_\ell=1$ in the LCB in the HQL, so that this state is a member of the {same} HQS triplet {with the $1(1^+)$ state having the component ~\eqref{eq:LCB11+Bound}}.
Thus, the $T_{bb}(1(2^+))$ resonance is the partner with the $T_{bb}(1(1^+))$ bound state, where the same origin is shared in the HQL.

Let us briefly summarize the discussion in this subsubsection. 
As Table~\ref{table;summary_HQSmultiplet} shows, we have found the HQS multiplet structures of $T_{bb}$ obtained in our analysis. 
The bound and resonant states of $T_{bb}$ with $0(1^+)$ and $1(0^+)$ belong to each HQS   singlet {(singlet-1-4)}, respectively. 
Meanwhile, the bound states of $T_{bb}(0(0^-))$ and $T_{bb}(0(2^-))$ and resonant states of $T_{bb}(0(1^-))$ belong to the HQS triplet {(triplet-1)}. 
The resonant state of $T_{bb}(0(2^-))$ belongs to the HQS triplet {(triplet-2)}, but the $0(1^-)$ and $0(3^-)$ states are not found and analyzed in our analysis. 
Finally, the bound state of $T_{bb}(1(1^+))$ and resonant state of $T_{bb}(1(2^+))$ belong to the same HQS triplet {(triplet-3)}, while a $T_{bb}(1(0^+))$ which is partner with these is not found.

\renewcommand{\arraystretch}{1.25}
\begin{table*}[tbp]
  \caption{HQS multiplet structures of the obtained $T_{bb}$. The first column shows the singlet or triplet using the same notation in Table~\ref{table;energy_bound_resonant}, the second and third columns shows the spin structure $(S_Q,j_\ell)$ and direct product $S_Q \otimes j_\ell$. The last column shows $T_{bb}$ belonging to the HQS multiplet written on the first column. The symbols B, R mean the bound states and resonant states of $T_{bb}$, respectively, while the symbols NF, NA mean the $T_{bb}$ state which is not found and not analyzed, respectively. }
  \centering\begin{tabular}{cccc}
    \toprule[0.3mm]
    HQS multiplet structure \ \ & \ \ Spin structure $(S_Q, j_\ell)$ \ \ &\ \  $S_Q\otimes j_\ell$\ \  &\ \  $T_{bb}$ belonging to the HQS multiplet \\ \midrule[0.1mm] 
    singlet-1~\cite{Sakai:2023syt} & $(0,1)$ & $1$ & $0(1^+)$[B] \\
    singlet-2 & $(1,0)$ & $1$ & $0(1^+)$[R] \\
    singlet-3 & $(0,0)$ & $0$ & $1(0^+)$[B] \\
    singlet-4~\cite{Sakai:2023syt} & $(0,0)$ & $0$ & $1(0^+)$[R]\\
    \midrule[0.1mm] 
    triplet-1~\cite{Sakai:2023syt} & $(1,1)$ & $0 \oplus 1 \oplus 2$ & $0(0^-)$[B], $0(1^-)$[R], $0(2^-)$[B] \\
    triplet-2 & $(1,2)$ & $1 \oplus 2 \oplus 3$ & $0(1^-)$[NF], $0(2^-)$[R], $0(3^-)$[NA] \\
    triplet-3~\cite{Sakai:2023syt} & $(1,1)$ & $0 \oplus 1 \oplus 2$ & $1(0^+)$[NF], $1(1^+)$[B], $1(2^+)$[R] \\
    \bottomrule[0.3mm]
 \end{tabular}\label{table;summary_HQSmultiplet}
\end{table*}
\renewcommand{\arraystretch}{1.00}

\section{Summary\label{Sec;Summary}}
In this paper, we have studied the bound and resonant states of $T_{cc}$ and $T_{bb}$ with spin $J\leq 2$, while we have analyzed the bound state with spin $J=0,1$ in our previous study~\cite{Sakai:2023syt}. 
{The observation of} $T_{cc}$ with $0(1^+)$ reported by the LHCb experiment {located just below} the $DD^\ast$ threshold has motivated us to analyze this state as a molecular state. 
Thus, in our previous study~\cite{Sakai:2023syt}, we considered $T_{cc}$ as a $D^{(\ast)}D^{(\ast)}$ {bound state}, where we respected the chiral and heavy quark spin {symmeties}. 
We {applied} our model to {studying} bound and resonant states of $T_{cc}$ and $T_{bb}$ {for $J\leq 2$.} 
We solved the coupled Schr\"{o}dinger equation using the Gaussian expansion method and complex scaling method {to handle bound and resonant states of $T_{cc}$ and $T_{bb}$. 

As for $T_{cc}$, 
only the bound state of $T_{cc}(0(1^+))$ and resonance of $T_{cc}(1(0^+))$ have been found. 
However, the bound state of $T_{bb}$ with $0(2^-)$ and the resonant states of $T_{bb}$ with $0(1^+)$, $0(1^-)$, $0(2^-)$, $1(0^+)$ and $1(2^+)$ have been found along with the bound states of $T_{bb}$ obtained in our previous study~\cite{Sakai:2023syt}.  
With respect to resonances, we found that {many} {{$T_{cc}$ and}} $T_{bb}$ {states appear as} Feshbach resonances, which are the quasi bound states of the upper channels {$D^\ast D^\ast$ and} $B^\ast B^\ast${, respectively}. 
{For} the resonance of $T_{bb}(0(1^-))$, however, it is generated by the centrifugal barrier. 

Finally, we have investigated the HQS multiplet structures of $T_{cc}$ and $T_{bb}$ obtained in our studies by introducing the light cloud basis. 
This basis is characterized by the heavy quark spin $S_Q$ and light cloud spin $j_\ell$, which is everything but the heavy quark spin, because the transition between the states with the different spins vanishes in the heavy quark limit. 
{The suppression of the transition indicates that} the Hamiltonian in the light cloud basis is block-diagonalized matrix. 
Therefore, we {investigated} the heavy quark spin structure of a hadronic molecule, which enables us to clarify the HQS multiplet structure. 
In order to classify $T_{cc}$ and $T_{bb}$ by the HQS multiplet structure, we calculated the mass dependence of the energy eigenvalue. 
{As discussed in Ref.~\cite{Sakai:2023syt}, the bound states of $T_{cc}(0(1^+))$, corresponding to the one reported by LHCb, and $T_{bb}(0(1^+))$ belong to the HQS singlet (called singlet-1).}
It turned out that the resonant state of $T_{bb}$ with $0(1^+)$ is a member of another HQS singlet (singlet-2) which is the different {from the singlet-1}. 
{For the other isoscalar states, we found that}
the bound states of $T_{bb}$ with $0(0^-)$ and $0(2^-)$, and the resonant state of $T_{bb}$ with $0(1^-)$ belong to the same HQS triplet (triplet-1). 
{The $T_{bb}(0(2^-))$ resonance belongs to the HQS triplet (triplet-2), together with $0(1^-)$ and $0(3^-)$. However, in the bottom sector, the $T_{bb}(0(1^-))$ state is not bound due to the lack of an attraction. For the isovector states, the bound and resonant states of $T_{bb}(1(0^+))$ belong to the different HQS singlets, singlet-3 and singlet-4, respectively. }
The bound state of $T_{bb}$ with $1(1^+)$ obtained in our previous study~\cite{Sakai:2023syt} and the resonance of $T_{bb}$ with $1(2^+)$ belong to the same HQS triplet {(triplet-3)}, {together with} the $1(0^+)$ state. {However, the $T_{bb}(1(0^+))$ state as their partner} was not found in our analysis. 

{We consider that these predictions of the $T_{bb}$ state and the classification of them help to search these staters in future experiments.}
{
Finally, we mention the isospin-breaking effect. 
In this study, we did not include isospin-breaking effects in our analysis, because we focus on the HQS multiplet structures in the heavy quark limit. 
However, the masses of the bound state of $T_{cc}(0(1^+))$ and resonance of $T_{cc}(1(0^+))$ located just below the thresholds are expected to be changed by the isospin-breaking effect. 
Then, we should consider the isospin-effect as a future plan. 
}

\section*{Acknowledgement}
The authors thank T.~Myo for fruitful discussions. This work is supported by the RCNP Collaboration Research Network program as the project number COREnet-056 (Y.Y.).

\appendix
\section{Kinetic and Potential Matrices}
In this section, we show the kinetic and potential matrices {in the hadronic molecule basis as $K_{I(J^P)}$ and $V^{\mathrm{HM}}_{m,I(J^{P})}$ $(m=\pi,v,\sigma)$}:

\begin{widetext}
\begin{itemize}

    \item $0(0^{-})$
    \begin{align}
     &K_{0(0^-)} = \diag\left(
        -\frac{1}{2\mu_{PP^\ast}}\triangle_1
     \right),\\
      &V^{\mathrm{HM}}_{\pi,0(0^{-})} 
      = (C_{\pi}+2T_{\pi}),\\
      &V^{\mathrm{HM}}_{v,0(0^{-})} 
      = (C^{\prime}_{v}+2C_{v}-2T_{v}),\\
      &V^{\mathrm{HM}}_{\sigma,0(0^{-})}
      = (C_{\sigma}) ; 
    \end{align}

    \item $0(1^{+})$
    \begin{align}
      K_{0(1^+)} &= \diag\left( 
    -\frac{1}{2\mu_{PP^\ast}}\triangle_0, 
    -\frac{1}{2\mu_{PP^\ast}}\triangle_2,  
    -\frac{1}{2\mu_{P^\ast P^\ast}}\triangle_0 + \Delta m_{PP^\ast},
    -\frac{1}{2\mu_{P^\ast P^\ast}}\triangle_2 + \Delta m_{PP^\ast}
   \right),\\
      V^{\mathrm{HM}}_{\pi,0(1^{+})} 
      &=\begin{pmatrix}
        -C_{\pi}&\!\!\sqrt{2}T_{\pi}&\!\!2C_{\pi}&\!\!\sqrt{2}T_{\pi}\\\vspace{1mm}
        \sqrt{2}T_{\pi}&\!\!-C_{\pi}-T_{\pi}&\!\!\sqrt{2}T_{\pi}&\!\!2C_{\pi}-T_{\pi}\\\vspace{1mm}
        2C_{\pi}&\!\!\sqrt{2}T_{\pi}&\!\!-C_{\pi}&\!\!\sqrt{2}T_{\pi}\\\vspace{1mm}
        \sqrt{2}T_{\pi}&\!\!2C_{\pi}-T_{\pi}&\!\!\sqrt{2}T_{\pi}&\!\!-C_{\pi}-T_{\pi}
      \end{pmatrix},\\
      V^{\mathrm{HM}}_{v,0(1^{+})}
      &=\begin{pmatrix}
        C^{\prime}_{v}-2C_{v}&\!\!-\sqrt{2}T_{v}&\!\!4C_{v}&\!\!-\sqrt{2}T_{v}\\\vspace{1mm}
        -\sqrt{2}T_{v}&\!\!C^{\prime}_{v}-2C_{v}+T_{v}&\!\!-\sqrt{2}T_{v}&\!\!4C_{v}+T_{v}\\\vspace{1mm}
        4C_{v}&\!\!-\sqrt{2}T_{v}&\!\!C^{\prime}_{v}-2C_{v}&\!\!-\sqrt{2}T_{v}\\\vspace{1mm}
        -\sqrt{2}T_{v}&\!\!4C_{v}+T_{v}&\!\!-\sqrt{2}T_{v}&\!\!C^{\prime}_{v}-2C_{v}+T_{v}
      \end{pmatrix},\\
      V^{\HM}_{\sigma,0(1^{+})} 
      &= \begin{pmatrix}
        C_{\sigma} & 0 & 0 & 0\\
        0 & C_{\sigma} & 0 & 0\\
        0 & 0 & C_{\sigma} & 0\\
        0 & 0 & 0 & C_{\sigma}
      \end{pmatrix} ; 
    \end{align}

    \item $0(1^{-})$
    \begin{align}
      K_{0(1^-)} &= \left( 
    -\frac{1}{2\mu_{PP}}\triangle_1, 
    -\frac{1}{2\mu_{PP^\ast}}\triangle_1 + \Delta m_{PP^\ast},
    -\frac{1}{2\mu_{P^\ast P^\ast}}\triangle_1 + 2\Delta m_{PP^\ast}, \right.\nonumber \\
    &\left. 
    -\frac{1}{2\mu_{P^\ast P^\ast}}\triangle_1 + 2\Delta m_{PP^\ast},
    -\frac{1}{2\mu_{P^\ast P^\ast}}\triangle_3 + 2 \Delta m_{PP^\ast}
   \right),\\
      V^{\mathrm{HM}}_{\pi,0(1^{-})} &= \begin{pmatrix}
        0&\!\!0&\!\!-\sqrt{3}C_{\pi}&\!\!-2\sqrt{\frac{3}{5}}T_{\pi}&3\sqrt{\frac{2}{5}}T_{\pi}\\\vspace{1mm}
        0&\!\!C_{\pi}-T_{\pi}&\!\!0&\!\!-3\sqrt{\frac{3}{5}}T_{\pi}&\!\!-3\sqrt{\frac{2}{5}}T_{\pi}\\\vspace{1mm}
        -\sqrt{3}C_{\pi}&\!\!0&\!\!-2C_{\pi}&\!\!\frac{2}{\sqrt{5}}T_{\pi}&\!\!-\sqrt{\frac{6}{5}}T_{\pi}\\\vspace{1mm}
        -2\sqrt{\frac{3}{5}}T_{\pi}&\!\!-3\sqrt{\frac{3}{5}}T_{\pi}&\!\!\frac{2}{\sqrt{5}}T_{\pi}&\!\!C_{\pi}-\frac{7}{5}T_{\pi}&\!\!\frac{\sqrt{6}}{5}T_{\pi}\\\vspace{1mm}
        3\sqrt{\frac{2}{5}}T_{\pi}&\!\!-3\sqrt{\frac{2}{5}}T_{\pi}&\!\!-\sqrt{\frac{6}{5}}T_{\pi}&\!\!\frac{\sqrt{6}}{5}T_{\pi}&\!\!C_{\pi}-\frac{8}{5}T_{\pi}
      \end{pmatrix},\\
      V^{\mathrm{HM}}_{v,0(1^{-})} &= \begin{pmatrix}
        C^{\prime}_{v}&\!\!0&\!\!-2\sqrt{3}C_{v}&\!\!2\sqrt{\frac{3}{5}}T_{v}&-3\sqrt{\frac{2}{5}}T_{v}\\\vspace{1mm}
        0&\!\!C^{\prime}_{v}+2C_{v}+T_{v}&\!\!0&\!\!3\sqrt{\frac{3}{5}}T_{v}&\!\!3\sqrt{\frac{2}{5}}T_{v}\\\vspace{1mm}
        -2\sqrt{3}C_{v}&\!\!0&\!\!C^{\prime}_{v}-4C_{v}&\!\!-\frac{2}{\sqrt{5}}T_{v}&\!\!\sqrt{\frac{6}{5}}T_{v}\\\vspace{1mm}
        2\sqrt{\frac{3}{5}}T_{v}&\!\!3\sqrt{\frac{3}{5}}T_{v}&\!\!-\frac{2}{\sqrt{5}}T_{v}&\!\!C^{\prime}_{v}+2C_{v}+\frac{7}{5}T_{v}&\!\!-\frac{\sqrt{6}}{5}T_{v}\\\vspace{1mm}
        -3\sqrt{\frac{2}{5}}T_{v}&\!\!3\sqrt{\frac{2}{5}}T_{v}&\!\!\sqrt{\frac{6}{5}}T_{v}&\!\!-\frac{\sqrt{6}}{5}T_{v}&\!\!C^{\prime}_{v}+2C_{v}+\frac{8}{5}T_{v}
      \end{pmatrix},\\
      V^{\HM}_{\sigma,0(1^{-})} &= \begin{pmatrix}
        C_{\sigma} & 0 & 0 & 0 & 0\\
        0 & C_{\sigma} & 0 & 0 & 0\\
        0 & 0 & C_{\sigma} & 0 & 0\\
        0 & 0 & 0 & C_{\sigma} & 0\\
        0 & 0 & 0 & 0 & C_{\sigma}
      \end{pmatrix} ; 
    \end{align}


    \item $0(2^+)$
    
    \begin{align}
    K_{0(2^+)} &= \left( 
    -\frac{1}{2\mu_{P P^\ast}}\triangle_2 ,
    -\frac{1}{2\mu_{P^\ast P^\ast}}\triangle_2 + \Delta m_{PP^\ast}
    \right),\\
        V^\HM_{\pi,0(2^+)} &= \begin{pmatrix}
            -C_\pi + T_\pi &\!\! 2C_\pi + T_\pi \\ \vspace{1mm}
            2C_\pi + T_\pi &\!\! -C_\pi + T_\pi
        \end{pmatrix},\\
        V^\HM_{v,0(2^+)} &= \begin{pmatrix}
          C^\prime_v - 2C_v - T_v &\!\!  4C_v - T_v \\ \vspace{1mm}
          4C_v - T_v &\!\! C_v^\prime - 2C_v - T_v 
        \end{pmatrix}, \\
        V^\HM_{\sigma,0(2^+)} &= \begin{pmatrix}
          C_\sigma & 0 \\\vspace{1mm}
          0 & C_\sigma
        \end{pmatrix} ; 
    \end{align}

    \item $0(2^-)$
    \begin{align}
    K_{0(2^-)} &= \left( 
    -\frac{1}{2\mu_{PP^\ast}}\triangle_1, 
    -\frac{1}{2\mu_{PP^\ast}}\triangle_3,
    -\frac{1}{2\mu_{P^\ast P^\ast}}\triangle_1 + \Delta m_{PP^\ast},
    -\frac{1}{2\mu_{P^\ast P^\ast}}\triangle_3 + \Delta m_{PP^\ast},
   \right),\\
        V^\HM_{\pi,0(2^-)} &= \begin{pmatrix}
            C_\pi + \frac{1}{5}T_\pi &\!\! -\frac{3\sqrt{6}}{5}T_\pi &\!\! \frac{3\sqrt{3}}{5}T_\pi &\!\! -\frac{6\sqrt{3}}{5}T_\pi \\ \vspace{1mm}
            -\frac{3\sqrt{6}}{5}T_\pi &\!\! C_\pi + \frac{4}{5}T_\pi &\!\! \frac{3\sqrt{2}}{5}T_\pi &\!\! -\frac{6\sqrt{2}}{5}T_\pi \\ \vspace{1mm}
            \frac{3\sqrt{3}}{5}T_\pi &\!\! \frac{3\sqrt{2}}{5}T_\pi &\!\! C_\pi + \frac{7}{5}T_\pi &\!\! \frac{6}{5}T_\pi \\ \vspace{1mm}
            -\frac{6\sqrt{3}}{5}T_\pi &\!\! -\frac{6\sqrt{2}}{5}T_\pi &\!\! \frac{6}{5}T_\pi &\!\! C_\pi - \frac{2}{5}T_\pi
        \end{pmatrix}, \\
        V^\HM_{v,0(2^-)} &= \begin{pmatrix}
          C_v^\prime + 2C_v - \frac{1}{5}T_v &\!\! \frac{3\sqrt{6}}{5}T_v &\!\! -\frac{3\sqrt{3}}{5}T_v &\!\! \frac{6\sqrt{3}}{5}T_v \\ \vspace{1mm}
          \frac{3\sqrt{6}}{5}T_v &\!\! C^\prime_v + 2C_v - \frac{4}{5}T_v &\!\! -\frac{3\sqrt{2}}{5}T_v &\!\! \frac{6\sqrt{2}}{5}T_v \\\vspace{1mm}
          -\frac{3\sqrt{3}}{5}T_v &\!\! - \frac{3\sqrt{2}}{5}T_v &\!\! C^\prime_v + 2C_v - \frac{7}{5}T_v &\!\! -\frac{6}{5}T_V \\ \vspace{1mm}
          \frac{6\sqrt{3}}{5}T_v &\!\! \frac{6\sqrt{2}}{5}T_v &\!\! -\frac{6}{5}T_v &\!\! C^\prime_v + 2C_v + \frac{2}{5}T_v
        \end{pmatrix},\\
        V^\HM_{\sigma,0(2^-)} &= \begin{pmatrix}
          C_\sigma & 0 & 0 & 0 \\\vspace{1mm}
          0 & C_\sigma & 0 & 0 \\\vspace{1mm}
          0 & 0 & C_\sigma & 0 \\\vspace{1mm}
          0 & 0 & 0 & C_\sigma
        \end{pmatrix} ;
    \end{align}

    \item $1(0^{+})$
    \begin{align}
    K_{1(0^+)} &= \left( 
    -\frac{1}{2\mu_{PP}}\triangle_0 ,
    -\frac{1}{2\mu_{P^\ast P^\ast}}\triangle_0 + 2\Delta m_{PP^\ast},
    -\frac{1}{2\mu_{P^\ast P^\ast}}\triangle_2 + 2\Delta m_{PP^\ast} 
    \right),\\
      V^{\mathrm{HM}}_{\pi,1(0^{+})} &= \begin{pmatrix}
        0&\!\!-\sqrt{3}C_{\pi}&\!\!\sqrt{6}T_{\pi}\\\vspace{1mm}
        -\sqrt{3}C_{\pi}&\!\!-2C_{\pi}&\!\!-\sqrt{2}T_{\pi}\\\vspace{1mm}
        \sqrt{6}T_{\pi}&\!\!-\sqrt{2}T_{\pi}&\!\!C_{\pi}-2T_{\pi}
      \end{pmatrix},\\
      V^{\mathrm{HM}}_{v,1(0^{+})} &= \begin{pmatrix}
        C^{\prime}_{v}&\!\!-2\sqrt{3}C_{v}&\!\!-\sqrt{6}T_{v}\\\vspace{1mm}
        -2\sqrt{3}C_{v}&\!\!C^{\prime}_{v}-4C_{v}&\!\!\sqrt{2}T_{v}\\\vspace{1mm}
        -\sqrt{6}T_{v}&\!\!\sqrt{2}T_{v}&\!\!C^{\prime}_{v}+2C_{v}+2T_{v}
      \end{pmatrix},\\
      V^{\HM}_{\sigma,1(0^{+})} &= \begin{pmatrix}
        C_{\sigma}&0&0\\
        0&C_{\sigma}&0\\
        0&0&C_{\sigma}
      \end{pmatrix} ; 
    \end{align}

    \item $1(0^{-})$
    \begin{align}
    K_{1(0^-)} &= \left( 
      -\frac{1}{2\mu_{PP^\ast}}\triangle_1,
      -\frac{1}{2\mu_{P^\ast P^\ast}}\triangle_1 + \Delta m_{PP^\ast}
     \right), \\
      V^{\mathrm{HM}}_{\pi,1(0^{-})} &= \begin{pmatrix}
        -C_{\pi}-2T_{\pi}&\!\!2C_{\pi}-2T_{\pi}\\\vspace{1mm}
        2C_{\pi}-2T_{\pi}&\!\!-C_{\pi}-2T_{\pi}
      \end{pmatrix},\\
      V^{\mathrm{HM}}_{\pi,1(0^{-})} &= \begin{pmatrix}
        C^{\prime}_{v}-2C_{v}+2T_{v}&\!\!4C_{v}+2T_{v}\\\vspace{1mm}
        4C_{v}+2T_{v}&\!\!C^{\prime}_{v}-2C_{v}+2T_{v}
      \end{pmatrix},\\
      V^{\HM}_{\sigma,1(0^{-})} &= \begin{pmatrix}
        C_{\sigma}&0\\
        0&C_{\sigma}
      \end{pmatrix} ; 
    \end{align}

    \item $1(1^{+})$
    \begin{align}
    K_{1(1^+)} &= \left( 
      -\frac{1}{2\mu_{PP^\ast}}\triangle_0,
      -\frac{1}{2\mu_{PP^\ast}}\triangle_2,
      -\frac{1}{2\mu_{P^\ast P^\ast}}\triangle_2 + \Delta m_{PP^\ast}
      \right), \\
        V^{\mathrm{HM}}_{\pi,1(1^{+})} &= \begin{pmatrix}
          C_{\pi}&\!\!-\sqrt{2}T_{\pi}&\!\!-\sqrt{6}T_{\pi}\\\vspace{1mm}
          -\sqrt{2}T_{\pi}&\!\!C_{\pi}+T_{\pi}&\!\!-\sqrt{3}T_{\pi}\\\vspace{1mm}
          -\sqrt{6}T_{\pi}&\!\!-\sqrt{3}T_{\pi}&\!\!C_{\pi}-T_{\pi}
        \end{pmatrix},\\
        V^{\mathrm{HM}}_{v,1(1^{+})} &= \begin{pmatrix}
          C^{\prime}_{v}+2C_{v}&\!\!\sqrt{2}T_{v}&\!\!\sqrt{6}T_{v}\\\vspace{1mm}
          \sqrt{2}T_v&\!\!C^{\prime}_{v}+2C_{v}-T_{v}&\!\!\sqrt{3}T_{v}\\\vspace{1mm}
          \sqrt{6}T_{v}&\!\!\sqrt{3}T_{v}&\!\!C^{\prime}_{v}+2C_{v}+T_{v}
        \end{pmatrix},\\
        V^{\HM}_{\sigma,1(1^{+})} &= \begin{pmatrix}
          C_{\sigma} & 0 & 0\\
          0 & C_{\sigma} & 0\\
          0 & 0 & C_{\sigma}
        \end{pmatrix} ; 
    \end{align}
    
    \item $1(1^{-})$
    \begin{align}
    K_{1(1^-)} &= \left( 
        -\frac{1}{2\mu_{PP^\ast}}\triangle_1,
        -\frac{1}{2\mu_{P^\ast P^\ast}}\triangle_1 + \Delta m_{PP^\ast}
       \right),\\
      V^{\mathrm{HM}}_{\pi,1(1^{-})} &= \begin{pmatrix}
        -C_{\pi}+T_{\pi}&\!\!2C_{\pi}+T_{\pi}\\\vspace{1mm}
        2C_{\pi}+T_{\pi}&\!\!-C_{\pi}+T_{\pi}
      \end{pmatrix},\\
      V^{\mathrm{HM}}_{\pi,1(1^{-})} &= \begin{pmatrix}
        C^{\prime}_{v}-2C_{v}-T_{v}&\!\!4C_{v}-T_{v}\\\vspace{1mm}
        4C_{v}-T_{v}&\!\!C^{\prime}_{v}-2C_{v}-T_{v}
      \end{pmatrix},\\
      V^{\HM}_{\sigma,1(1^{-})} &= \begin{pmatrix}
        C_{\sigma}&0\\
        0&C_{\sigma}
      \end{pmatrix} ; 
    \end{align}

    \item $1(2^+)$ 

    \begin{align}
    &K_{1(2^+)} = \left( 
        -\frac{1}{2\mu_{PP}}\triangle_2, 
        -\frac{1}{2\mu_{PP^\ast}}\triangle_2 + \Delta m_{PP^\ast},
        -\frac{1}{2\mu_{P^\ast P^\ast}}\triangle_2 + 2\Delta m_{PP^\ast}, \right. \nonumber\\
        &\left.
          -\frac{1}{2\mu_{P^\ast P^\ast}}\triangle_0 + 2\Delta m_{PP^\ast}, 
          -\frac{1}{2\mu_{P^\ast P^\ast}}\triangle_2 + 2\Delta m_{PP^\ast},
          -\frac{1}{2\mu_{P^\ast P^\ast}}\triangle_4 + 2\Delta m_{PP^\ast}  
      \right),\\
        &V^\HM_{\pi, 1(2^+)} = \begin{pmatrix}
            0 &\!\! 0 &\!\! -\sqrt{3}C_\pi &\!\! \sqrt{\frac{6}{5}}T_\pi &\!\! -2 \sqrt{\frac{3}{7}}T_\pi &\!\! 6\sqrt{\frac{3}{35}}T_\pi \\\vspace{1mm} 
            0 &\!\! C_\pi - T_\pi &\!\! 0 &\!\! 3\sqrt{\frac{2}{5}}T_\pi &\!\! -\frac{3}{\sqrt{7}}T_\pi &\!\! - \frac{12}{\sqrt{35}}T_\pi \\ \vspace{1mm} 
            -\sqrt{3}C_\pi &\!\! 0 &\!\! 2C_\pi &\!\! -\sqrt{\frac{2}{5}}T_\pi &\!\! \frac{2}{\sqrt{7}}T_\pi &\!\! -\frac{6}{\sqrt{35}}T_\pi \\ \vspace{1mm}
            \sqrt{\frac{6}{5}}T_\pi &\!\! 3\sqrt{\frac{2}{5}}T_\pi &\!\! -\sqrt{\frac{2}{5}}T_\pi &\!\! C_\pi &\!\! \sqrt{\frac{14}{5}}T_\pi &\!\! 0 \\ \vspace{1mm} 
            -2\sqrt{\frac{3}{7}}T_\pi &\!\! -\frac{3}{\sqrt{7}}T_\pi &\!\! \frac{2}{\sqrt{7}}T_\pi &\!\! \sqrt{\frac{14}{5}}T_\pi &\!\! C_\pi + \frac{3}{7}T_\pi &\!\! \frac{12}{7\sqrt{5}}T_\pi \\ \vspace{1mm} 
            6\sqrt{\frac{3}{35}}T_\pi &\!\! -\frac{12}{\sqrt{35}}T_\pi &\!\! -\frac{6}{\sqrt{35}}T_\pi &\!\! 0 &\!\! \frac{12}{7\sqrt{5}}T_\pi &\!\! C_\pi - \frac{10}{7}T_\pi
        \end{pmatrix}, \\
        &V^\HM_{v,1(2^+)} \nonumber\\ &=  \begin{pmatrix}
          C^\prime_v &\!\! 0 &\!\! -2\sqrt{3}C_v &\!\! - \sqrt{\frac{6}{5}}T_v &\!\! {2\sqrt{\frac{3}{7}}T_v} &\!\! - 6\sqrt{\frac{3}{35}}T_v \\ \vspace{1mm}
          0 &\!\! C^\prime_v + 2C_v + T_v &\!\! 0 &\!\! -3\sqrt{\frac{2}{5}}T_v &\!\! \frac{3}{\sqrt{7}}T_v &\!\! \frac{12}{\sqrt{35}}T_v \\\vspace{1mm}
          -2\sqrt{3}C_v &\!\! 0 &\!\! C^\prime_v + 4C_v &\!\! \sqrt{\frac{2}{5}}T_v &\!\! -\frac{2}{\sqrt{7}}T_v &\!\! \frac{6}{\sqrt{35}}T_v \\\vspace{1mm}
          -\sqrt{\frac{6}{5}}T_v &\!\! -3\sqrt{\frac{2}{5}}T_v &\!\! \sqrt{\frac{2}{5}}T_v &\!\!C^\prime_v + 2C_v &\!\! -\sqrt{\frac{14}{5}}T_v &\!\! 0 \\\vspace{1mm}
          2\sqrt{\frac{3}{7}}T_v &\!\! \frac{3}{\sqrt{7}}T_v &\!\! \frac{2}{\sqrt{7}}T_v &\!\! -\sqrt{\frac{14}{5}}T_v &\!\! C^\prime_v + 2C_v - \frac{3}{7}T_v &\!\! -\frac{12}{7\sqrt{5}}T_v \\\vspace{1mm}
          -6\sqrt{\frac{3}{35}}T_v &\!\! \frac{12}{\sqrt{35}}T_v &\!\! \frac{6}{\sqrt{35}}T_v &\!\! 0 &\!\! -\frac{12}{7\sqrt{5}}T_v &\!\! C^\prime_v + 2C_v + \frac{10}{7}T_v
        \end{pmatrix}\nonumber\\
        &V^\HM_{\sigma,1(2^+)}  = \begin{pmatrix}
          C_\sigma & 0 & 0 & 0 & 0 & 0 \\\vspace{1mm}
          0 & C_\sigma & 0 & 0 & 0 & 0 \\\vspace{1mm}
          0 & 0 & C_\sigma & 0 & 0 & 0 \\\vspace{1mm}
          0 & 0 & 0 & C_\sigma & 0 & 0 \\\vspace{1mm}
          0 & 0 & 0 & 0 & C_\sigma & 0 \\\vspace{1mm}
          0 & 0 & 0 & 0 & 0 & C_\sigma 
        \end{pmatrix} ; 
    \end{align}
    \item $1(2^-)$

    \begin{align}
    K_{1(2^-)} &= \left(
        -\frac{1}{2\mu_{PP^\ast}}\triangle_1, 
        -\frac{1}{2\mu_{PP^\ast}}\triangle_3,
        -\frac{1}{2\mu_{P^\ast P^\ast}}\triangle_1 + \Delta m_{PP^\ast}, 
        -\frac{1}{2\mu_{P^\ast P^\ast}}\triangle_3 + \Delta m_{PP^\ast}
      \right),\\
        V^\HM_{\pi,1(2^-)} &= \begin{pmatrix}
            -C_\pi - \frac{1}{5}T_\pi &\!\! \frac{3\sqrt{6}}{5}T_\pi &\!\! 2C_\pi - \frac{1}{5}T_\pi &\!\! \frac{3\sqrt{6}}{5}T_\pi \\ \vspace{1mm}
            \frac{3\sqrt{6}}{5}T_\pi &\!\! -C_\pi - \frac{4}{5}T_\pi &\!\! \frac{3\sqrt{6}}{5}T_\pi &\!\! 2C_\pi - \frac{4}{5}T_\pi \\ \vspace{1mm}
            2C_\pi - \frac{1}{5}T_\pi &\!\! \frac{3\sqrt{6}}{5}T_\pi &\!\! -C_\pi - \frac{1}{5}T_\pi &\!\! \frac{3\sqrt{6}}{5}T_\pi \\ \vspace{1mm}
            \frac{3\sqrt{6}}{5}T_\pi &\!\! 2C_\pi - \frac{4}{5}T_\pi &\!\! \frac{3\sqrt{6}}{5}T_\pi &\!\!
            -C_\pi - \frac{4}{5}T_\pi
        \end{pmatrix},\\
        V^\HM_{v,1(2^-)} &= \begin{pmatrix}
          C^\prime_v - 2C_v + \frac{1}{5}T_\pi &\!\! -\frac{3\sqrt{6}}{5}T_v &\!\! 4C_\pi + \frac{1}{5}T_v &\!\! -\frac{3\sqrt{6}}{5}T_v \\
          -\frac{3\sqrt{6}}{5}T_v &\!\! C^\prime_v -2C_v + \frac{4}{5}T_v &\!\! -\frac{3\sqrt{6}}{5}T_v &\!\! 4C_v + \frac{4}{5}T_v \\
          4C_v + \frac{1}{5}T_v &\!\! - \frac{3\sqrt{6}}{5}T_v &\!\! C^\prime_v - 2C_v + \frac{1}{5}T_v &\!\! -\frac{3\sqrt{6}}{5}T_v \\
          -\frac{3\sqrt{6}}{5}T_v &\!\! 4C_v + \frac{4}{5}T_v &\!\! -\frac{3\sqrt{6}}{5}T_v &\!\! C^\prime_v - 2C_v + \frac{4}{5}T_v
        \end{pmatrix},\\
        V^\HM_{\sigma,1(2^-)} &= \begin{pmatrix}
          C_\sigma & 0 & 0 & 0 \\
          0 & C_\sigma & 0 & 0 \\
          0 & 0 & C_\sigma & 0 \\
          0 & 0 & 0 & C_\sigma
        \end{pmatrix} ; 
    \end{align}
    \end{itemize}

    \begin{align}
        \mu_{P^{(\ast)}P^{(\ast)}} &= \frac{m_{P^{(\ast)}}m_{P^{(\ast)}}}{m_{P^{(\ast)}}+m_{P^{(\ast)}}},\\
        \triangle_l &= \frac{d^2}{dr^2} - \frac{l(l+1)}{r^2},\\
        C_{\pi} &= \frac{1}{3}{\left(\frac{g_\pi}{2f_{\pi}}\right)}^2 C(r;m_{\pi})\vec{\tau}_{1}\cdot\vec{\tau}_{2},\\
        T_{\pi} &= \frac{1}{3}{\left(\frac{g_\pi}{2f_{\pi}}\right)}^2 T(r;m_{\pi})\vec{\tau}_{1}\cdot\vec{\tau}_{2},\\
        C^{\prime}_{v} &= {\left(\frac{\beta g_{V}}{2m_{v}}\right)}^2 C(r;m_{v})\vec{\tau}_{1}\cdot\vec{\tau}_{2},\\
        C_{v} &= \frac{1}{3}{(\lambda g_{V})}^2 C(r;m_{v})\vec{\tau}_{1}\cdot\vec{\tau}_{2},\\
        T_{v} &= \frac{1}{3}{(\lambda g_{V})}^2 T(r;m_{v})\vec{\tau}_{1}\cdot\vec{\tau}_{2},\\
        C_{\sigma} &= -{\left(\frac{g_\sigma}{m_{\sigma}}\right)}^2 C(r;m_{\sigma}).
      \end{align}

\end{widetext}

\section{Light-cloud basis}
We show the Light-Cloud basis {denoted by
\begin{equation}
    \Bigg[
    \Big[
    QQ
    \Big]_{S_Q}\ 
    \Big[L\ [
    \bar{q}\bar{q}
    ]_{S_q}  
    \Big]_{j_\ell}
    \Bigg]_J.
\end{equation}
}

\begin{widetext}
    \begin{itemize}

    \item $0(0^{-})$
    \begin{align}
      \psi^{\mathrm{HM}}_{0(0^{-})} &= \begin{pmatrix}
        \ket{{[PP^{*}]}_{+}({}^3P_{0})}
      \end{pmatrix},\\
      \psi^{\mathrm{LC}}_{0(0^{-})} &= \begin{pmatrix}
        -\Ket{{\Big[{\big[QQ\big]}_{1}\ {\big[P\ {[\bar{q}\bar{q}]}_{1}\big]}_{1}\Big]}_{0}}
      \end{pmatrix},\\
      V^{\mathrm{LC}}_{\pi,0(0^{-})} 
      &= (C_{\pi}+2T_{\pi}),\\
      V^{\mathrm{LC}}_{v,0(0^{-})} 
      &= (C^{\prime}_{v}+2C_{v}-2T_{v}),\\
      V^{\LC}_{\sigma,0(0^{-})}
      &= (C_{\sigma});
    \end{align}
    
    \item $0(1^{+})$
    \begin{align}
      {\psi^{\mathrm{HM}}_{0(1^{+})}} &= 
      \begin{pmatrix}
        \ket{{[PP^{*}]}_{-}({}^3S_{1})}\vspace{1mm}\\\ket{{[PP^{*}]}_{-}({}^3D_{1})}\vspace{1mm}\\
        \ket{P^{*}P^{*}({}^3S_{1})}\vspace{1mm}\\\ket{P^{*}P^{*}({}^3D_{1})}
      \end{pmatrix},\\
      \psi^{\mathrm{LC}}_{0(1^{+})} 
      &= U^{-1}_{0(1^{+})} \psi^{\mathrm{HM}}_{0(1^{+})}\notag \\
      &= 
      \begin{pmatrix}
        {\Ket{{\Big[{\big[QQ\big]}_{1}\ {\big[S\ {[\bar{q}\bar{q}]}_{0}\big]}_{0}\Big]}_{1}}}\vspace{1mm}\\
        {\Ket{{\Big[{\big[QQ\big]}_{0}\ {\big[S\ {[\bar{q}\bar{q}]}_{1}\big]}_{1}\Big]}_{1}}}\vspace{1mm}\\
        {\Ket{{\Big[{\big[QQ\big]}_{0}\ {\big[D\ {[\bar{q}\bar{q}]}_{1}\big]}_{1}\Big]}_{1}}}\vspace{1mm}\\
        {\Ket{{\Big[{\big[QQ\big]}_{1}\ {\big[D\ {[\bar{q}\bar{q}]}_{0}\big]}_{2}\Big]}_{1}}}\vspace{1mm}
      \end{pmatrix},\\
      U_{0(1^{+})} &= \begin{pmatrix}
        -\frac{1}{\sqrt{2}}&\frac{1}{\sqrt{2}}&0&0\vspace{1mm}\\
        0&0&\frac{1}{\sqrt{2}}&-\frac{1}{\sqrt{2}}\vspace{1mm}\\
        \frac{1}{\sqrt{2}}&\frac{1}{\sqrt{2}}&0&0\vspace{1mm}\\
        0&0&\frac{1}{\sqrt{2}}&\frac{1}{\sqrt{2}}
      \end{pmatrix}, \label{eq:U_01+}\\
      V^{\mathrm{LC}}_{\pi,0(1^{+})} 
      &=\left(\begin{array}{c|cc|c}
        -3C_{\pi} & 0 & 0 & 0 \\
        \hline
        0 & C_{\pi} & 2\sqrt{2}T_{\pi} & 0\\
        0 & 2\sqrt{2}T_{\pi} & C_{\pi}-2T_{\pi} & 0\\
        \hline
        0 & 0 & 0 & -3C_{\pi}
      \end{array}\right),\\
      V^{\mathrm{LC}}_{v,0(1^{+})} 
      &=\left(
        \begin{array}{c|cc|c}
          C^{\prime}_{v}-6C_{v} & 0 & 0 & 0\\\hline
          0 & C^{\prime}_{v}+2C_{v} & -2\sqrt{2}T_{v} & 0\\
          0 & -2\sqrt{2}T_{v} & C^{\prime}_{v}+2C_{v}+2T_{v} & 0\\\hline
          0 & 0 & 0 & C^{\prime}_{v}-6C_{v}
        \end{array}
      \right),\\
      V^{\mathrm{LC}}_{\sigma,0(1^{+})} 
      &=\left(
        \begin{array}{c|cc|c}
          C_{\sigma} & 0 & 0 & 0\\\hline
          0 & C_{\sigma} & 0 & 0\\
          0 & 0 & C_{\sigma} & 0\\\hline
          0 & 0 & 0 & C_{\sigma}
        \end{array}
      \right);
    \end{align}

    \item $0(1^{-})$
    \begin{align}
      \psi^{\mathrm{HM}}_{0(1^{-})} &= \begin{pmatrix}
        \ket{PP({}^1P_{1})}\vspace{1mm}\\
        \ket{{[PP^{*}]}_{+}({}^3P_{1})}\vspace{1mm}\\
        \ket{P^{*}P^{*}({}^1P_{1})}\vspace{1mm}\\
        \ket{P^{*}P^{*}({}^5P_{1})}\vspace{1mm}\\
        \ket{P^{*}P^{*}({}^5F_{1})}
      \end{pmatrix},\\
      \psi^{\mathrm{LC}}_{0(1^{-})} &= U^{-1}_{0(1^{-})}\psi^{\mathrm{HM}}_{0(1^{-})}\\
      &= \begin{pmatrix}
        \Ket{{\Big[{\big[QQ\big]}_{0}\ {\big[P\ {[\bar{q}\bar{q}]}_{0}\big]}_{1}\Big]}_{1}}\vspace{1mm}\\
        \Ket{{\Big[{\big[QQ\big]}_{1}\ {\big[P\ {[\bar{q}\bar{q}]}_{1}\big]}_{0}\Big]}_{1}}\vspace{1mm}\\
        \Ket{{\Big[{\big[QQ\big]}_{1}\ {\big[P\ {[\bar{q}\bar{q}]}_{1}\big]}_{1}\Big]}_{1}}\vspace{1mm}\\
        \Ket{{\Big[{\big[QQ\big]}_{1}\ {\big[P\ {[\bar{q}\bar{q}]}_{1}\big]}_{2}\Big]}_{1}}\vspace{1mm}\\
        \Ket{{\Big[{\big[QQ\big]}_{1}\ {\big[F\ {[\bar{q}\bar{q}]}_{1}\big]}_{2}\Big]}_{1}}
      \end{pmatrix},\\
      U_{0(1^{-})} 
      &= \begin{pmatrix}
        \frac{1}{2}&\frac{\sqrt{3}}{6}&\frac{1}{2}&\frac{\sqrt{15}}{6}&0\vspace{1mm}\\
        0&\frac{\sqrt{3}}{3}&\frac{1}{2}&-\frac{\sqrt{15}}{6}&0\vspace{1mm}\\
        \frac{\sqrt{3}}{2}&-\frac{1}{6}&-\frac{\sqrt{3}}{6}&-\frac{\sqrt{5}}{6}&0\vspace{1mm}\\
        0&\frac{\sqrt{5}}{3}&-\frac{\sqrt{15}}{6}&\frac{1}{6}&0\vspace{1mm}\\
        0 & 0 & 0 & 0 & 1
      \end{pmatrix}\label{eq:U_01-},\\
      V^{\mathrm{LC}}_{\pi,0(1^{-})} 
      &= \left(\begin{array}{c|c|c|cc}
        -3C_{\pi} & 0 & 0 & 0 & 0\\
        \hline
        0 & C_{\pi}-4T_{\pi} &  0 & 0 & 0\\
        \hline
        0 & 0 & C_{\pi}+2T_{\pi} & 0 & 0\\
        \hline
        0 & 0 & 0 & C_{\pi}-\frac{2}{5}T_{\pi} & \frac{6\sqrt{6}}{5}T_{\pi}\\
        0 & 0 & 0 & \frac{6\sqrt{6}}{5}T_{\pi} & C_{\pi}-\frac{8}{5}T_{\pi}
      \end{array}\right),\\
      V^{\mathrm{LC}}_{v,0(1^{-})} 
      &= \left(
        \begin{array}{c|c|c|cc}
          C^{\prime}_{v}-6C_{v} & 0 & 0 & 0 & 0\\\hline
          0 & C^{\prime}_{v}+2C_{v}+4T_{v} & 0 & 0 & 0\\\hline
          0 & 0 & C^{\prime}_{v}+2C_{v}-2T_{v} & 0 & 0\\\hline
          0 & 0 & 0 & C^{\prime}_{v}+2C_{v}+\frac{2}{5}T_{v} & -\frac{6\sqrt{6}}{5}T_{v}\\
          0 & 0 & 0 & -\frac{6\sqrt{6}}{5}T_{v}&\!\!C^{\prime}_{v}+2C_{v}+\frac{8}{5}T_{v}
        \end{array}
      \right),\\
      V^{\LC}_{\sigma,0(1^{-})} 
      &= \left(
        \begin{array}{c|c|c|cc}
          C_{\sigma} & 0 & 0 & 0 & 0\\\hline
          0 & C_{\sigma} & 0 & 0 & 0\\\hline
          0 & 0 & C_{\sigma} & 0 & 0\\\hline
          0 & 0 & 0 & C_{\sigma} & 0\\
          0 & 0 & 0 & 0 & C_{\sigma}
        \end{array}
      \right);
    \end{align}

    \item $0(2^+)$
    \begin{align}
      \psi^{\mathrm{HM}}_{0(2^{+})} &= \begin{pmatrix}
        \ket{[PP^\ast]_- (^3D_2)}\vspace{1mm}\\
        \ket{P^\ast P^\ast (^3D_2)}
      \end{pmatrix},\\
      \psi^{\mathrm{LC}}_{0(2^+)} &= U^{-1}_{0(2^+)}\psi^{\mathrm{HM}}_{0(2^+)}\\
      &= \begin{pmatrix}
        \Ket{{\Big[{\big[QQ\big]}_{0}\ {\big[D\ {[\bar{q}\bar{q}]}_{1}\big]}_{2}\Big]}_{2}}\vspace{1mm}\\
        \Ket{{\Big[{\big[QQ\big]}_{1}\ {\big[D\ {[\bar{q}\bar{q}]}_0 \big]}_{2}\Big]}_{2}}
      \end{pmatrix},\\
      U_{0(2^+)} &= \begin{pmatrix}
        \frac{1}{\sqrt{2}} &\!\! \frac{1}{\sqrt{2}} \\ \vspace{1mm}
        -\frac{1}{\sqrt{2}} &\!\! - \frac{1}{\sqrt{2}}
      \end{pmatrix},\\
      V^{\mathrm{LC}}_{\pi,0(2^+)}  
      &= \left(\begin{array}{c|c}
        C_\pi + 2T_\pi&0\\
        \hline
        0&-3C_\pi
      \end{array}\right),\\
      V^{\mathrm{LC}}_{\pi,0(2^+)} 
      &= \left(\begin{array}{c|c}
        C^\prime_v + 2C_v - 2T_v & 0 \\
        \hline
        0& C^\prime_v - 6C_v 
      \end{array}\right),\\
      V^{\mathrm{LC}}_{\sigma,0(2^+)} 
      &= \left(\begin{array}{c|c}
        C_\sigma & 0 \\
        \hline
        0 & C_\sigma
      \end{array}\right);
    \end{align}


    \item $0(2^-)$
    \begin{align}
      {\psi^{\mathrm{HM}}_{0(2^-)}} &= 
      \begin{pmatrix}
        \ket{{[PP^{*}]}_+({}^3P_2)}\vspace{1mm}\\
        \ket{{[PP^{*}]}_+({}^3F_2)}\vspace{1mm}\\
        \ket{P^{*}P^{*}({}^5P_2)}\vspace{1mm}\\
        \ket{P^{*}P^{*}(^5F_2)}
      \end{pmatrix},\\
      \psi^{\mathrm{LC}}_{0(2^-)} 
      &= U^{-1}_{0(2^{-})} \psi^{\mathrm{HM}}_{0(2^-)}\notag \\
      &= 
      \begin{pmatrix}
        {\Ket{{\Big[{\big[QQ\big]}_{1}\ {\big[P\ {[\bar{q}\bar{q}]}_{1}\big]}_{1}\Big]}_{2}}}\vspace{1mm}\\
        {\Ket{{\Big[{\big[QQ\big]}_{1}\ {\big[P\ {[\bar{q}\bar{q}]}_{1}\big]}_{2}\Big]}_{2}}}\vspace{1mm}\\
        {\Ket{{\Big[{\big[QQ\big]}_{1}\ {\big[F\ {[\bar{q}\bar{q}]}_{1}\big]}_2\Big]}_{2}}}\vspace{1mm}\\
        {\Ket{{\Big[{\big[QQ\big]}_{1}\ {\big[F\ {[\bar{q}\bar{q}]}_1\big]}_3\Big]}_2}}\vspace{1mm}
      \end{pmatrix},\\
      U_{0(2^-)} &= \begin{pmatrix}
        \frac{1}{2} &\!\! \frac{\sqrt{3}}{2} &\!\! 0 &\!\! 0 \\ \vspace{1mm}
        0 &\!\! 0 &\!\! -\frac{1}{\sqrt{3}} &\!\! - \sqrt{\frac{2}{3}} \\ \vspace{1mm}
        \frac{\sqrt{3}}{2} &\!\! - \frac{1}{2} &\!\! 0 &\!\! 0 \\ \vspace{1mm}
        0 &\!\! 0 &\!\! -\sqrt{\frac{2}{3}} &\!\! \frac{1}{\sqrt{3}}
      \end{pmatrix}\label{eq:U_02-},\\
      V^{\mathrm{LC}}_{\pi,0(2^-)} 
      &=\left(\begin{array}{c|cc|c}
        C_\pi + 2T_\pi &0&0&0\\
        \hline
        0&C_\pi - \frac{2}{5}T_\pi& \frac{6\sqrt{6}}{5}T_\pi & 0\\
        0&\frac{6\sqrt{6}}{5}T_\pi & C_\pi - \frac{8}{5}T_\pi & 0\\
        \hline
        0&0&0&C_\pi + 2T_\pi
      \end{array}\right),\\
      V^{\mathrm{LC}}_{v,0(2^-)} 
      &=\left(\begin{array}{c|cc|c}
        C^\prime_v + 2C_\pi - 2T_\pi & 0 & 0 & 0\\
        \hline
        0 & C^\prime_v + 2C_v + \frac{2}{5}T_v & -\frac{6\sqrt{6}}{5}T_v & 0\\
        0 & -\frac{6\sqrt{6}}{5}T_v & C^\prime_v + 2C_v + \frac{8}{5}T_v & 0\\
        \hline
        0 & 0 & 0 & C^\prime_v + 2C_v - 2T_v
      \end{array}\right),\\
      V^{\mathrm{LC}}_{\sigma,0(2^-)} 
      &=\left(\begin{array}{c|cc|c}
        C_\sigma & 0 & 0 & 0 \\
        \hline
        0 & C_\sigma & 0 & 0\\
        0 & 0 & C_\sigma & 0\\
        \hline
        0 & 0 & 0 & C_\sigma
      \end{array}\right);
    \end{align}

    \item $1(0^{+})$
    \begin{align}
      \psi^{\mathrm{HM}}_{1(0^{+})} &= \begin{pmatrix}
        \ket{PP({}^1S_{0})}\vspace{1mm}\\
        \ket{P^{*}P^{*}({}^1S_{0})}\vspace{1mm}\\
        \ket{P^{*}P^{*}({}^5D_{0})}\vspace{1mm}\\
      \end{pmatrix},\\
      \psi^{\mathrm{LC}}_{1(0^{+})} &= U^{-1}_{1(0^{+})}\psi^{\mathrm{HM}}_{1(0^{+})}\notag\\
      &=\begin{pmatrix}
        \Ket{{\Big[{\big[QQ\big]}_{0}\ {\big[S\ {[\bar{q}\bar{q}]}_{0}\big]}_{0}\Big]}_{0}}\vspace{1mm}\\
        \Ket{{\Big[{\big[QQ\big]}_{1}\ {\big[S\ {[\bar{q}\bar{q}]}_{1}\big]}_{1}\Big]}_{0}}\vspace{1mm}\\
        \Ket{{\Big[{\big[QQ\big]}_{1}\ {\big[D\ {[\bar{q}\bar{q}]}_{1}\big]}_{1}\Big]}_{0}}
      \end{pmatrix},\\
      U_{1(0^{+})} &= 
        \begin{pmatrix}
          \frac{1}{2}&\frac{\sqrt{3}}{2}&0\vspace{1mm}\\
          \frac{\sqrt{3}}{2}&-\frac{1}{2}&0\vspace{1mm}\\
          0&0&1
        \end{pmatrix},\\
      V^{\mathrm{LC}}_{\pi,1(0^{+})} 
      &= \left(\begin{array}{c|cc}
        -3C_{\pi}&0&0\\\hline
        0&C_{\pi}&2\sqrt{2}T_{\pi}\\
        0&2\sqrt{2}T_{\pi}&C_{\pi}-2T_{\pi}
      \end{array}\right),\\
      V^{\mathrm{LC}}_{v,1(0^{+})} 
      &= \left(
        \begin{array}{c|cc}
          C^{\prime}_{v}-6C_{v}&0&0\\\hline
          0&C^{\prime}_{v}+2C_{v}&-2\sqrt{2}T_{v}\\
          0&-2\sqrt{2}T_{v}&C^{\prime}_{v}+2C_{v}+2T_{v}
        \end{array}
      \right),\\
      V^{\LC}_{\sigma,1(0^{+})} 
      &= \left(
        \begin{array}{c|cc}
          C_{\sigma}&0&0\\\hline
          0&C_{\sigma}&0\\
          0&0&C_{\sigma}
        \end{array}
      \right);
    \end{align}
    
    \item $1(0^{-})$
    \begin{align}
      \psi^{\mathrm{HM}}_{1(0^{-})} &= \begin{pmatrix}
        \ket{{[PP^{*}]}_{-}({}^3P_{0})}\\
        \ket{P^{*}P^{*}({}^3P_{0})}
      \end{pmatrix},\\
      \psi^{\mathrm{LC}}_{1(0^{-})} &= U^{-1}_{1(0^{-})}\psi^{\mathrm{HM}}_{1(0^{-})}\notag\\
      &=\begin{pmatrix}
        \Ket{{\Big[{\big[QQ\big]}_{0}\ {\big[P\ {[\bar{q}\bar{q}]}_{1}\big]}_{0}\Big]}_{0}}\vspace{1mm}\\
        \Ket{{\Big[{\big[QQ\big]}_{1}\ {\big[P\ {[\bar{q}\bar{q}]}_{0}\big]}_{1}\Big]}_{0}}
      \end{pmatrix},\\
      U_{1(0^{-})} &= 
      \begin{pmatrix}
        \frac{1}{\sqrt{2}}&-\frac{1}{\sqrt{2}}\vspace{1mm}\\
        \frac{1}{\sqrt{2}}&\frac{1}{\sqrt{2}}
      \end{pmatrix},\\
      V^{\mathrm{LC}}_{\pi,1(0^{-})}
      &= \left(
        \begin{array}{c|c}
          C_{\pi}-4T_{\pi}&0\\\hline
          0&-3C_{\pi}
        \end{array}
      \right),\\
      V^{\mathrm{LC}}_{\pi,1(0^{-})}
      &= \left(
        \begin{array}{c|c}
          C^{\prime}_{v}+2C_{v}+4T_{v}&0\\\hline
          0&C^{\prime}_{v}-6C_{v}
        \end{array}
      \right),\\
      V^{\LC}_{\sigma,1(0^{-})} 
      &= \left(
        \begin{array}{c|c}
          C_{\sigma}&0\\\hline
          0&C_{\sigma}
        \end{array}
      \right);
    \end{align}
    \item $1(1^{+})$
    \begin{align}
      \psi^{\mathrm{HM}}_{1(1^{+})} &= \begin{pmatrix}
        \ket{{[PP^{*}]}_{+}({}^3S_{1})}\vspace{1mm}\\
        \ket{{[PP^{*}]}_{+}({}^3D_{1})}\vspace{1mm}\\
        \ket{P^{*}P^{*}({}^5D_{1})}\vspace{1mm}\\
      \end{pmatrix},\\
      \psi^{\mathrm{LC}}_{1(1^{+})} &= U^{-1}_{1(1^{+})}\psi^{\mathrm{HM}}_{1(1^{+})}\notag\\
      &=\begin{pmatrix}
        \Ket{{\Big[{\big[QQ\big]}_{1}\ {\big[S\ {[\bar{q}\bar{q}]}_{1}\big]}_{1}\Big]}_{1}}\vspace{1mm}\\
        \Ket{{\Big[{\big[QQ\big]}_{1}\ {\big[D\ {[\bar{q}\bar{q}]}_{1}\big]}_{1}\Big]}_{1}}\vspace{1mm}\\
        \Ket{{\Big[{\big[QQ\big]}_{1}\ {\big[D\ {[\bar{q}\bar{q}]}_{1}\big]}_{2}\Big]}_{1}}
      \end{pmatrix},\\
      U_{1(1^{+})} &= 
        \begin{pmatrix}
          1&0&0\vspace{1mm}\\
          0&-\frac{1}{2}&-\frac{\sqrt{3}}{2}\vspace{1mm}\\
          0&-\frac{\sqrt{3}}{2}&\frac{1}{2}
        \end{pmatrix},\\
      V^{\mathrm{LC}}_{\pi,1(1^{+})} 
      &= \left(\begin{array}{cc|c}
        C_{\pi}&2\sqrt{2}T_{\pi}&0\\
        2\sqrt{2}T_{\pi}&C_{\pi}-2T_{\pi}&0\\\hline
        0&0&C_{\pi}+2T_{\pi}
      \end{array}\right),\\
      V^{\mathrm{LC}}_{v,1(1^{+})}
      &= \left(
        \begin{array}{cc|c}
          C^{\prime}_{v}+2C_{v}&-2\sqrt{2}T_{v}&0\\
          -2\sqrt{2}T_{v}&C^{\prime}_{v}+2C_{v}+2T_{v}&0\\\hline
          0&0&C^{\prime}_{v}+2C_{v}-2T_{v}
        \end{array}
      \right),\\
      V^{\LC}_{\sigma,1(1^{+})} 
      &= \left(
        \begin{array}{cc|c}
          C_{\sigma} & 0 & 0\\
          0 & C_{\sigma} & 0\\\hline
          0 & 0 & C_{\sigma}
        \end{array}
      \right);
    \end{align}
    \item $1(1^{-})$
    \begin{align}
      \psi^{\mathrm{HM}}_{1(1^{-})} &= \begin{pmatrix}
        \ket{{[PP^{*}]}_{-}({}^3P_{1})}\\
        \ket{P^{*}P^{*}({}^3P_{1})}
      \end{pmatrix},\\
      \psi^{\mathrm{LC}}_{1(1^{-})} &= U^{-1}_{1(1^{-})}\psi^{\mathrm{HM}}_{1(1^{-})}\notag\\
      &=\begin{pmatrix}
        \Ket{{\Big[{\big[QQ\big]}_{0}\ {\big[P\ {[\bar{q}\bar{q}]}_{1}\big]}_{1}\Big]}_{1}}\vspace{1mm}\\
        \Ket{{\Big[{\big[QQ\big]}_{1}\ {\big[P\ {[\bar{q}\bar{q}]}_{0}\big]}_{1}\Big]}_{1}}
      \end{pmatrix},\\
      U_{1(1^{-})} &= 
      \begin{pmatrix}
        \frac{1}{\sqrt{2}}&\frac{1}{\sqrt{2}}\vspace{1mm}\\
        \frac{1}{\sqrt{2}}&-\frac{1}{\sqrt{2}}
      \end{pmatrix},\\
      V^{\mathrm{LC}}_{\pi,1(1^{-})} 
      &= \left(
        \begin{array}{c|c}
          C_{\pi}+2T_{\pi}&0\\\hline
          0&-3C_{\pi}
        \end{array}
      \right),\\
      V^{\mathrm{LC}}_{\pi,1(1^{-})} 
      &= \left(
        \begin{array}{c|c}
          C^{\prime}_{v}+2C_{v}-2T_{v}&0\\\hline
          0&C^{\prime}_{v}-6C_{v}
        \end{array}
      \right),\\
      V^{\LC}_{\sigma,1(1^{-})} 
      &= \left(
        \begin{array}{c|c}
          C_{\sigma}&0\\\hline
          0&C_{\sigma}
        \end{array}
      \right);
    \end{align}

    \item $1(2^+)$
    \begin{align}
      &\psi^{\mathrm{HM}}_{1(2^+)} = \begin{pmatrix}
        \ket{PP(^1 D_2)}\vspace{1mm}\\
        \ket{{[PP^{*}]}_{+}(^3D_2)}\vspace{1mm}\\
        \ket{P^{*}P^{*}(^1D_2)}\vspace{1mm}\\
        \ket{P^{*}P^{*}(^5S_2)}\vspace{1mm}\\
        \ket{P^{*}P^{*}(^5D_2)} \vspace{1mm}\\
        \ket{P^\ast P^\ast (^5G_2)} \vspace{1mm}
      \end{pmatrix},\\
      &\psi^{\mathrm{LC}}_{1(2^+)} 
      = U^{-1}_{1(2^+)}\psi^{\mathrm{HM}}_{1(2^+)}\\
      &= \begin{pmatrix}
        \Ket{{\Big[{\big[QQ\big]}_{0}\ {\big[D\ {[\bar{q}\bar{q}]}_{0}\big]}_{2}\Big]}_{2}}\vspace{1mm}\\
        \Ket{{\Big[{\big[QQ\big]}_{1}\ {\big[S\ {[\bar{q}\bar{q}]}_{1}\big]}_{1}\Big]}_{2}}\vspace{1mm}\\
        \Ket{{\Big[{\big[QQ\big]}_{1}\ {\big[D\ {[\bar{q}\bar{q}]}_{1}\big]}_{1}\Big]}_{2}}\vspace{1mm}\\
        \Ket{{\Big[{\big[QQ\big]}_{1}\ {\big[D\ {[\bar{q}\bar{q}]}_{1}\big]}_{2}\Big]}_{2}}\vspace{1mm}\\
        \Ket{{\Big[{\big[QQ\big]}_{1}\ {\big[D\ {[\bar{q}\bar{q}]}_{1}\big]}_{3}\Big]}_{2}} \vspace{1mm}\\
        \Ket{{\Big[{\big[QQ\big]}_{1}\ {\big[G\ {[\bar{q}\bar{q}]}_{1}\big]}_{3}\Big]}_{2}}
      \end{pmatrix},\\
      &U_{1(2^+)} = \begin{pmatrix}
        \frac{1}{2} &\!\! 0 &\!\! \frac{\sqrt{15}}{10} &\!\! \frac{1}{2} &\!\! \frac{\sqrt{35}}{10} &\!\! 0 \\ \vspace{1mm}
        0 &\!\! 0 &\!\! \frac{3\sqrt{5}}{10} &\!\! \frac{\sqrt{3}}{6} &\!\! - \sqrt{\frac{7}{15}} &\!\! 0\\ \vspace{1mm}
        \frac{\sqrt{3}}{2} &\!\! 0 &\!\! - \frac{\sqrt{5}}{10} &\!\! -\frac{\sqrt{3}}{6} &\!\! -\frac{\sqrt{105}}{30} &0\\ \vspace{1mm} 
        0 &\!\! 1 &\!\! 0 &\!\!0 &\!\!0 &\!\!0 \\ \vspace{1mm}
        0 &\!\!0 &\!\!\frac{\sqrt{35}}{10} &\!\! - \frac{\sqrt{21}}{6} &\!\! \frac{1}{\sqrt{15}} &\!\! 0 \\ \vspace{1mm} 
        0 &\!\! 0 &\!\!0 &\!\!0 &\!\!0 &\!\!1
      \end{pmatrix},\\
      &V^{\mathrm{LC}}_{\pi,1(2^+)} 
      = \left(\begin{array}{c|cc|c|cc}
        -3C_{\pi} & 0 & 0 & 0 & 0 & 0\\
        \hline
        0 & C_{\pi} & 2\sqrt{2}T_\pi & 0 & 0 & 0\\
        0 & 2\sqrt{2}T_\pi & C_{\pi}-2T_{\pi} & 0 & 0 & 0\\
        \hline 
        0 & 0 & 0 & C_\pi+2T_\pi & 0 & 0 \\
        \hline
        0 &\!\! 0 &\!\!0 &\!\! 0 &\!\! C_\pi - \frac{4}{7}T_\pi &\!\! \frac{12\sqrt{3}}{7}T_\pi \\ 
        0 &\!\! 0 &\!\!0 &\!\! 0 &\!\! \frac{12\sqrt{3}}{7}T_\pi &\!\! C_\pi - \frac{10}{7}T_\pi
      \end{array}\right),\\
      &V^{\mathrm{LC}}_{v,1(2^{+})} \nonumber\\
      &= \left(\begin{array}{c|cc|c|cc}
      C^\prime_v-6C_v & 0 & 0 & 0 & 0 & 0 \\
      \hline
      0 & C_v^\prime + 4C_v & -2\sqrt{2}T_v& 0 & 0 & 0 \\
      0 & -2\sqrt{2}T_\pi & C^\prime_v + C_v - 2T_v & 0 & 0 & 0\\
      \hline
      0 & 0 & 0 & C^\prime + 2C_v - 2T_v & 0 & 0 \\
      \hline
      0 &\!\! 0 &\!\!0 &\!\! 0 &\!\! C^\prime_v + 2C_\pi + \frac{4}{7}T_v &\!\! -\frac{12\sqrt{3}}{7}T_v \\ 
      0 &\!\! 0 &\!\!0 &\!\! 0 &\!\! -\frac{12\sqrt{3}}{7}T_v &\!\! C^\prime - 2C_v + \frac{10}{7}T_v
    \end{array}\right),\\
      &V^{\LC}_{\sigma,1(2^+)} = \left(
        \begin{array}{c|cc|c|cc}
          C_{\sigma} & 0 & 0 & 0 & 0 & 0\\\hline
          0 & C_{\sigma} & 0 & 0 & 0 & 0\\
          0 & 0 & C_{\sigma} & 0 & 0 & 0\\\hline
          0 & 0 & 0 & C_{\sigma} & 0 & 0\\\hline
          0 & 0 & 0 & 0 & C_{\sigma} & 0\\
          0 & 0 & 0 & 0 & 0 & C_\sigma
        \end{array}
      \right);
    \end{align}

    \item $1(2^-)$
    \begin{align}
      {\psi^{\mathrm{HM}}_{1(2^-)}} &= 
      \begin{pmatrix}
        \ket{{[PP^{*}]}_-({}^3P_2)}\vspace{1mm}\\
        \ket{{[PP^{*}]}_-({}^3F_2)}\vspace{1mm}\\
        \ket{P^{*}P^{*}({}^3P_2)}\vspace{1mm}\\
        \ket{P^{*}P^{*}(^3F_2)}
      \end{pmatrix},\\
      \psi^{\mathrm{LC}}_{1(2^-)} 
      &= U^{-1}_{1(2^{-})} \psi^{\mathrm{HM}}_{1(2^-)}\notag \\
      &= 
      \begin{pmatrix}
        {\Ket{{\Big[{\big[QQ\big]}_{0}\ {\big[P\ {[\bar{q}\bar{q}]}_{1}\big]}_{2}\Big]}_{2}}}\vspace{1mm}\\
        {\Ket{{\Big[{\big[QQ\big]}_{0}\ {\big[F\ {[\bar{q}\bar{q}]}_{1}\big]}_{2}\Big]}_{2}}}\vspace{1mm}\\
        {\Ket{{\Big[{\big[QQ\big]}_{1}\ {\big[P\ {[\bar{q}\bar{q}]}_0\big]}_1\Big]}_{2}}}\vspace{1mm}\\
        {\Ket{{\Big[{\big[QQ\big]}_{1}\ {\big[F\ {[\bar{q}\bar{q}]}_0\big]}_3\Big]}_2}}\vspace{1mm}
      \end{pmatrix},\\
      U_{1(2^-)} &= \begin{pmatrix}
        \frac{1}{\sqrt{2}} &\!\! 0 &\!\! -\frac{1}{\sqrt{2}} &\!\! 0 \\\vspace{1mm}
        0 &\!\! \frac{1}{\sqrt{2}} &\!\! 0 &\!\! -\frac{1}{\sqrt{2}} \\\vspace{1mm} 
        \frac{1}{\sqrt{2}} &\!\! 0 &\!\! \frac{1}{\sqrt{2}} &\!\! 0 \\\vspace{1mm}
        0 &\!\! \frac{1}{\sqrt{2}} &\!\! 0 &\!\! \frac{1}{\sqrt{2}}
      \end{pmatrix},\\
      V^{\mathrm{LC}}_{\pi,1(2^-)} 
      &=\left(\begin{array}{cc|c|c}
        C_\pi - \frac{2}{5}T_\pi & \frac{6\sqrt{6}}{5}T_\pi & 0 & 0\\ 
        \frac{6\sqrt{6}}{5}T_\pi & C_\pi - \frac{8}{5}T_\pi & 0 & 0\\
        \hline
        0 & 0 & -3C_\pi & 0 \\\hline
        0 & 0 & 0 & -3C_\pi 
      \end{array}\right),\\
      V^{\mathrm{LC}}_{v,1(2^-)} 
      &=\left(\begin{array}{cc|c|c}
        C^\prime_v + 2C_v + \frac{2}{5}T_v & -\frac{6\sqrt{6}}{5}T_v & 0 & 0\\ 
        -\frac{6\sqrt{6}}{5}T_v & C^\prime_v + C_v + \frac{8}{5}T_v & 0 & 0\\
        \hline
        0 & 0 & C^\prime_v - 6C_v & 0 \\\hline
        0 & 0 & 0 & C^\prime_v - 6C_v 
      \end{array}\right),\\
      V^{\mathrm{LC}}_{\sigma,1(2^-)} 
      &=\left(
        \begin{array}{cc|c|c}
          C_{\sigma} & 0 & 0 & 0\\
          0 & C_{\sigma} & 0 & 0\\\hline
          0 & 0 & C_{\sigma} & 0\\\hline
          0 & 0 & 0 & C_{\sigma}
        \end{array}
      \right).
    \end{align}

  \end{itemize}
\end{widetext}

\bibliography{apssamp}

\begin{thebibliography}{63}%
\makeatletter
\providecommand \@ifxundefined [1]{%
 \@ifx{#1\undefined}
}%
\providecommand \@ifnum [1]{%
 \ifnum #1\expandafter \@firstoftwo
 \else \expandafter \@secondoftwo
 \fi
}%
\providecommand \@ifx [1]{%
 \ifx #1\expandafter \@firstoftwo
 \else \expandafter \@secondoftwo
 \fi
}%
\providecommand \natexlab [1]{#1}%
\providecommand \enquote  [1]{``#1''}%
\providecommand \bibnamefont  [1]{#1}%
\providecommand \bibfnamefont [1]{#1}%
\providecommand \citenamefont [1]{#1}%
\providecommand \href@noop [0]{\@secondoftwo}%
\providecommand \href [0]{\begingroup \@sanitize@url \@href}%
\providecommand \@href[1]{\@@startlink{#1}\@@href}%
\providecommand \@@href[1]{\endgroup#1\@@endlink}%
\providecommand \@sanitize@url [0]{\catcode `\\12\catcode `\$12\catcode `\&12\catcode `\#12\catcode `\^12\catcode `\_12\catcode `\%12\relax}%
\providecommand \@@startlink[1]{}%
\providecommand \@@endlink[0]{}%
\providecommand \url  [0]{\begingroup\@sanitize@url \@url }%
\providecommand \@url [1]{\endgroup\@href {#1}{\urlprefix }}%
\providecommand \urlprefix  [0]{URL }%
\providecommand \Eprint [0]{\href }%
\providecommand \doibase [0]{https://doi.org/}%
\providecommand \selectlanguage [0]{\@gobble}%
\providecommand \bibinfo  [0]{\@secondoftwo}%
\providecommand \bibfield  [0]{\@secondoftwo}%
\providecommand \translation [1]{[#1]}%
\providecommand \BibitemOpen [0]{}%
\providecommand \bibitemStop [0]{}%
\providecommand \bibitemNoStop [0]{.\EOS\space}%
\providecommand \EOS [0]{\spacefactor3000\relax}%
\providecommand \BibitemShut  [1]{\csname bibitem#1\endcsname}%
\let\auto@bib@innerbib\@empty
\bibitem [{\citenamefont {Choi}\ \emph {et~al.}(2003)\citenamefont {Choi} \emph {et~al.}}]{Belle:2003nnu}%
  \BibitemOpen
  \bibfield  {author} {\bibinfo {author} {\bibfnamefont {S.~K.}\ \bibnamefont {Choi}} \emph {et~al.} (\bibinfo {collaboration} {Belle}),\ }\bibfield  {title} {\bibinfo {title} {{Observation of a narrow charmonium-like state in exclusive $B^\pm \to K^\pm \pi^+ \pi^- J/\psi$ decays}},\ }\href {https://doi.org/10.1103/PhysRevLett.91.262001} {\bibfield  {journal} {\bibinfo  {journal} {Phys. Rev. Lett.}\ }\textbf {\bibinfo {volume} {91}},\ \bibinfo {pages} {262001} (\bibinfo {year} {2003})},\ \Eprint {https://arxiv.org/abs/hep-ex/0309032} {arXiv:hep-ex/0309032} \BibitemShut {NoStop}%
\bibitem [{\citenamefont {Aubert}\ \emph {et~al.}(2005)\citenamefont {Aubert} \emph {et~al.}}]{BaBar:2005hhc}%
  \BibitemOpen
  \bibfield  {author} {\bibinfo {author} {\bibfnamefont {B.}~\bibnamefont {Aubert}} \emph {et~al.} (\bibinfo {collaboration} {BaBar}),\ }\bibfield  {title} {\bibinfo {title} {{Observation of a broad structure in the $\pi^+ \pi^- J/\psi$ mass spectrum around 4.26-GeV/c$^2$}},\ }\href {https://doi.org/10.1103/PhysRevLett.95.142001} {\bibfield  {journal} {\bibinfo  {journal} {Phys. Rev. Lett.}\ }\textbf {\bibinfo {volume} {95}},\ \bibinfo {pages} {142001} (\bibinfo {year} {2005})},\ \Eprint {https://arxiv.org/abs/hep-ex/0506081} {arXiv:hep-ex/0506081} \BibitemShut {NoStop}%
\bibitem [{\citenamefont {Yuan}\ \emph {et~al.}(2007)\citenamefont {Yuan} \emph {et~al.}}]{Belle:2007dxy}%
  \BibitemOpen
  \bibfield  {author} {\bibinfo {author} {\bibfnamefont {C.~Z.}\ \bibnamefont {Yuan}} \emph {et~al.} (\bibinfo {collaboration} {Belle}),\ }\bibfield  {title} {\bibinfo {title} {{Measurement of e+ e- ---\ensuremath{>} pi+ pi- J/psi cross-section via initial state radiation at Belle}},\ }\href {https://doi.org/10.1103/PhysRevLett.99.182004} {\bibfield  {journal} {\bibinfo  {journal} {Phys. Rev. Lett.}\ }\textbf {\bibinfo {volume} {99}},\ \bibinfo {pages} {182004} (\bibinfo {year} {2007})},\ \Eprint {https://arxiv.org/abs/0707.2541} {arXiv:0707.2541 [hep-ex]} \BibitemShut {NoStop}%
\bibitem [{\citenamefont {Choi}\ \emph {et~al.}(2008)\citenamefont {Choi} \emph {et~al.}}]{Belle:2007hrb}%
  \BibitemOpen
  \bibfield  {author} {\bibinfo {author} {\bibfnamefont {S.~K.}\ \bibnamefont {Choi}} \emph {et~al.} (\bibinfo {collaboration} {Belle}),\ }\bibfield  {title} {\bibinfo {title} {{Observation of a resonance-like structure in the $pi^\pm \psi^\prime$ mass distribution in exclusive $B \to K \pi^\pm \psi^\prime$ decays}},\ }\href {https://doi.org/10.1103/PhysRevLett.100.142001} {\bibfield  {journal} {\bibinfo  {journal} {Phys. Rev. Lett.}\ }\textbf {\bibinfo {volume} {100}},\ \bibinfo {pages} {142001} (\bibinfo {year} {2008})},\ \Eprint {https://arxiv.org/abs/0708.1790} {arXiv:0708.1790 [hep-ex]} \BibitemShut {NoStop}%
\bibitem [{\citenamefont {Aubert}\ \emph {et~al.}(2009)\citenamefont {Aubert} \emph {et~al.}}]{BaBar:2008bxw}%
  \BibitemOpen
  \bibfield  {author} {\bibinfo {author} {\bibfnamefont {B.}~\bibnamefont {Aubert}} \emph {et~al.} (\bibinfo {collaboration} {BaBar}),\ }\bibfield  {title} {\bibinfo {title} {{Search for the Z(4430)- at BABAR}},\ }\href {https://doi.org/10.1103/PhysRevD.79.112001} {\bibfield  {journal} {\bibinfo  {journal} {Phys. Rev. D}\ }\textbf {\bibinfo {volume} {79}},\ \bibinfo {pages} {112001} (\bibinfo {year} {2009})},\ \Eprint {https://arxiv.org/abs/0811.0564} {arXiv:0811.0564 [hep-ex]} \BibitemShut {NoStop}%
\bibitem [{\citenamefont {Aaij}\ \emph {et~al.}(2015)\citenamefont {Aaij} \emph {et~al.}}]{LHCb:2015yax}%
  \BibitemOpen
  \bibfield  {author} {\bibinfo {author} {\bibfnamefont {R.}~\bibnamefont {Aaij}} \emph {et~al.} (\bibinfo {collaboration} {LHCb}),\ }\bibfield  {title} {\bibinfo {title} {{Observation of $J/\psi p$ Resonances Consistent with Pentaquark States in $\Lambda_b^0 \to J/\psi K^- p$ Decays}},\ }\href {https://doi.org/10.1103/PhysRevLett.115.072001} {\bibfield  {journal} {\bibinfo  {journal} {Phys. Rev. Lett.}\ }\textbf {\bibinfo {volume} {115}},\ \bibinfo {pages} {072001} (\bibinfo {year} {2015})},\ \Eprint {https://arxiv.org/abs/1507.03414} {arXiv:1507.03414 [hep-ex]} \BibitemShut {NoStop}%
\bibitem [{\citenamefont {Aaij}\ \emph {et~al.}(2019)\citenamefont {Aaij} \emph {et~al.}}]{LHCb:2019kea}%
  \BibitemOpen
  \bibfield  {author} {\bibinfo {author} {\bibfnamefont {R.}~\bibnamefont {Aaij}} \emph {et~al.} (\bibinfo {collaboration} {LHCb}),\ }\bibfield  {title} {\bibinfo {title} {{Observation of a narrow pentaquark state, $P_c(4312)^+$, and of two-peak structure of the $P_c(4450)^+$}},\ }\href {https://doi.org/10.1103/PhysRevLett.122.222001} {\bibfield  {journal} {\bibinfo  {journal} {Phys. Rev. Lett.}\ }\textbf {\bibinfo {volume} {122}},\ \bibinfo {pages} {222001} (\bibinfo {year} {2019})},\ \Eprint {https://arxiv.org/abs/1904.03947} {arXiv:1904.03947 [hep-ex]} \BibitemShut {NoStop}%
\bibitem [{\citenamefont {Aaij}\ \emph {et~al.}(2020)\citenamefont {Aaij} \emph {et~al.}}]{LHCb:2020bwg}%
  \BibitemOpen
  \bibfield  {author} {\bibinfo {author} {\bibfnamefont {R.}~\bibnamefont {Aaij}} \emph {et~al.} (\bibinfo {collaboration} {LHCb}),\ }\bibfield  {title} {\bibinfo {title} {{Observation of structure in the $J /\psi$ -pair mass spectrum}},\ }\href {https://doi.org/10.1016/j.scib.2020.08.032} {\bibfield  {journal} {\bibinfo  {journal} {Sci. Bull.}\ }\textbf {\bibinfo {volume} {65}},\ \bibinfo {pages} {1983} (\bibinfo {year} {2020})},\ \Eprint {https://arxiv.org/abs/2006.16957} {arXiv:2006.16957 [hep-ex]} \BibitemShut {NoStop}%
\bibitem [{\citenamefont {Ablikim}\ \emph {et~al.}(2021)\citenamefont {Ablikim} \emph {et~al.}}]{BESIII:2020qkh}%
  \BibitemOpen
  \bibfield  {author} {\bibinfo {author} {\bibfnamefont {M.}~\bibnamefont {Ablikim}} \emph {et~al.} (\bibinfo {collaboration} {BESIII}),\ }\bibfield  {title} {\bibinfo {title} {{Observation of a Near-Threshold Structure in the $K^+$ Recoil-Mass Spectra in $e^+e^- \rightarrow K^+(D_s^-D^{*0}+D_s^{*-}D^0$)}},\ }\href {https://doi.org/10.1103/PhysRevLett.126.102001} {\bibfield  {journal} {\bibinfo  {journal} {Phys. Rev. Lett.}\ }\textbf {\bibinfo {volume} {126}},\ \bibinfo {pages} {102001} (\bibinfo {year} {2021})},\ \Eprint {https://arxiv.org/abs/2011.07855} {arXiv:2011.07855 [hep-ex]} \BibitemShut {NoStop}%
\bibitem [{\citenamefont {Aaij}\ \emph {et~al.}(2021)\citenamefont {Aaij} \emph {et~al.}}]{LHCb:2020jpq}%
  \BibitemOpen
  \bibfield  {author} {\bibinfo {author} {\bibfnamefont {R.}~\bibnamefont {Aaij}} \emph {et~al.} (\bibinfo {collaboration} {LHCb}),\ }\bibfield  {title} {\bibinfo {title} {{Evidence of a $J/\psi\Lambda$ structure and observation of excited $\Xi^-$ states in the $\Xi^-_b \to J/\psi\Lambda K^-$ decay}},\ }\href {https://doi.org/10.1016/j.scib.2021.02.030} {\bibfield  {journal} {\bibinfo  {journal} {Sci. Bull.}\ }\textbf {\bibinfo {volume} {66}},\ \bibinfo {pages} {1278} (\bibinfo {year} {2021})},\ \Eprint {https://arxiv.org/abs/2012.10380} {arXiv:2012.10380 [hep-ex]} \BibitemShut {NoStop}%
\bibitem [{\citenamefont {Aaij}\ \emph {et~al.}(2022{\natexlab{a}})\citenamefont {Aaij} \emph {et~al.}}]{LHCb:2021chn}%
  \BibitemOpen
  \bibfield  {author} {\bibinfo {author} {\bibfnamefont {R.}~\bibnamefont {Aaij}} \emph {et~al.} (\bibinfo {collaboration} {LHCb}),\ }\bibfield  {title} {\bibinfo {title} {{Evidence for a new structure in the $J/\psi p$ and $J/\psi \bar{p}$ systems in $B_s^0 \to J/\psi p \bar{p}$ decays}},\ }\href {https://doi.org/10.1103/PhysRevLett.128.062001} {\bibfield  {journal} {\bibinfo  {journal} {Phys. Rev. Lett.}\ }\textbf {\bibinfo {volume} {128}},\ \bibinfo {pages} {062001} (\bibinfo {year} {2022}{\natexlab{a}})},\ \Eprint {https://arxiv.org/abs/2108.04720} {arXiv:2108.04720 [hep-ex]} \BibitemShut {NoStop}%
\bibitem [{\citenamefont {Aaij}\ \emph {et~al.}(2022{\natexlab{b}})\citenamefont {Aaij} \emph {et~al.}}]{LHCb:2021vvq}%
  \BibitemOpen
  \bibfield  {author} {\bibinfo {author} {\bibfnamefont {R.}~\bibnamefont {Aaij}} \emph {et~al.} (\bibinfo {collaboration} {LHCb}),\ }\bibfield  {title} {\bibinfo {title} {{Observation of an exotic narrow doubly charmed tetraquark}},\ }\href {https://doi.org/10.1038/s41567-022-01614-y} {\bibfield  {journal} {\bibinfo  {journal} {Nature Phys.}\ }\textbf {\bibinfo {volume} {18}},\ \bibinfo {pages} {751} (\bibinfo {year} {2022}{\natexlab{b}})},\ \Eprint {https://arxiv.org/abs/2109.01038} {arXiv:2109.01038 [hep-ex]} \BibitemShut {NoStop}%
\bibitem [{\citenamefont {Aaij}\ \emph {et~al.}(2022{\natexlab{c}})\citenamefont {Aaij} \emph {et~al.}}]{LHCb:2021auc}%
  \BibitemOpen
  \bibfield  {author} {\bibinfo {author} {\bibfnamefont {R.}~\bibnamefont {Aaij}} \emph {et~al.} (\bibinfo {collaboration} {LHCb}),\ }\bibfield  {title} {\bibinfo {title} {{Study of the doubly charmed tetraquark $T_{cc}^{+}$}},\ }\href {https://doi.org/10.1038/s41467-022-30206-w} {\bibfield  {journal} {\bibinfo  {journal} {Nature Commun.}\ }\textbf {\bibinfo {volume} {13}},\ \bibinfo {pages} {3351} (\bibinfo {year} {2022}{\natexlab{c}})},\ \Eprint {https://arxiv.org/abs/2109.01056} {arXiv:2109.01056 [hep-ex]} \BibitemShut {NoStop}%
\bibitem [{\citenamefont {Ader}\ \emph {et~al.}(1982)\citenamefont {Ader}, \citenamefont {Richard},\ and\ \citenamefont {Taxil}}]{Ader:1981db}%
  \BibitemOpen
  \bibfield  {author} {\bibinfo {author} {\bibfnamefont {J.~P.}\ \bibnamefont {Ader}}, \bibinfo {author} {\bibfnamefont {J.~M.}\ \bibnamefont {Richard}},\ and\ \bibinfo {author} {\bibfnamefont {P.}~\bibnamefont {Taxil}},\ }\bibfield  {title} {\bibinfo {title} {{DO NARROW HEAVY MULTI - QUARK STATES EXIST?}},\ }\href {https://doi.org/10.1103/PhysRevD.25.2370} {\bibfield  {journal} {\bibinfo  {journal} {Phys. Rev. D}\ }\textbf {\bibinfo {volume} {25}},\ \bibinfo {pages} {2370} (\bibinfo {year} {1982})}\BibitemShut {NoStop}%
\bibitem [{\citenamefont {Ballot}\ and\ \citenamefont {Richard}(1983)}]{Ballot:1983iv}%
  \BibitemOpen
  \bibfield  {author} {\bibinfo {author} {\bibfnamefont {J.~l.}\ \bibnamefont {Ballot}}\ and\ \bibinfo {author} {\bibfnamefont {J.~M.}\ \bibnamefont {Richard}},\ }\bibfield  {title} {\bibinfo {title} {{FOUR QUARK STATES IN ADDITIVE POTENTIALS}},\ }\href {https://doi.org/10.1016/0370-2693(83)90991-7} {\bibfield  {journal} {\bibinfo  {journal} {Phys. Lett. B}\ }\textbf {\bibinfo {volume} {123}},\ \bibinfo {pages} {449} (\bibinfo {year} {1983})}\BibitemShut {NoStop}%
\bibitem [{\citenamefont {Zouzou}\ \emph {et~al.}(1986)\citenamefont {Zouzou}, \citenamefont {Silvestre-Brac}, \citenamefont {Gignoux},\ and\ \citenamefont {Richard}}]{Zouzou:1986qh}%
  \BibitemOpen
  \bibfield  {author} {\bibinfo {author} {\bibfnamefont {S.}~\bibnamefont {Zouzou}}, \bibinfo {author} {\bibfnamefont {B.}~\bibnamefont {Silvestre-Brac}}, \bibinfo {author} {\bibfnamefont {C.}~\bibnamefont {Gignoux}},\ and\ \bibinfo {author} {\bibfnamefont {J.~M.}\ \bibnamefont {Richard}},\ }\bibfield  {title} {\bibinfo {title} {{FOUR QUARK BOUND STATES}},\ }\href {https://doi.org/10.1007/BF01557611} {\bibfield  {journal} {\bibinfo  {journal} {Z. Phys. C}\ }\textbf {\bibinfo {volume} {30}},\ \bibinfo {pages} {457} (\bibinfo {year} {1986})}\BibitemShut {NoStop}%
\bibitem [{\citenamefont {Meng}\ \emph {et~al.}(2021)\citenamefont {Meng}, \citenamefont {Hiyama}, \citenamefont {Hosaka}, \citenamefont {Oka}, \citenamefont {Gubler}, \citenamefont {Can}, \citenamefont {Takahashi},\ and\ \citenamefont {Zong}}]{Meng:2020knc}%
  \BibitemOpen
  \bibfield  {author} {\bibinfo {author} {\bibfnamefont {Q.}~\bibnamefont {Meng}}, \bibinfo {author} {\bibfnamefont {E.}~\bibnamefont {Hiyama}}, \bibinfo {author} {\bibfnamefont {A.}~\bibnamefont {Hosaka}}, \bibinfo {author} {\bibfnamefont {M.}~\bibnamefont {Oka}}, \bibinfo {author} {\bibfnamefont {P.}~\bibnamefont {Gubler}}, \bibinfo {author} {\bibfnamefont {K.~U.}\ \bibnamefont {Can}}, \bibinfo {author} {\bibfnamefont {T.~T.}\ \bibnamefont {Takahashi}},\ and\ \bibinfo {author} {\bibfnamefont {H.~S.}\ \bibnamefont {Zong}},\ }\bibfield  {title} {\bibinfo {title} {{Stable double-heavy tetraquarks: spectrum and structure}},\ }\href {https://doi.org/10.1016/j.physletb.2021.136095} {\bibfield  {journal} {\bibinfo  {journal} {Phys. Lett. B}\ }\textbf {\bibinfo {volume} {814}},\ \bibinfo {pages} {136095} (\bibinfo {year} {2021})},\ \Eprint {https://arxiv.org/abs/2009.14493} {arXiv:2009.14493 [nucl-th]} \BibitemShut {NoStop}%
\bibitem [{\citenamefont {Meng}\ \emph {et~al.}(2022)\citenamefont {Meng}, \citenamefont {Harada}, \citenamefont {Hiyama}, \citenamefont {Hosaka},\ and\ \citenamefont {Oka}}]{Meng:2021yjr}%
  \BibitemOpen
  \bibfield  {author} {\bibinfo {author} {\bibfnamefont {Q.}~\bibnamefont {Meng}}, \bibinfo {author} {\bibfnamefont {M.}~\bibnamefont {Harada}}, \bibinfo {author} {\bibfnamefont {E.}~\bibnamefont {Hiyama}}, \bibinfo {author} {\bibfnamefont {A.}~\bibnamefont {Hosaka}},\ and\ \bibinfo {author} {\bibfnamefont {M.}~\bibnamefont {Oka}},\ }\bibfield  {title} {\bibinfo {title} {{Doubly heavy tetraquark resonant states}},\ }\href {https://doi.org/10.1016/j.physletb.2021.136800} {\bibfield  {journal} {\bibinfo  {journal} {Phys. Lett. B}\ }\textbf {\bibinfo {volume} {824}},\ \bibinfo {pages} {136800} (\bibinfo {year} {2022})},\ \Eprint {https://arxiv.org/abs/2106.11868} {arXiv:2106.11868 [hep-ph]} \BibitemShut {NoStop}%
\bibitem [{\citenamefont {Yan}\ \emph {et~al.}(2018)\citenamefont {Yan}, \citenamefont {Zhong},\ and\ \citenamefont {Zhu}}]{Yan:2018gik}%
  \BibitemOpen
  \bibfield  {author} {\bibinfo {author} {\bibfnamefont {X.}~\bibnamefont {Yan}}, \bibinfo {author} {\bibfnamefont {B.}~\bibnamefont {Zhong}},\ and\ \bibinfo {author} {\bibfnamefont {R.}~\bibnamefont {Zhu}},\ }\bibfield  {title} {\bibinfo {title} {{Doubly charmed tetraquarks in a diquark\textendash{}antidiquark model}},\ }\href {https://doi.org/10.1142/S0217751X18500963} {\bibfield  {journal} {\bibinfo  {journal} {Int. J. Mod. Phys. A}\ }\textbf {\bibinfo {volume} {33}},\ \bibinfo {pages} {1850096} (\bibinfo {year} {2018})},\ \Eprint {https://arxiv.org/abs/1804.06761} {arXiv:1804.06761 [hep-ph]} \BibitemShut {NoStop}%
\bibitem [{\citenamefont {Mutuk}(2024{\natexlab{a}})}]{Mutuk:2023oyz}%
  \BibitemOpen
  \bibfield  {author} {\bibinfo {author} {\bibfnamefont {H.}~\bibnamefont {Mutuk}},\ }\bibfield  {title} {\bibinfo {title} {{Masses and magnetic moments of doubly heavy tetraquarks via diffusion Monte Carlo method}},\ }\href {https://doi.org/10.1140/epjc/s10052-024-12736-3} {\bibfield  {journal} {\bibinfo  {journal} {Eur. Phys. J. C}\ }\textbf {\bibinfo {volume} {84}},\ \bibinfo {pages} {395} (\bibinfo {year} {2024}{\natexlab{a}})},\ \Eprint {https://arxiv.org/abs/2312.13383} {arXiv:2312.13383 [hep-ph]} \BibitemShut {NoStop}%
\bibitem [{\citenamefont {Mutuk}(2024{\natexlab{b}})}]{Mutuk:2024vzv}%
  \BibitemOpen
  \bibfield  {author} {\bibinfo {author} {\bibfnamefont {H.}~\bibnamefont {Mutuk}},\ }\bibfield  {title} {\bibinfo {title} {{Doubly-charged Tcc++ states in the dynamical diquark model}},\ }\href {https://doi.org/10.1103/PhysRevD.110.034025} {\bibfield  {journal} {\bibinfo  {journal} {Phys. Rev. D}\ }\textbf {\bibinfo {volume} {110}},\ \bibinfo {pages} {034025} (\bibinfo {year} {2024}{\natexlab{b}})},\ \Eprint {https://arxiv.org/abs/2401.02788} {arXiv:2401.02788 [hep-ph]} \BibitemShut {NoStop}%
\bibitem [{\citenamefont {Tornqvist}(1994)}]{Tornqvist:1993ng}%
  \BibitemOpen
  \bibfield  {author} {\bibinfo {author} {\bibfnamefont {N.~A.}\ \bibnamefont {Tornqvist}},\ }\bibfield  {title} {\bibinfo {title} {{From the deuteron to deusons, an analysis of deuteron - like meson meson bound states}},\ }\href {https://doi.org/10.1007/BF01413192} {\bibfield  {journal} {\bibinfo  {journal} {Z. Phys. C}\ }\textbf {\bibinfo {volume} {61}},\ \bibinfo {pages} {525} (\bibinfo {year} {1994})},\ \Eprint {https://arxiv.org/abs/hep-ph/9310247} {arXiv:hep-ph/9310247} \BibitemShut {NoStop}%
\bibitem [{\citenamefont {Ohkoda}\ \emph {et~al.}(2012)\citenamefont {Ohkoda}, \citenamefont {Yamaguchi}, \citenamefont {Yasui}, \citenamefont {Sudoh},\ and\ \citenamefont {Hosaka}}]{Ohkoda:2012hv}%
  \BibitemOpen
  \bibfield  {author} {\bibinfo {author} {\bibfnamefont {S.}~\bibnamefont {Ohkoda}}, \bibinfo {author} {\bibfnamefont {Y.}~\bibnamefont {Yamaguchi}}, \bibinfo {author} {\bibfnamefont {S.}~\bibnamefont {Yasui}}, \bibinfo {author} {\bibfnamefont {K.}~\bibnamefont {Sudoh}},\ and\ \bibinfo {author} {\bibfnamefont {A.}~\bibnamefont {Hosaka}},\ }\bibfield  {title} {\bibinfo {title} {{Exotic mesons with double charm and bottom flavor}},\ }\href {https://doi.org/10.1103/PhysRevD.86.034019} {\bibfield  {journal} {\bibinfo  {journal} {Phys. Rev. D}\ }\textbf {\bibinfo {volume} {86}},\ \bibinfo {pages} {034019} (\bibinfo {year} {2012})},\ \Eprint {https://arxiv.org/abs/1202.0760} {arXiv:1202.0760 [hep-ph]} \BibitemShut {NoStop}%
\bibitem [{\citenamefont {Li}\ \emph {et~al.}(2013)\citenamefont {Li}, \citenamefont {Sun}, \citenamefont {Liu},\ and\ \citenamefont {Zhu}}]{Li:2012ss}%
  \BibitemOpen
  \bibfield  {author} {\bibinfo {author} {\bibfnamefont {N.}~\bibnamefont {Li}}, \bibinfo {author} {\bibfnamefont {Z.-F.}\ \bibnamefont {Sun}}, \bibinfo {author} {\bibfnamefont {X.}~\bibnamefont {Liu}},\ and\ \bibinfo {author} {\bibfnamefont {S.-L.}\ \bibnamefont {Zhu}},\ }\bibfield  {title} {\bibinfo {title} {{Coupled-channel analysis of the possible $D^{(*)}D^{(*)}, \overline{B}^{(*)}\overline{B}^{(*)}$ and $D^{(*)}\overline{B}^{(*)}$ molecular states}},\ }\href {https://doi.org/10.1103/PhysRevD.88.114008} {\bibfield  {journal} {\bibinfo  {journal} {Phys. Rev. D}\ }\textbf {\bibinfo {volume} {88}},\ \bibinfo {pages} {114008} (\bibinfo {year} {2013})},\ \Eprint {https://arxiv.org/abs/1211.5007} {arXiv:1211.5007 [hep-ph]} \BibitemShut {NoStop}%
\bibitem [{\citenamefont {Chen}\ \emph {et~al.}(2023)\citenamefont {Chen}, \citenamefont {Chen}, \citenamefont {Liu}, \citenamefont {Liu},\ and\ \citenamefont {Zhu}}]{Chen:2022asf}%
  \BibitemOpen
  \bibfield  {author} {\bibinfo {author} {\bibfnamefont {H.-X.}\ \bibnamefont {Chen}}, \bibinfo {author} {\bibfnamefont {W.}~\bibnamefont {Chen}}, \bibinfo {author} {\bibfnamefont {X.}~\bibnamefont {Liu}}, \bibinfo {author} {\bibfnamefont {Y.-R.}\ \bibnamefont {Liu}},\ and\ \bibinfo {author} {\bibfnamefont {S.-L.}\ \bibnamefont {Zhu}},\ }\bibfield  {title} {\bibinfo {title} {{An updated review of the new hadron states}},\ }\href {https://doi.org/10.1088/1361-6633/aca3b6} {\bibfield  {journal} {\bibinfo  {journal} {Rept. Prog. Phys.}\ }\textbf {\bibinfo {volume} {86}},\ \bibinfo {pages} {026201} (\bibinfo {year} {2023})},\ \Eprint {https://arxiv.org/abs/2204.02649} {arXiv:2204.02649 [hep-ph]} \BibitemShut {NoStop}%
\bibitem [{\citenamefont {Wang}\ and\ \citenamefont {Liu}(2021)}]{Wang:2021yld}%
  \BibitemOpen
  \bibfield  {author} {\bibinfo {author} {\bibfnamefont {F.-L.}\ \bibnamefont {Wang}}\ and\ \bibinfo {author} {\bibfnamefont {X.}~\bibnamefont {Liu}},\ }\bibfield  {title} {\bibinfo {title} {{Investigating new type of doubly charmed molecular tetraquarks composed of charmed mesons in the H and T doublets}},\ }\href {https://doi.org/10.1103/PhysRevD.104.094030} {\bibfield  {journal} {\bibinfo  {journal} {Phys. Rev. D}\ }\textbf {\bibinfo {volume} {104}},\ \bibinfo {pages} {094030} (\bibinfo {year} {2021})},\ \Eprint {https://arxiv.org/abs/2108.09925} {arXiv:2108.09925 [hep-ph]} \BibitemShut {NoStop}%
\bibitem [{\citenamefont {Wang}\ \emph {et~al.}(2022)\citenamefont {Wang}, \citenamefont {Chen},\ and\ \citenamefont {Liu}}]{Wang:2021ajy}%
  \BibitemOpen
  \bibfield  {author} {\bibinfo {author} {\bibfnamefont {F.-L.}\ \bibnamefont {Wang}}, \bibinfo {author} {\bibfnamefont {R.}~\bibnamefont {Chen}},\ and\ \bibinfo {author} {\bibfnamefont {X.}~\bibnamefont {Liu}},\ }\bibfield  {title} {\bibinfo {title} {{A new group of doubly charmed molecule with T-doublet charmed meson pair}},\ }\href {https://doi.org/10.1016/j.physletb.2022.137502} {\bibfield  {journal} {\bibinfo  {journal} {Phys. Lett. B}\ }\textbf {\bibinfo {volume} {835}},\ \bibinfo {pages} {137502} (\bibinfo {year} {2022})},\ \Eprint {https://arxiv.org/abs/2111.00208} {arXiv:2111.00208 [hep-ph]} \BibitemShut {NoStop}%
\bibitem [{\citenamefont {Ren}\ \emph {et~al.}(2022)\citenamefont {Ren}, \citenamefont {Wu},\ and\ \citenamefont {Zhu}}]{Ren:2021dsi}%
  \BibitemOpen
  \bibfield  {author} {\bibinfo {author} {\bibfnamefont {H.}~\bibnamefont {Ren}}, \bibinfo {author} {\bibfnamefont {F.}~\bibnamefont {Wu}},\ and\ \bibinfo {author} {\bibfnamefont {R.}~\bibnamefont {Zhu}},\ }\bibfield  {title} {\bibinfo {title} {{Hadronic Molecule Interpretation of Tcc+ and Its Beauty Partners}},\ }\href {https://doi.org/10.1155/2022/9103031} {\bibfield  {journal} {\bibinfo  {journal} {Adv. High Energy Phys.}\ }\textbf {\bibinfo {volume} {2022}},\ \bibinfo {pages} {9103031} (\bibinfo {year} {2022})},\ \Eprint {https://arxiv.org/abs/2109.02531} {arXiv:2109.02531 [hep-ph]} \BibitemShut {NoStop}%
\bibitem [{\citenamefont {Asanuma}\ \emph {et~al.}(2024)\citenamefont {Asanuma}, \citenamefont {Yamaguchi},\ and\ \citenamefont {Harada}}]{Asanuma:2023atv}%
  \BibitemOpen
  \bibfield  {author} {\bibinfo {author} {\bibfnamefont {T.}~\bibnamefont {Asanuma}}, \bibinfo {author} {\bibfnamefont {Y.}~\bibnamefont {Yamaguchi}},\ and\ \bibinfo {author} {\bibfnamefont {M.}~\bibnamefont {Harada}},\ }\bibfield  {title} {\bibinfo {title} {{Analysis of DD* and D\textasciimacron{}(*)\ensuremath{\Xi}cc(*) molecule by one boson exchange model based on heavy quark symmetry}},\ }\href {https://doi.org/10.1103/PhysRevD.110.074030} {\bibfield  {journal} {\bibinfo  {journal} {Phys. Rev. D}\ }\textbf {\bibinfo {volume} {110}},\ \bibinfo {pages} {074030} (\bibinfo {year} {2024})},\ \Eprint {https://arxiv.org/abs/2311.04695} {arXiv:2311.04695 [hep-ph]} \BibitemShut {NoStop}%
\bibitem [{\citenamefont {Sakai}\ and\ \citenamefont {Yamaguchi}(2024)}]{Sakai:2023syt}%
  \BibitemOpen
  \bibfield  {author} {\bibinfo {author} {\bibfnamefont {M.}~\bibnamefont {Sakai}}\ and\ \bibinfo {author} {\bibfnamefont {Y.}~\bibnamefont {Yamaguchi}},\ }\bibfield  {title} {\bibinfo {title} {{Analysis of Tcc and Tbb based on the hadronic molecular model and their spin multiplets}},\ }\href {https://doi.org/10.1103/PhysRevD.109.054016} {\bibfield  {journal} {\bibinfo  {journal} {Phys. Rev. D}\ }\textbf {\bibinfo {volume} {109}},\ \bibinfo {pages} {054016} (\bibinfo {year} {2024})},\ \Eprint {https://arxiv.org/abs/2312.08663} {arXiv:2312.08663 [hep-ph]} \BibitemShut {NoStop}%
\bibitem [{\citenamefont {Ren}\ \emph {et~al.}(2024)\citenamefont {Ren}, \citenamefont {Wang}, \citenamefont {Yang},\ and\ \citenamefont {Wu}}]{Ren:2023pip}%
  \BibitemOpen
  \bibfield  {author} {\bibinfo {author} {\bibfnamefont {Y.-S.}\ \bibnamefont {Ren}}, \bibinfo {author} {\bibfnamefont {G.-J.}\ \bibnamefont {Wang}}, \bibinfo {author} {\bibfnamefont {Z.}~\bibnamefont {Yang}},\ and\ \bibinfo {author} {\bibfnamefont {J.-J.}\ \bibnamefont {Wu}},\ }\bibfield  {title} {\bibinfo {title} {{Investigation on the bottom analogs Tbb- of Tcc+}},\ }\href {https://doi.org/10.1103/PhysRevD.110.074007} {\bibfield  {journal} {\bibinfo  {journal} {Phys. Rev. D}\ }\textbf {\bibinfo {volume} {110}},\ \bibinfo {pages} {074007} (\bibinfo {year} {2024})},\ \Eprint {https://arxiv.org/abs/2310.09836} {arXiv:2310.09836 [hep-ph]} \BibitemShut {NoStop}%
\bibitem [{\citenamefont {Cheng}\ \emph {et~al.}(2022)\citenamefont {Cheng}, \citenamefont {Lin},\ and\ \citenamefont {Zhu}}]{Cheng:2022qcm}%
  \BibitemOpen
  \bibfield  {author} {\bibinfo {author} {\bibfnamefont {J.-B.}\ \bibnamefont {Cheng}}, \bibinfo {author} {\bibfnamefont {Z.-Y.}\ \bibnamefont {Lin}},\ and\ \bibinfo {author} {\bibfnamefont {S.-L.}\ \bibnamefont {Zhu}},\ }\bibfield  {title} {\bibinfo {title} {{Double-charm tetraquark under the complex scaling method}},\ }\href {https://doi.org/10.1103/PhysRevD.106.016012} {\bibfield  {journal} {\bibinfo  {journal} {Phys. Rev. D}\ }\textbf {\bibinfo {volume} {106}},\ \bibinfo {pages} {016012} (\bibinfo {year} {2022})},\ \Eprint {https://arxiv.org/abs/2205.13354} {arXiv:2205.13354 [hep-ph]} \BibitemShut {NoStop}%
\bibitem [{\citenamefont {Lin}\ \emph {et~al.}(2024)\citenamefont {Lin}, \citenamefont {Cheng},\ and\ \citenamefont {Zhu}}]{Lin:2022wmj}%
  \BibitemOpen
  \bibfield  {author} {\bibinfo {author} {\bibfnamefont {Z.-Y.}\ \bibnamefont {Lin}}, \bibinfo {author} {\bibfnamefont {J.-B.}\ \bibnamefont {Cheng}},\ and\ \bibinfo {author} {\bibfnamefont {S.-L.}\ \bibnamefont {Zhu}},\ }\bibfield  {title} {\bibinfo {title} {{Tcc+ and \ensuremath{\chi}c1(3872) with the complex scaling method and DD(D\textasciimacron{})\ensuremath{\pi} three-body effect}},\ }\href {https://doi.org/10.1103/PhysRevD.110.054008} {\bibfield  {journal} {\bibinfo  {journal} {Phys. Rev. D}\ }\textbf {\bibinfo {volume} {110}},\ \bibinfo {pages} {054008} (\bibinfo {year} {2024})},\ \Eprint {https://arxiv.org/abs/2205.14628} {arXiv:2205.14628 [hep-ph]} \BibitemShut {NoStop}%
\bibitem [{\citenamefont {Ikeda}\ \emph {et~al.}(2014)\citenamefont {Ikeda}, \citenamefont {Charron}, \citenamefont {Aoki}, \citenamefont {Doi}, \citenamefont {Hatsuda}, \citenamefont {Inoue}, \citenamefont {Ishii}, \citenamefont {Murano}, \citenamefont {Nemura},\ and\ \citenamefont {Sasaki}}]{Ikeda:2013vwa}%
  \BibitemOpen
  \bibfield  {author} {\bibinfo {author} {\bibfnamefont {Y.}~\bibnamefont {Ikeda}}, \bibinfo {author} {\bibfnamefont {B.}~\bibnamefont {Charron}}, \bibinfo {author} {\bibfnamefont {S.}~\bibnamefont {Aoki}}, \bibinfo {author} {\bibfnamefont {T.}~\bibnamefont {Doi}}, \bibinfo {author} {\bibfnamefont {T.}~\bibnamefont {Hatsuda}}, \bibinfo {author} {\bibfnamefont {T.}~\bibnamefont {Inoue}}, \bibinfo {author} {\bibfnamefont {N.}~\bibnamefont {Ishii}}, \bibinfo {author} {\bibfnamefont {K.}~\bibnamefont {Murano}}, \bibinfo {author} {\bibfnamefont {H.}~\bibnamefont {Nemura}},\ and\ \bibinfo {author} {\bibfnamefont {K.}~\bibnamefont {Sasaki}},\ }\bibfield  {title} {\bibinfo {title} {{Charmed tetraquarks $T_{cc}$ and $T_{cs}$ from dynamical lattice QCD simulations}},\ }\href {https://doi.org/10.1016/j.physletb.2014.01.002} {\bibfield  {journal} {\bibinfo  {journal} {Phys. Lett. B}\ }\textbf {\bibinfo {volume} {729}},\ \bibinfo {pages} {85} (\bibinfo {year} {2014})},\ \Eprint {https://arxiv.org/abs/1311.6214}
  {arXiv:1311.6214 [hep-lat]} \BibitemShut {NoStop}%
\bibitem [{\citenamefont {Padmanath}\ and\ \citenamefont {Prelovsek}(2022)}]{Padmanath:2022cvl}%
  \BibitemOpen
  \bibfield  {author} {\bibinfo {author} {\bibfnamefont {M.}~\bibnamefont {Padmanath}}\ and\ \bibinfo {author} {\bibfnamefont {S.}~\bibnamefont {Prelovsek}},\ }\bibfield  {title} {\bibinfo {title} {{Signature of a Doubly Charm Tetraquark Pole in DD* Scattering on the Lattice}},\ }\href {https://doi.org/10.1103/PhysRevLett.129.032002} {\bibfield  {journal} {\bibinfo  {journal} {Phys. Rev. Lett.}\ }\textbf {\bibinfo {volume} {129}},\ \bibinfo {pages} {032002} (\bibinfo {year} {2022})},\ \Eprint {https://arxiv.org/abs/2202.10110} {arXiv:2202.10110 [hep-lat]} \BibitemShut {NoStop}%
\bibitem [{\citenamefont {Lyu}\ \emph {et~al.}(2023)\citenamefont {Lyu}, \citenamefont {Aoki}, \citenamefont {Doi}, \citenamefont {Hatsuda}, \citenamefont {Ikeda},\ and\ \citenamefont {Meng}}]{Lyu:2023xro}%
  \BibitemOpen
  \bibfield  {author} {\bibinfo {author} {\bibfnamefont {Y.}~\bibnamefont {Lyu}}, \bibinfo {author} {\bibfnamefont {S.}~\bibnamefont {Aoki}}, \bibinfo {author} {\bibfnamefont {T.}~\bibnamefont {Doi}}, \bibinfo {author} {\bibfnamefont {T.}~\bibnamefont {Hatsuda}}, \bibinfo {author} {\bibfnamefont {Y.}~\bibnamefont {Ikeda}},\ and\ \bibinfo {author} {\bibfnamefont {J.}~\bibnamefont {Meng}},\ }\bibfield  {title} {\bibinfo {title} {{Doubly Charmed Tetraquark Tcc+ from Lattice QCD near Physical Point}},\ }\href {https://doi.org/10.1103/PhysRevLett.131.161901} {\bibfield  {journal} {\bibinfo  {journal} {Phys. Rev. Lett.}\ }\textbf {\bibinfo {volume} {131}},\ \bibinfo {pages} {161901} (\bibinfo {year} {2023})},\ \Eprint {https://arxiv.org/abs/2302.04505} {arXiv:2302.04505 [hep-lat]} \BibitemShut {NoStop}%
\bibitem [{\citenamefont {Eichten}\ and\ \citenamefont {Quigg}(2017)}]{Eichten:2017ffp}%
  \BibitemOpen
  \bibfield  {author} {\bibinfo {author} {\bibfnamefont {E.~J.}\ \bibnamefont {Eichten}}\ and\ \bibinfo {author} {\bibfnamefont {C.}~\bibnamefont {Quigg}},\ }\bibfield  {title} {\bibinfo {title} {{Heavy-quark symmetry implies stable heavy tetraquark mesons $Q_iQ_j \bar q_k \bar q_l$}},\ }\href {https://doi.org/10.1103/PhysRevLett.119.202002} {\bibfield  {journal} {\bibinfo  {journal} {Phys. Rev. Lett.}\ }\textbf {\bibinfo {volume} {119}},\ \bibinfo {pages} {202002} (\bibinfo {year} {2017})},\ \Eprint {https://arxiv.org/abs/1707.09575} {arXiv:1707.09575 [hep-ph]} \BibitemShut {NoStop}%
\bibitem [{\citenamefont {Cheng}\ \emph {et~al.}(2021)\citenamefont {Cheng}, \citenamefont {Li}, \citenamefont {Liu}, \citenamefont {Si},\ and\ \citenamefont {Yao}}]{Cheng:2020wxa}%
  \BibitemOpen
  \bibfield  {author} {\bibinfo {author} {\bibfnamefont {J.-B.}\ \bibnamefont {Cheng}}, \bibinfo {author} {\bibfnamefont {S.-Y.}\ \bibnamefont {Li}}, \bibinfo {author} {\bibfnamefont {Y.-R.}\ \bibnamefont {Liu}}, \bibinfo {author} {\bibfnamefont {Z.-G.}\ \bibnamefont {Si}},\ and\ \bibinfo {author} {\bibfnamefont {T.}~\bibnamefont {Yao}},\ }\bibfield  {title} {\bibinfo {title} {{Double-heavy tetraquark states with heavy diquark-antiquark symmetry}},\ }\href {https://doi.org/10.1088/1674-1137/abde2f} {\bibfield  {journal} {\bibinfo  {journal} {Chin. Phys. C}\ }\textbf {\bibinfo {volume} {45}},\ \bibinfo {pages} {043102} (\bibinfo {year} {2021})},\ \Eprint {https://arxiv.org/abs/2008.00737} {arXiv:2008.00737 [hep-ph]} \BibitemShut {NoStop}%
\bibitem [{\citenamefont {Tanaka}\ \emph {et~al.}(2024)\citenamefont {Tanaka}, \citenamefont {Yamaguchi},\ and\ \citenamefont {Harada}}]{Tanaka:2024siw}%
  \BibitemOpen
  \bibfield  {author} {\bibinfo {author} {\bibfnamefont {M.}~\bibnamefont {Tanaka}}, \bibinfo {author} {\bibfnamefont {Y.}~\bibnamefont {Yamaguchi}},\ and\ \bibinfo {author} {\bibfnamefont {M.}~\bibnamefont {Harada}},\ }\bibfield  {title} {\bibinfo {title} {{Mass and decay width of Tccs from symmetries}},\ }\href {https://doi.org/10.1103/PhysRevD.110.016024} {\bibfield  {journal} {\bibinfo  {journal} {Phys. Rev. D}\ }\textbf {\bibinfo {volume} {110}},\ \bibinfo {pages} {016024} (\bibinfo {year} {2024})},\ \Eprint {https://arxiv.org/abs/2403.03548} {arXiv:2403.03548 [hep-ph]} \BibitemShut {NoStop}%
\bibitem [{\citenamefont {Andreev}(2022)}]{Andreev:2021eyj}%
  \BibitemOpen
  \bibfield  {author} {\bibinfo {author} {\bibfnamefont {O.}~\bibnamefont {Andreev}},\ }\bibfield  {title} {\bibinfo {title} {{$QQ\bar{q}\bar{q}$ potential in string models}},\ }\href {https://doi.org/10.1103/PhysRevD.105.086025} {\bibfield  {journal} {\bibinfo  {journal} {Phys. Rev. D}\ }\textbf {\bibinfo {volume} {105}},\ \bibinfo {pages} {086025} (\bibinfo {year} {2022})},\ \Eprint {https://arxiv.org/abs/2111.14418} {arXiv:2111.14418 [hep-ph]} \BibitemShut {NoStop}%
\bibitem [{\citenamefont {Navarra}\ \emph {et~al.}(2007)\citenamefont {Navarra}, \citenamefont {Nielsen},\ and\ \citenamefont {Lee}}]{Navarra:2007yw}%
  \BibitemOpen
  \bibfield  {author} {\bibinfo {author} {\bibfnamefont {F.~S.}\ \bibnamefont {Navarra}}, \bibinfo {author} {\bibfnamefont {M.}~\bibnamefont {Nielsen}},\ and\ \bibinfo {author} {\bibfnamefont {S.~H.}\ \bibnamefont {Lee}},\ }\bibfield  {title} {\bibinfo {title} {{QCD sum rules study of QQ - anti-u anti-d mesons}},\ }\href {https://doi.org/10.1016/j.physletb.2007.04.010} {\bibfield  {journal} {\bibinfo  {journal} {Phys. Lett. B}\ }\textbf {\bibinfo {volume} {649}},\ \bibinfo {pages} {166} (\bibinfo {year} {2007})},\ \Eprint {https://arxiv.org/abs/hep-ph/0703071} {arXiv:hep-ph/0703071} \BibitemShut {NoStop}%
\bibitem [{\citenamefont {Du}\ \emph {et~al.}(2013)\citenamefont {Du}, \citenamefont {Chen}, \citenamefont {Chen},\ and\ \citenamefont {Zhu}}]{Du:2012wp}%
  \BibitemOpen
  \bibfield  {author} {\bibinfo {author} {\bibfnamefont {M.-L.}\ \bibnamefont {Du}}, \bibinfo {author} {\bibfnamefont {W.}~\bibnamefont {Chen}}, \bibinfo {author} {\bibfnamefont {X.-L.}\ \bibnamefont {Chen}},\ and\ \bibinfo {author} {\bibfnamefont {S.-L.}\ \bibnamefont {Zhu}},\ }\bibfield  {title} {\bibinfo {title} {{Exotic $QQ\bar{q}\bar{q}$, $QQ\bar{q}\bar{s}$ and $QQ\bar{s}\bar{s}$ states}},\ }\href {https://doi.org/10.1103/PhysRevD.87.014003} {\bibfield  {journal} {\bibinfo  {journal} {Phys. Rev. D}\ }\textbf {\bibinfo {volume} {87}},\ \bibinfo {pages} {014003} (\bibinfo {year} {2013})},\ \Eprint {https://arxiv.org/abs/1209.5134} {arXiv:1209.5134 [hep-ph]} \BibitemShut {NoStop}%
\bibitem [{\citenamefont {Agaev}\ \emph {et~al.}(2022)\citenamefont {Agaev}, \citenamefont {Azizi},\ and\ \citenamefont {Sundu}}]{Agaev:2021vur}%
  \BibitemOpen
  \bibfield  {author} {\bibinfo {author} {\bibfnamefont {S.~S.}\ \bibnamefont {Agaev}}, \bibinfo {author} {\bibfnamefont {K.}~\bibnamefont {Azizi}},\ and\ \bibinfo {author} {\bibfnamefont {H.}~\bibnamefont {Sundu}},\ }\bibfield  {title} {\bibinfo {title} {{Newly observed exotic doubly charmed meson Tcc+}},\ }\href {https://doi.org/10.1016/j.nuclphysb.2022.115650} {\bibfield  {journal} {\bibinfo  {journal} {Nucl. Phys. B}\ }\textbf {\bibinfo {volume} {975}},\ \bibinfo {pages} {115650} (\bibinfo {year} {2022})},\ \Eprint {https://arxiv.org/abs/2108.00188} {arXiv:2108.00188 [hep-ph]} \BibitemShut {NoStop}%
\bibitem [{\citenamefont {Xin}\ and\ \citenamefont {Wang}(2022)}]{Xin:2021wcr}%
  \BibitemOpen
  \bibfield  {author} {\bibinfo {author} {\bibfnamefont {Q.}~\bibnamefont {Xin}}\ and\ \bibinfo {author} {\bibfnamefont {Z.-G.}\ \bibnamefont {Wang}},\ }\bibfield  {title} {\bibinfo {title} {{Analysis of the doubly-charmed tetraquark molecular states with the QCD sum rules}},\ }\href {https://doi.org/10.1140/epja/s10050-022-00752-4} {\bibfield  {journal} {\bibinfo  {journal} {Eur. Phys. J. A}\ }\textbf {\bibinfo {volume} {58}},\ \bibinfo {pages} {110} (\bibinfo {year} {2022})},\ \Eprint {https://arxiv.org/abs/2108.12597} {arXiv:2108.12597 [hep-ph]} \BibitemShut {NoStop}%
\bibitem [{\citenamefont {Xin}\ \emph {et~al.}(2022)\citenamefont {Xin}, \citenamefont {Wang},\ and\ \citenamefont {Yang}}]{Xin:2022bzt}%
  \BibitemOpen
  \bibfield  {author} {\bibinfo {author} {\bibfnamefont {Q.}~\bibnamefont {Xin}}, \bibinfo {author} {\bibfnamefont {Z.-G.}\ \bibnamefont {Wang}},\ and\ \bibinfo {author} {\bibfnamefont {X.-S.}\ \bibnamefont {Yang}},\ }\bibfield  {title} {\bibinfo {title} {{Analysis of the X(3960) and related tetraquark molecular states via the QCD sum rules}},\ }\href {https://doi.org/10.1007/s43673-022-00070-3} {\bibfield  {journal} {\bibinfo  {journal} {AAPPS Bull.}\ }\textbf {\bibinfo {volume} {32}},\ \bibinfo {pages} {37} (\bibinfo {year} {2022})},\ \Eprint {https://arxiv.org/abs/2207.09910} {arXiv:2207.09910 [hep-ph]} \BibitemShut {NoStop}%
\bibitem [{\citenamefont {Neubert}(1994)}]{Neubert:1993mb}%
  \BibitemOpen
  \bibfield  {author} {\bibinfo {author} {\bibfnamefont {M.}~\bibnamefont {Neubert}},\ }\bibfield  {title} {\bibinfo {title} {{Heavy quark symmetry}},\ }\href {https://doi.org/10.1016/0370-1573(94)90091-4} {\bibfield  {journal} {\bibinfo  {journal} {Phys. Rept.}\ }\textbf {\bibinfo {volume} {245}},\ \bibinfo {pages} {259} (\bibinfo {year} {1994})},\ \Eprint {https://arxiv.org/abs/hep-ph/9306320} {arXiv:hep-ph/9306320} \BibitemShut {NoStop}%
\bibitem [{\citenamefont {Grinstein}(1995)}]{Grinstein:1995uv}%
  \BibitemOpen
  \bibfield  {author} {\bibinfo {author} {\bibfnamefont {B.}~\bibnamefont {Grinstein}},\ }\bibfield  {title} {\bibinfo {title} {{An Introduction to heavy mesons}},\ }in\ \href@noop {} {\emph {\bibinfo {booktitle} {{6th Mexican School of Particles and Fields}}}}\ (\bibinfo {year} {1995})\ pp.\ \bibinfo {pages} {122--184},\ \Eprint {https://arxiv.org/abs/hep-ph/9508227} {arXiv:hep-ph/9508227} \BibitemShut {NoStop}%
\bibitem [{\citenamefont {Casalbuoni}\ \emph {et~al.}(1997)\citenamefont {Casalbuoni}, \citenamefont {Deandrea}, \citenamefont {Di~Bartolomeo}, \citenamefont {Gatto}, \citenamefont {Feruglio},\ and\ \citenamefont {Nardulli}}]{Casalbuoni:1996pg}%
  \BibitemOpen
  \bibfield  {author} {\bibinfo {author} {\bibfnamefont {R.}~\bibnamefont {Casalbuoni}}, \bibinfo {author} {\bibfnamefont {A.}~\bibnamefont {Deandrea}}, \bibinfo {author} {\bibfnamefont {N.}~\bibnamefont {Di~Bartolomeo}}, \bibinfo {author} {\bibfnamefont {R.}~\bibnamefont {Gatto}}, \bibinfo {author} {\bibfnamefont {F.}~\bibnamefont {Feruglio}},\ and\ \bibinfo {author} {\bibfnamefont {G.}~\bibnamefont {Nardulli}},\ }\bibfield  {title} {\bibinfo {title} {{Phenomenology of heavy meson chiral Lagrangians}},\ }\href {https://doi.org/10.1016/S0370-1573(96)00027-0} {\bibfield  {journal} {\bibinfo  {journal} {Phys. Rept.}\ }\textbf {\bibinfo {volume} {281}},\ \bibinfo {pages} {145} (\bibinfo {year} {1997})},\ \Eprint {https://arxiv.org/abs/hep-ph/9605342} {arXiv:hep-ph/9605342} \BibitemShut {NoStop}%
\bibitem [{\citenamefont {Yasui}\ \emph {et~al.}(2013)\citenamefont {Yasui}, \citenamefont {Sudoh}, \citenamefont {Yamaguchi}, \citenamefont {Ohkoda}, \citenamefont {Hosaka},\ and\ \citenamefont {Hyodo}}]{Yasui:2013vca}%
  \BibitemOpen
  \bibfield  {author} {\bibinfo {author} {\bibfnamefont {S.}~\bibnamefont {Yasui}}, \bibinfo {author} {\bibfnamefont {K.}~\bibnamefont {Sudoh}}, \bibinfo {author} {\bibfnamefont {Y.}~\bibnamefont {Yamaguchi}}, \bibinfo {author} {\bibfnamefont {S.}~\bibnamefont {Ohkoda}}, \bibinfo {author} {\bibfnamefont {A.}~\bibnamefont {Hosaka}},\ and\ \bibinfo {author} {\bibfnamefont {T.}~\bibnamefont {Hyodo}},\ }\bibfield  {title} {\bibinfo {title} {{Spin degeneracy in multi-hadron systems with a heavy quark}},\ }\href {https://doi.org/10.1016/j.physletb.2013.10.019} {\bibfield  {journal} {\bibinfo  {journal} {Phys. Lett. B}\ }\textbf {\bibinfo {volume} {727}},\ \bibinfo {pages} {185} (\bibinfo {year} {2013})},\ \Eprint {https://arxiv.org/abs/1304.5293} {arXiv:1304.5293 [hep-ph]} \BibitemShut {NoStop}%
\bibitem [{\citenamefont {Yamaguchi}\ \emph {et~al.}(2015)\citenamefont {Yamaguchi}, \citenamefont {Ohkoda}, \citenamefont {Hosaka}, \citenamefont {Hyodo},\ and\ \citenamefont {Yasui}}]{Yamaguchi:2014era}%
  \BibitemOpen
  \bibfield  {author} {\bibinfo {author} {\bibfnamefont {Y.}~\bibnamefont {Yamaguchi}}, \bibinfo {author} {\bibfnamefont {S.}~\bibnamefont {Ohkoda}}, \bibinfo {author} {\bibfnamefont {A.}~\bibnamefont {Hosaka}}, \bibinfo {author} {\bibfnamefont {T.}~\bibnamefont {Hyodo}},\ and\ \bibinfo {author} {\bibfnamefont {S.}~\bibnamefont {Yasui}},\ }\bibfield  {title} {\bibinfo {title} {{Heavy quark symmetry in multihadron systems}},\ }\href {https://doi.org/10.1103/PhysRevD.91.034034} {\bibfield  {journal} {\bibinfo  {journal} {Phys. Rev. D}\ }\textbf {\bibinfo {volume} {91}},\ \bibinfo {pages} {034034} (\bibinfo {year} {2015})},\ \Eprint {https://arxiv.org/abs/1402.5222} {arXiv:1402.5222 [hep-ph]} \BibitemShut {NoStop}%
\bibitem [{\citenamefont {Shimizu}\ \emph {et~al.}(2018)\citenamefont {Shimizu}, \citenamefont {Yamaguchi},\ and\ \citenamefont {Harada}}]{Shimizu:2018ran}%
  \BibitemOpen
  \bibfield  {author} {\bibinfo {author} {\bibfnamefont {Y.}~\bibnamefont {Shimizu}}, \bibinfo {author} {\bibfnamefont {Y.}~\bibnamefont {Yamaguchi}},\ and\ \bibinfo {author} {\bibfnamefont {M.}~\bibnamefont {Harada}},\ }\bibfield  {title} {\bibinfo {title} {{Heavy quark spin multiplet structure of $\bar{P}^{(*)}\Sigma_Q^{(*)}$ molecular states}},\ }\href {https://doi.org/10.1103/PhysRevD.98.014021} {\bibfield  {journal} {\bibinfo  {journal} {Phys. Rev. D}\ }\textbf {\bibinfo {volume} {98}},\ \bibinfo {pages} {014021} (\bibinfo {year} {2018})},\ \Eprint {https://arxiv.org/abs/1805.05740} {arXiv:1805.05740 [hep-ph]} \BibitemShut {NoStop}%
\bibitem [{\citenamefont {Ahmed}\ \emph {et~al.}(2001)\citenamefont {Ahmed} \emph {et~al.}}]{CLEO:2001foe}%
  \BibitemOpen
  \bibfield  {author} {\bibinfo {author} {\bibfnamefont {S.}~\bibnamefont {Ahmed}} \emph {et~al.} (\bibinfo {collaboration} {CLEO}),\ }\bibfield  {title} {\bibinfo {title} {{First measurement of Gamma(D*+)}},\ }\href {https://doi.org/10.1103/PhysRevLett.87.251801} {\bibfield  {journal} {\bibinfo  {journal} {Phys. Rev. Lett.}\ }\textbf {\bibinfo {volume} {87}},\ \bibinfo {pages} {251801} (\bibinfo {year} {2001})},\ \Eprint {https://arxiv.org/abs/hep-ex/0108013} {arXiv:hep-ex/0108013} \BibitemShut {NoStop}%
\bibitem [{\citenamefont {Liu}\ \emph {et~al.}(2019)\citenamefont {Liu}, \citenamefont {Wu}, \citenamefont {Pavon~Valderrama}, \citenamefont {Xie},\ and\ \citenamefont {Geng}}]{Liu:2019stu}%
  \BibitemOpen
  \bibfield  {author} {\bibinfo {author} {\bibfnamefont {M.-Z.}\ \bibnamefont {Liu}}, \bibinfo {author} {\bibfnamefont {T.-W.}\ \bibnamefont {Wu}}, \bibinfo {author} {\bibfnamefont {M.}~\bibnamefont {Pavon~Valderrama}}, \bibinfo {author} {\bibfnamefont {J.-J.}\ \bibnamefont {Xie}},\ and\ \bibinfo {author} {\bibfnamefont {L.-S.}\ \bibnamefont {Geng}},\ }\bibfield  {title} {\bibinfo {title} {{Heavy-quark spin and flavor symmetry partners of the X(3872) revisited: What can we learn from the one boson exchange model?}},\ }\href {https://doi.org/10.1103/PhysRevD.99.094018} {\bibfield  {journal} {\bibinfo  {journal} {Phys. Rev. D}\ }\textbf {\bibinfo {volume} {99}},\ \bibinfo {pages} {094018} (\bibinfo {year} {2019})},\ \Eprint {https://arxiv.org/abs/1902.03044} {arXiv:1902.03044 [hep-ph]} \BibitemShut {NoStop}%
\bibitem [{\citenamefont {Isola}\ \emph {et~al.}(2003)\citenamefont {Isola}, \citenamefont {Ladisa}, \citenamefont {Nardulli},\ and\ \citenamefont {Santorelli}}]{Isola:2003fh}%
  \BibitemOpen
  \bibfield  {author} {\bibinfo {author} {\bibfnamefont {C.}~\bibnamefont {Isola}}, \bibinfo {author} {\bibfnamefont {M.}~\bibnamefont {Ladisa}}, \bibinfo {author} {\bibfnamefont {G.}~\bibnamefont {Nardulli}},\ and\ \bibinfo {author} {\bibfnamefont {P.}~\bibnamefont {Santorelli}},\ }\bibfield  {title} {\bibinfo {title} {{Charming penguins in B ---\ensuremath{>} K* pi, K(rho, omega, phi) decays}},\ }\href {https://doi.org/10.1103/PhysRevD.68.114001} {\bibfield  {journal} {\bibinfo  {journal} {Phys. Rev. D}\ }\textbf {\bibinfo {volume} {68}},\ \bibinfo {pages} {114001} (\bibinfo {year} {2003})},\ \Eprint {https://arxiv.org/abs/hep-ph/0307367} {arXiv:hep-ph/0307367} \BibitemShut {NoStop}%
\bibitem [{\citenamefont {Hiyama}\ \emph {et~al.}(2003)\citenamefont {Hiyama}, \citenamefont {Kino},\ and\ \citenamefont {Kamimura}}]{Hiyama:2003cu}%
  \BibitemOpen
  \bibfield  {author} {\bibinfo {author} {\bibfnamefont {E.}~\bibnamefont {Hiyama}}, \bibinfo {author} {\bibfnamefont {Y.}~\bibnamefont {Kino}},\ and\ \bibinfo {author} {\bibfnamefont {M.}~\bibnamefont {Kamimura}},\ }\bibfield  {title} {\bibinfo {title} {{Gaussian expansion method for few-body systems}},\ }\href {https://doi.org/10.1016/S0146-6410(03)90015-9} {\bibfield  {journal} {\bibinfo  {journal} {Prog. Part. Nucl. Phys.}\ }\textbf {\bibinfo {volume} {51}},\ \bibinfo {pages} {223} (\bibinfo {year} {2003})}\BibitemShut {NoStop}%
\bibitem [{\citenamefont {Hiyama}\ and\ \citenamefont {Kamimura}(2018)}]{Hiyama:2018ivm}%
  \BibitemOpen
  \bibfield  {author} {\bibinfo {author} {\bibfnamefont {E.}~\bibnamefont {Hiyama}}\ and\ \bibinfo {author} {\bibfnamefont {M.}~\bibnamefont {Kamimura}},\ }\bibfield  {title} {\bibinfo {title} {{Study of various few-body systems using Gaussian expansion method (GEM)}},\ }\href {https://doi.org/10.1007/s11467-018-0828-5} {\bibfield  {journal} {\bibinfo  {journal} {Front. Phys. (Beijing)}\ }\textbf {\bibinfo {volume} {13}},\ \bibinfo {pages} {132106} (\bibinfo {year} {2018})},\ \Eprint {https://arxiv.org/abs/1809.02619} {arXiv:1809.02619 [nucl-th]} \BibitemShut {NoStop}%
\bibitem [{\citenamefont {Suzuki}\ \emph {et~al.}(2005)\citenamefont {Suzuki}, \citenamefont {Myo},\ and\ \citenamefont {Kato}}]{Suzuki:2005wv}%
  \BibitemOpen
  \bibfield  {author} {\bibinfo {author} {\bibfnamefont {R.}~\bibnamefont {Suzuki}}, \bibinfo {author} {\bibfnamefont {T.}~\bibnamefont {Myo}},\ and\ \bibinfo {author} {\bibfnamefont {K.}~\bibnamefont {Kato}},\ }\bibfield  {title} {\bibinfo {title} {{Level density in complex scaling method}},\ }\href {https://doi.org/10.1063/1.1933001} {\bibfield  {journal} {\bibinfo  {journal} {AIP Conf. Proc.}\ }\textbf {\bibinfo {volume} {768}},\ \bibinfo {pages} {455} (\bibinfo {year} {2005})},\ \Eprint {https://arxiv.org/abs/nucl-th/0502012} {arXiv:nucl-th/0502012} \BibitemShut {NoStop}%
\bibitem [{\citenamefont {Myo}\ \emph {et~al.}(2014)\citenamefont {Myo}, \citenamefont {Kikuchi}, \citenamefont {Masui},\ and\ \citenamefont {Kat\={o}}}]{Myo:2014ypa}%
  \BibitemOpen
  \bibfield  {author} {\bibinfo {author} {\bibfnamefont {T.}~\bibnamefont {Myo}}, \bibinfo {author} {\bibfnamefont {Y.}~\bibnamefont {Kikuchi}}, \bibinfo {author} {\bibfnamefont {H.}~\bibnamefont {Masui}},\ and\ \bibinfo {author} {\bibfnamefont {K.}~\bibnamefont {Kat\={o}}},\ }\bibfield  {title} {\bibinfo {title} {{Recent development of complex scaling method for many-body resonances and continua in light nuclei}},\ }\href {https://doi.org/10.1016/j.ppnp.2014.08.001} {\bibfield  {journal} {\bibinfo  {journal} {Prog. Part. Nucl. Phys.}\ }\textbf {\bibinfo {volume} {79}},\ \bibinfo {pages} {1} (\bibinfo {year} {2014})},\ \Eprint {https://arxiv.org/abs/1410.4356} {arXiv:1410.4356 [nucl-th]} \BibitemShut {NoStop}%
\bibitem [{\citenamefont {Myo}\ and\ \citenamefont {Kato}(2020)}]{Myo:2020rni}%
  \BibitemOpen
  \bibfield  {author} {\bibinfo {author} {\bibfnamefont {T.}~\bibnamefont {Myo}}\ and\ \bibinfo {author} {\bibfnamefont {K.}~\bibnamefont {Kato}},\ }\bibfield  {title} {\bibinfo {title} {{Complex scaling: Physics of unbound light nuclei and perspective}},\ }\href {https://doi.org/10.1093/ptep/ptaa101} {\bibfield  {journal} {\bibinfo  {journal} {PTEP}\ }\textbf {\bibinfo {volume} {2020}},\ \bibinfo {pages} {12A101} (\bibinfo {year} {2020})},\ \Eprint {https://arxiv.org/abs/2007.12172} {arXiv:2007.12172 [nucl-th]} \BibitemShut {NoStop}%
\bibitem [{\citenamefont {Aguilar}\ and\ \citenamefont {Combes}(1971)}]{Aguilar:1971ve}%
  \BibitemOpen
  \bibfield  {author} {\bibinfo {author} {\bibfnamefont {J.}~\bibnamefont {Aguilar}}\ and\ \bibinfo {author} {\bibfnamefont {J.~M.}\ \bibnamefont {Combes}},\ }\bibfield  {title} {\bibinfo {title} {{A class of analytic perturbations for one-body schroedinger hamiltonians}},\ }\href {https://doi.org/10.1007/BF01877510} {\bibfield  {journal} {\bibinfo  {journal} {Commun. Math. Phys.}\ }\textbf {\bibinfo {volume} {22}},\ \bibinfo {pages} {269} (\bibinfo {year} {1971})}\BibitemShut {NoStop}%
\bibitem [{\citenamefont {Balslev}\ and\ \citenamefont {Combes}(1971)}]{Balslev:1971vb}%
  \BibitemOpen
  \bibfield  {author} {\bibinfo {author} {\bibfnamefont {E.}~\bibnamefont {Balslev}}\ and\ \bibinfo {author} {\bibfnamefont {J.~M.}\ \bibnamefont {Combes}},\ }\bibfield  {title} {\bibinfo {title} {{Spectral properties of many-body schroedinger operators with dilatation-analytic interactions}},\ }\href {https://doi.org/10.1007/BF01877511} {\bibfield  {journal} {\bibinfo  {journal} {Commun. Math. Phys.}\ }\textbf {\bibinfo {volume} {22}},\ \bibinfo {pages} {280} (\bibinfo {year} {1971})}\BibitemShut {NoStop}%
\bibitem [{\citenamefont {Taylor}(1972)}]{Taylor:1972pty}%
  \BibitemOpen
  \bibfield  {author} {\bibinfo {author} {\bibfnamefont {J.~R.}\ \bibnamefont {Taylor}},\ }\href@noop {} {\emph {\bibinfo {title} {{Scattering Theory: The Quantum Theory of Nonrelativistic Collisions}}}}\ (\bibinfo  {publisher} {John Wiley \& Sons, Inc.},\ \bibinfo {address} {New York},\ \bibinfo {year} {1972})\BibitemShut {NoStop}%
\bibitem [{\citenamefont {Masui}\ \emph {et~al.}(2000)\citenamefont {Masui}, \citenamefont {Aoyama}, \citenamefont {Myo}, \citenamefont {Kat\={o}},\ and\ \citenamefont {Ikeda}}]{Masui:2000mug}%
  \BibitemOpen
  \bibfield  {author} {\bibinfo {author} {\bibfnamefont {H.}~\bibnamefont {Masui}}, \bibinfo {author} {\bibfnamefont {S.}~\bibnamefont {Aoyama}}, \bibinfo {author} {\bibfnamefont {T.}~\bibnamefont {Myo}}, \bibinfo {author} {\bibfnamefont {K.}~\bibnamefont {Kat\={o}}},\ and\ \bibinfo {author} {\bibfnamefont {K.}~\bibnamefont {Ikeda}},\ }\bibfield  {title} {\bibinfo {title} {{Study of virtual states in 5 He and 10 Li with the Jost function method}},\ }\href {https://doi.org/10.1016/S0375-9474(00)00148-2} {\bibfield  {journal} {\bibinfo  {journal} {Nucl. Phys. A}\ }\textbf {\bibinfo {volume} {673}},\ \bibinfo {pages} {207} (\bibinfo {year} {2000})}\BibitemShut {NoStop}%
\end{thebibliography}%

\end{document}